\documentclass[]{aa} 
\usepackage{graphicx}
\usepackage{subfigure}
\usepackage[utf8]{inputenc}
\graphicspath{ {images/} }

\usepackage{subfigure}
\usepackage{txfonts}
\usepackage{booktabs}
\usepackage{url} 
\usepackage{amsmath}
\usepackage{enumitem}
\usepackage{multicol}
\usepackage{multirow}
\usepackage{graphicx}
\usepackage{subfigure}
\graphicspath{ {images/} }

\usepackage{txfonts}
\usepackage{tablefootnote}
\usepackage{lscape}
\usepackage{chngcntr}
\usepackage{chngcntr}
\usepackage{soul}
\usepackage{color}
\usepackage{rotating} 
\usepackage{ulem}
\newcommand{\kms}{km s$^{-1}$\xspace}
\newcommand{\HI}{{\rm H\,{\scriptsize I}}\xspace}
\newcommand{\HII}{{\rm H\,{\scriptsize II}}\xspace}
\newcommand{\ho}{$\rm H_{2}O$ }
\usepackage{natbib}

\usepackage[colorlinks, citecolor={blue}]{hyperref}

\begin{document}

\title{Radio continuum and \HI 21-cm line observations of a nearby luminous infrared galaxy IRAS 17526+3253 }

   \author{Jianfeng Wu
          \inst{1}
          \and
          Zhongzu Wu\inst{1}\fnmsep\thanks{zzwu08@gmail.com}
           \and
          Yulia Sotnikova,\inst{2}
           \and
         Bo Zhang\inst{3}
          \and
        Yongjun Chen\inst{3}
         \and
         Timur Mufakharov\inst{2,4}
          \and
         Zhiqiang Shen\inst{3}
        }

   \institute{College of Physics, Guizhou University, 550025 Guiyang, PR China \email{zzwu08@gmail.com}
              \and
Special Astrophysical Observatory of RAS, Nizhny Arkhyz 369167, Russia
              \and
              Shanghai Astronomical Observatory,
Chinese Academy of Sciences, 80 Nandan Road, Shanghai 200030, PR China \and
Kazan Federal University, 18 Kremlyovskaya St, Kazan 420008, Russia 
         }
   \date{  }


  \abstract
   {We present results from our EVN and GMRT observations of the radio continuum and spectral line emission in IRAS 17526+3253, along with an analysis of its arcsecond-scale radio properties using archival VLA data. The EVN observations detected radio continuum emission from both the northwest (NW) and southeast (SE) nuclei. The NW nucleus shows two components with high brightness temperatures and radio luminosities, likely indicating the presence of an AGN core and jet. Meanwhile, our EVN observation failed to detect the OH line emission, possibly due to radio frequency interference and/or the emission being partly resolved out and below our detection limit. The multi-band radio spectral energy distribution (SED) deviates from a single power-law at low frequencies, suggesting low-frequency absorption. The GMRT spectral line data reveal both \HI absorption and emission. The \HI emission is diffuse and shows a velocity gradient from about 7500 \kms in the NW to 7800 \kms in the SE nucleus. On larger scales, the \HI emission extends about 4' along the NW–SE direction, with the southeastern extension matching the optical tidal tail. In addition, the weak \HI absorption features show broad line profiles, possibly due to overlapping \HI gas from the two nuclei. The aforementioned results are consistent with properties of intermediate-stage mergers reported in the literature.  
   }

  \authorrunning{Wu et al. }            
   \titlerunning{radio properties of IRAS 17526+3253}  
   \maketitle
%
\section{Introduction}


The Infrared Astronomical Satellite \citep[IRAS,][]{1984ApJ...278L...1N} survey has revealed a class of (ultra)luminous infrared galaxies ([U]LIRGs), characterized by infrared luminosities (LIR) between $10^{11}$ and $10^{12}$ $L_{\odot}$ for LIRGs, and exceeding $10^{12}$ $L_{\odot}$ for ULIRGs. Observational evidence indicates that these galaxies often undergo strong tidal interactions and mergers, and they likely represent a critical evolutionary stage linking starburst galaxies with the AGN activity observed in QSOs and radio galaxies \citep{1996ARA&A..34..749S,2011AJ....141..100H}. During mergers, gas is funneled into the central few kiloparsecs, triggering intense star formation and/or fueling a supermassive black hole \citep{2006ApJS..163....1H}. (U)LIRGs are therefore essential for understanding the role of mergers in galaxy evolution and black hole growth \citep{2019MNRAS.486.3350S}.


IRAS 17526+3253 is a nearby merging system with a far-infrared (FIR) luminosity that qualifies it as a LIRG. This source has a redshift of z = 0.025, corresponding to a luminosity distance of approximately 108 Mpc. It contains northwestern (NW) and southeastern (SE) nuclei undergoing a mid-to-advanced stage major merger, with a projected separation of approximately 8.5 kpc \citep[see Fig. \ref{fig1} and ][]{2019MNRAS.486.3350S}. Notably, it is one of the few galaxies known to host both luminous OH and \ho masers \citep{2013A&A...560A..12W,2016ApJ...816...55W}, and it shows features consistent with the dual megamaser host scenario: it is an \HII-type galaxy and a merger with spatially distinct nuclei \citep{2016ApJ...816...55W}. It is generally accepted that extragalactic OH and \ho masers require different physical conditions for masing \citep{2005ARA&A..43..625L}. \cite{2019MNRAS.486.3350S} suggested that the OH and \ho maser emissions are associated with the NW and SE nuclei of IRAS 17526+3253, respectively. This implies that the two nuclei, which host these distinct maser types, likely have different physical properties. 


\cite{2019MNRAS.486.3350S} investigated the properties of IRAS 17526+3253 using multiwavelength observations. They found ongoing star formation throughout the envelope surrounding the two nuclei, supporting the interpretation that the radio emission originates from star-forming regions. They also identified two distinct velocity structures in the NW nucleus, which may indicate a distorted disk possibly mixed with tidal features from the SE nucleus. Tidally ejected disk material in merging galaxies is typically rich in neutral hydrogen and can extend to large distances, which are commonly observed in starburst galaxies \citep{2000AJ....119.1130H,1997RMxAC...6...55K}. \HI mapping is therefore well-suited to study the tidal tails and bridges formed during strong gravitational interactions. The angular separation (18") between the two nuclei of IRAS 17526+3253 makes it ideal for arcsecond-scale resolution observations, which are essential for examining the merger stage and kinematic properties of the system. Galaxies undergoing mergers or interactions are known to show a high detection rate of \HI absorption lines \citep[about 84\%, see][]{2018MNRAS.480..947D,2019MNRAS.489.1099D}. Thus, \HI mapping is a valuable tool for studying the environment of the merging system IRAS 17526+3253.
In general, the central regions of (U)LIRGs are obscured at most wavelengths due to large amounts of dust. As a result, the true nature of the dominant energy source in their centers has been the focus of extensive study over the years \citep{2015A&A...574A...4V}. IRAS 17526+3253 shows multiple dust lanes and is heavily obscured in the optical, suggesting that the visible optical features may not come from the nucleus itself, but rather trace the surrounding star-forming regions \citep[see][]{2019MNRAS.486.3350S}. Unlike optical and, to some extent, infrared observations, VLBI is unaffected by dust and can be used to detect compact AGN-related core–jet structures with high brightness temperatures, as well as diffuse, lower $T_{\mathrm{B}}$ emission from starburst activity \citep[see][]{2021A&ARv..29....2P}.

The primary aim of this paper is to investigate the presence of a radio AGN, examine the galaxy's environment through possible spectral line emission and absorption, and identify similarities with other well-studied LIRGs. Details of the radio data collection, reduction, and analysis are provided in Sect.\hyperref[data_collect]{\ref*{data_collect}}. The results and discussion are presented in Sects.\hyperref[sect:Obs]{\ref*{sect:Obs}} and \hyperref[sec:discussion]{\ref*{sec:discussion}}, respectively. Sect.~\hyperref[summary]{\ref*{summary}} summarizes the main findings and conclusions of this study. In this paper, we adopt the convention that the flux density follows $S_{\nu} \propto \nu^{\alpha}$, where $\alpha$ is the spectral index. Throughout this work, we assume a cosmology with $H_{0} = 73$ km\,s$^{-1}$\,Mpc$^{-1}$, $\Omega_{\mathrm{m}} = 0.27$, and $\Omega_{\Lambda} = 0.73$.

\section{Data collection, reduction and analysis}
\label{data_collect}
  \setlength{\tabcolsep}{0.05in}
  \begin{table*}
       \caption{Parameters of the high resolution spectral line observations. }
     \label{table1:linedata}
  \begin{center}
  \begin{tabular}{c c c c l c c c c c }     
  \hline\hline
   Observing Date & Frequencies & Line & Array     & Phase       & Program   & $\Delta_V$& Beam                   &  P.A.      & rms \\
                  &    (GHz)    &      &           & Calibrator  &           &  \kms     & (") $\times$(")        &  ($\circ$) & (mJy/beam) \\
   \hline
     2022Mar08   &  1.6         &   OH &   EVN     & J1748+3404  &  EW022A   & 11.9      &  0.006 $\times$0.005   & 57       &0.9  \\
     2023Sep10   &  1.4         &  \HI &   GMRT    & J1753+2848  &  44-048   & 10.8      &  4.14$\times$2.76      & -10      &0.5 \\
     2023Sep10   &  1.4         &  \HI &   GMRT    & J1753+2848  &  44-048   & 10.8      &  12.07$\times$7.27     & 1        &0.8 \\
     2023Sep10   &  1.4         &  \HI &   GMRT    & J1753+2848  &  44-048   & 10.8      &  32.47$\times$19.49    & 36       &1.0 \\
   \hline
   \end{tabular}
   \end{center}
   \vskip 0.1 true cm \noindent Column (3): The atomic or molecular lines. Column (4): The observational array. Column (5): The phase calibrator. Column (6): The program name. Column (7): The adopted spectral velocity resolution in \kms after binning. For the EVN and GMRT line projects, we binned every 2 channels, and for the EVLA OH line project, we binned 4 channels. Columns (8) and (9): The beam FWHM and position angle. Column (10): The 1$\sigma$ noise level for the channel image with the $\Delta_V$ listed in Column (7).
   \end{table*}

\setlength{\tabcolsep}{1pt}
 \begin{table*}
       \caption{EVN observations of the radio continuum observations. }
     \label{table2:evncontinuum}
  \begin{center}
\begin{tabular}{c c c c l c c c c c c c c c}     
  \hline\hline
    Epoch          & Freq. & Program   & beam             &$PA$&rms         & Comp. & Angular size &$PA$& $F_{int}$  & $F_{peak}$     & $Log T_{b}$ & $L_{r}$ \\
                   & (GHz) &           &(mas)&($\circ$)  &(mJy/b)  &       & (mas)        & ($\circ$)    & (mJy)      &(mJy/b)      & [K]         &($10^{27}erg/s/Hz$ )\\
      \hline
    2005Mar09      & 1.66  & EH010     &6.3$\times$5.2&68&0.065      & NW1    &  6.0$\times$1.7 &      101  &0.66 $\pm$ 0.11    & 0.44  $\pm$ 0.04  &7.63     &9.2\\
                   &       &           &                 &    &      & NW2    &  9.2$\times$3.8 &      55   &0.99 $\pm$ 0.18    & 0.45  $\pm$ 0.06  &7.28     &13.8\\
    2022Mar08      & 1.62  & EW022A    &5.5$\times$3.5&16&0.062      & NW1    &  4.9$\times$1.8 &      158  &0.34 $\pm$ 0.05    & 0.21  $\pm$ 0.02  &7.43     &4.7\\
                   &       &           &                 &    &      & NW2    &  4.9$\times$2.6 &      39  &0.42 $\pm$ 0.08    & 0.24  $\pm$ 0.03  &7.36     &5.8\\
                   &       &           &5.3$\times$3.7&21&0.060      & SE     &        -        &         -        &         -         &         <0.18     & -       &-\\ 
    2009Mar01      & 4.99  & EP065     &10.5$\times$4.9&49&0.027     & NW1    &  10.5$\times$5.6&      54&1.55 $\pm$ 0.05    & 1.31  $\pm$ 0.02  &6.28     &21.6\\
                   &       &           &                 &    &      & NW2    &  10.8$\times$5.9&      58&0.42 $\pm$ 0.04    & 0.33  $\pm$ 0.02  &5.68     &5.8\\
    2022Mar15      & 4.99  & EW022B    &1.9$\times$1.1&11&0.037      & NW1    &  2.8$\times$0.8&        98 &0.85 $\pm$ 0.13    & 0.29  $\pm$ 0.03  &7.45     &11.8\\
                   &       &           &                 &     &     & NW2    & 1.8$\times$1.2&      30&0.12 $\pm$ 0.02    & 0.12  $\pm$ 0.01  &6.82     &1.6\\
                   &       &           &1.9$\times$1.1&9 &0.035      & SE     & 1.9$\times$1.2&      29&0.19 $\pm$ 0.05    & 0.18  $\pm$ 0.03  &6.78      &2.6\\

  \hline
  \end{tabular}\\
  \end{center}
     \vskip 0.1 true cm \noindent Column (3): The program name. Columns (4) and (5): The beam FWHM and position angle. Column (6): The 1$\sigma$ noise level for the radio continuum image as shown in Fig. \ref{evnNW} and Fig. \ref{evnSE}. Column (7): The component name. Columns (8)-(11): Angular size and position angle, as well as the integrated and peak flux densities, obtained from Gaussian fitting using CASA software. Column (12) and (13): The logarithm of the brightness temperature and the radio luminosity of the components.
    \end{table*}

\subsection{Data collection}
In this work, we present EVN observations of IRAS 17526+3253, focusing on radio continuum and OH line emissions. Additionally, we include results from GMRT L- and P-band observations, which cover \HI line emission and radio continuum emission. Furthermore, we incorporate archival data from the VLA and EVN for additional radio continuum emission. Details of these observations and projects are provided below, with basic information summarized in Tables \ref{table1:linedata}, \ref{table2:evncontinuum}, and \ref{tablea1:multibandata}.

\subsubsection{EVN observations}
On March 8 and 15, 2022, we observed IRAS 17526+3253 in the L and C bands using the EVN array, with each observation lasting 8 hours. The L-band observation targeted both continuum and OH line emission, while the C-band observation focused solely on continuum studies. J1748+3404, located approximately 1$^{\circ}$ from IRAS 17526+3253, was used as the phase reference calibrator. The total observation time on IRAS 17526+3253 was 8 hours, with an on-source time of approximately 5–6 hours. The L-band data included 11 antennas: Westerbork (WB), Effelsberg (EF), Medicina (MC), Noto (NT), Onsala85 (ON85), Tianma65 (T6), Urumqi (UR), Torun (TR), Hartebeesthoek (HH), Irbene (IR), and Jodrell Bank2 (JB2). For the C-band data, the WB and IR antennas participated in the observations but did not provide effective data, while a new antenna, Yebes (YS), was included. As a result, the L-band data utilized 11 antennas, whereas the C-band data involved 10 antennas.

At L-band, we used a frequency range of 1616.2–1632.2 MHz with 512 channels, covering the expected frequency of the redshifted OH 1665 and 1667 MHz lines (around 1626 MHz), and providing a frequency resolution of approximately 31.2 kHz ($\sim$5.9 km/s). The radio continuum setup was instead composed of 8 $\times$ 16 MHz IFs, covering a total band between 1544 MHz and 1672 MHz. At C-band, the frequency range was 4910–5038 MHz, consisting of 4 $\times$ 32 MHz IFs. Both the L- and C-band continuum observations were recorded at a data rate of 1 Gbps.

The separation between the NW and SE nuclei is about seconds, which exceeds the estimated field of view of our observations. Therefore, multi-phase center correlation was performed for the L- and C-band EVN observations. This technique allowed the generation of two separate datasets from a single observation, each centered on the NW and SE nuclei, providing OH line data as well as L- and C-band radio continuum data for both nuclei.

\subsubsection{GMRT L and P band observation}

We observed IRAS 17526+3253 for six hours (280 minutes on-source) in Band 5 using the GMRT on September 10, 2023. The total bandwidth was 400 MHz, covering the range from 1060 to 1460 MHz. This bandwidth was divided into 16,384 channels, providing a frequency resolution of 24.4 kHz ($\sim$5.4 km/s). The selected band allows for the study of radio continuum emission and high-resolution redshifted \HI line profiles centered at 1385.2 MHz.
On September 11, 2023, we observed IRAS 17526+3253 with the GMRT for 50 minutes in Band 4 (approximately 25 minutes on-source). The total bandwidth was 200 MHz, spanning from 550 to 750 MHz, used for radio continuum studies. For both observations, the standard flux density calibrator 3C286 was observed at regular intervals to correct for amplitude and bandpass variations. Each observational scan lasted approximately six minutes and was used for flux density, delay, and bandpass calibration. The compact radio source J1748+3404 served as the phase calibrator. A summary of the GMRT observations is presented in Table \ref{table1:linedata} and \ref{tablea1:multibandata}.

\subsubsection{Archival data}
We also obtained archival EVN and VLA data for IRAS 17526+3253, covering both spectral line and radio continuum studies. Detailed information about these datasets is provided in Tables \ref{table1:linedata}, \ref{table2:evncontinuum}, and \ref{tablea1:multibandata}.

\subsection{Data Reduction}
\label{sect2}
For the EVN data, we used the NRAO Astronomical Image Processing System (AIPS) software package for calibration. The main steps in the EVN data reduction included importing online calibration tables: the CL2 table for a priori amplitude calibration and parallactic angle correction, a flagging table (FG1) and a bandpass table (BP1). Additional steps included ionospheric correction, data editing, instrumental phase calibration, fringe fitting with an antenna-based phase calibrator, and applying the solutions to the target source.

The historical VLA data (AB275, see Table \ref{tablea1:multibandata}) were calibrated using AIPS software following standard procedures. For the EVLA data (16B-063), we directly used the pipeline-calibrated data from the VLA archive. The GMRT data for \HI and P-band observations were calibrated using the Common Astronomy Software Application (CASA) \citep{2007ASPC..376..127M}. For \HI data calibration and continuum subtraction, we followed the GMRT online tutorial for data analysis \footnote{\href{http://www.ncra.tifr.res.in/~ruta/files/CASA_spectral-line_analysis.pdf}{GMRT Spectral Line Analysis}}. The calibration procedures for both P-band and L-band GMRT observations were similar. The data processing steps involved identifying and flagging bad data, then performing delay and bandpass calibration using flux calibrators. Next, we calibrated the phase and amplitude for the standard flux calibrators (3C286) and the phase calibrator, ensuring the flux density scale was transferred to the phase calibrator. After self-calibration of the flux and phase calibrators, additional manual flagging was done, and the dataset was recalibrated. Finally, the calibration tables were applied to both the calibrators and the target source.
For the L-band GMRT data, significant radio frequency interference (RFI) was present in the 1160 to 1260 MHz range. As a result, we generated two separate radio continuum datasets using two frequency bands: 1260–1460 MHz and 1060–1160 MHz. For the P-band GMRT observations, we similarly excluded the RFI-affected band and adopted a bandwidth of about 100 MHz, ranging from 650 MHz to 750 MHz.

For the calibrated OH line and radio continuum data, we imported them into the DIFMAP package \citep{1995BAAS...27..903S}  using natural weighting to create radio continuum and spectral line channel images. Due to the low signal-to-noise ratio (S/N) of the target sources, no self-calibration was applied to any continuum or line images. Since the Multi-band VLA and GMRT observations have different resolutions (see Fig. \ref{continuum-contour1}), we reprocessed the images with the same cell size and restored the beam to 5" × 5" (similar to the lowest-resolution GMRT P-band image). The radio continuum emission appears continuously distributed around both the NW and SE nuclei. Therefore, we measured the total flux densities by integrating the flux densities within circular regions of approximately 15" and 12" for the NW and SE nuclei, respectively, using the 'imstat' task in CASA (mean intensity multiplied by the number of beam areas). The results are summarized in Table \ref{tablea1:multibandata}.

The calibrated \HI data were imaged using the 'TCLEAN' task in CASA with the following key parameters: 'weighting=natural' and 'specmode=cube'. Three different cell sizes were adopted, as the \HI emission is diffusely distributed on a large scale in this galaxy. The 0.5"/pix and 4"/pix images were used to study the spatial distributions of \HI absorption and emission, respectively, while the 10"/pix image was used to analyze the total \HI emission line profile. This process produced three naturally weighted \HI images  with different beam sizes (see Table \ref{table1:linedata}, Figs. \ref{HI-FS+XS} and \ref{HItotal}): high-resolution (4" $\times$ 3"), medium-resolution (12" $\times$ 7"), and low-resolution (33" $\times$ 20"). 

To further investigate the velocity structure in these \HI images, we employed the SoFiA software package \citep{2015MNRAS.448.1922S,2021MNRAS.506.3962W}, following the official online tutorial\footnote{\href{https://gitlab.com/SoFiA-Admin/SoFiA-2/-/wikis/SoFiA-Tutorial}{SoFiA-Tutorial}}. Three key parameters were configured. First, the spatial smoothing filter sizes (\texttt{scfind.kernelsXY}) were set based on the number of pixels per beam, using scales of 0, 1, 2, 3, etc., times the beam FWHM (i.e., 0, 3, 6, ... pixels). Second, the spectral smoothing filter sizes (\texttt{scfind.kernelsZ}) were chosen according to the typical widths of the \HI emission and absorption line profiles, with additional values increasing logarithmically (e.g., 0, 3, 7, 15, 31, ...). Third, the source detection threshold (\texttt{scfind.threshold}) was set to 4, which lies within the recommended range of 3.5 to 4.5 and corresponds to approximately four times the noise level, typically yielding reliable results.  Using these settings, SoFiA identified significant signal regions in each cube and generated three moment maps: moment 0 (total intensity), moment 1 (velocity centroid), and moment 2 (velocity dispersion). Meanwhile, we also enabled the output parameter "output.writePV = True", which produced two position–velocity (PV) diagrams for each \HI image, extracted along the identified kinematic major and minor axes, respectively.

\section{Results}
\label{sect:Obs}

\subsection{EVN observations of the radio continuum emission and OH line emission}
 
We present the L- and C-band EVN images of the NW and SE nuclei in Fig. \ref{evnNW} and Fig. \ref{evnSE}, respectively. The results show successful detection of radio continuum emission from both nuclei at L and C bands, with data from two epochs. The NW nucleus contains two components (NW1 and NW2, see Fig. \ref{evnNW}) in both L- and C-band observations. The SE nucleus contains only one component, detected in the C-band EVN image, while the L-band observations reached a noise level of approximately 0.06 mJy/beam. The parameters of these components are presented in Table \ref{table2:evncontinuum}.

The L- and C-band EVN archive data (EH010 and EP065) only provide images of the NW nucleus, as these data are centered on the NW nucleus, and the SE nucleus falls outside their field of view. The two L-band observations show that both the NW1 and NW2 components have a brightness temperature Log $T_b$ of > 7. The high-resolution C-band project (EW022B) also measures a high $T_b$ for the NW1 component, around 7.5, while the C-band observation gives a $T_b$ of about 6.8 for NW2 and SE nucleus.

Since the L- and C-band EVN observations (EW022A and EW022B) were separated by only about 7 days, the flux densities of the components are unlikely to be affected by intrinsic source variations. The results show that the NW1 component of the NW nucleus exhibits an inverted spectrum, likely due to low-frequency absorption, as the L-band flux density is significantly lower than the C-band flux density. We have restored the L- and C-band images to the same cell size and beam size. The resulting spectral index of the integrated flux for NW1 and NW2, $\alpha^{NW1}_{\mathrm{int}}$= $0.49 \pm 0.16$ -- consistent with $\alpha_{\mathrm{int}} > 0$ and indicating a inverted spectra index, $\alpha^{NW2}_{\mathrm{int}}$= $-0.39 \pm 0.29$, a possible flat spectral index \citep[see][]{2024ApJ...961..109S}. The spectral index and its uncertainty were calculated using equations from \cite{2019RAA....19...96T}:

\begin{equation}
\alpha = \frac{\ln(F_2 / F_1)}{\ln(\nu_2 / \nu_1)}
\end{equation}

\begin{equation}
\sigma_\alpha = \frac{1}{\ln(\nu_2 / \nu_1)} \sqrt{ \left( \frac{\sigma_{F_1}}{F_1} \right)^2 + \left( \frac{\sigma_{F_2}}{F_2} \right)^2 }
\end{equation}
,where $\sigma_{F_1}$ and $\sigma_{F_2}$ denote the flux density uncertainties, with $\nu_1$, $\nu_2$ and $F_1$, $F_2$ representing the L-band and C-band frequencies and flux densities, respectively. Similarly, we restored the C-band image for the SE nucleus to the same cellsize and beam size as the L-band image. This yielded a peak flux density of $\sim$0.28 mJy/beam at C-band. Since the L-band observation resulted in a non-detection (3$\sigma$ upper limit: 0.18 mJy/beam), we derived $\alpha^{\mathrm{SE}}_{\mathrm{peak}} > 0.38$, indicating possible spectral inversion \citep{2024ApJ...961..109S}.

For the EVN observations of the OH line emission, we binned every two channels, resulting in a spectral resolution of approximately 61 kHz ($\sim$12 \kms). The basic observational setup is summarized in Table \ref{table1:linedata}. No significant OH line emission was detected, even after additional binning to further reduce the noise level (see Figs. \ref{EVNOHdirty} and \ref{EVNOHline}). The OH emission from IRAS 17526+3253 was likely affected by strong radio frequency interference (RFI), particularly from the Iridium satellite \citep[see][for details]{2013ApJ...763....8M}. In our data, the noise level in RFI-free channel images ($\sim$12 km/s width) is about 0.9 mJy/beam, while mildly RFI-affected channels show noise levels ranging from 1.4 mJy/beam to 1.7 mJy/beam. Some heavily RFI-contaminated channels reach noise levels as high as 2–5 mJy/beam (see Fig. \ref{EVNOHline}).

 \begin{figure}
   \centering
    \includegraphics[trim=80 0 100 45, clip, width=9cm, height=6cm]{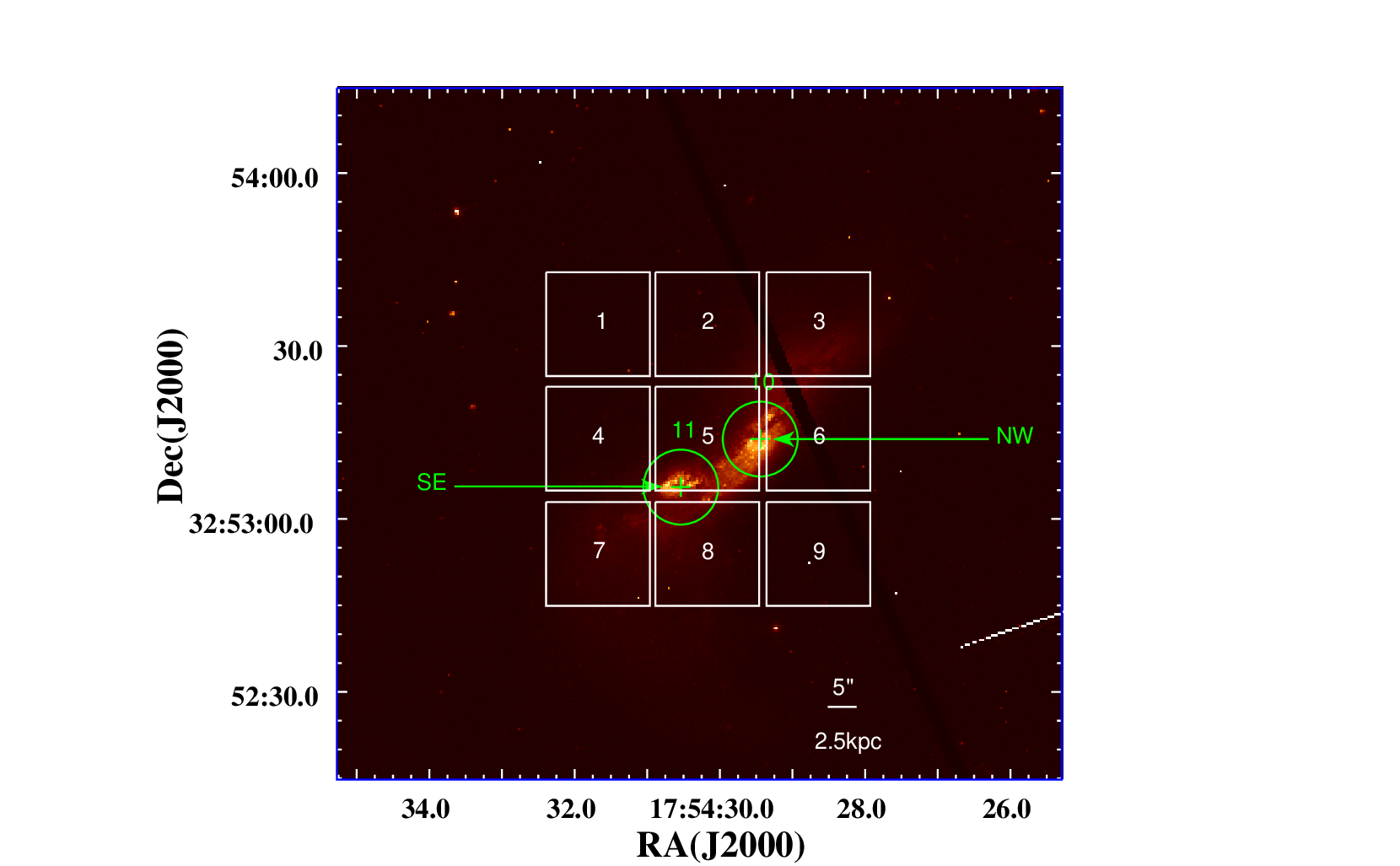}
      \caption{HST-ACS F814W(I) image of IRAS 17526+3253. The white boxes represent the nine regions where the \HI emission lines were extracted, each with a size of 18" $\times$ 18". The green circles indicate the two regions where the \HI emission lines were obtained, each with a size of 13" $\times$ 13".}
    \label{fig1}
\end{figure}
\subsection{The \HI emission and absorption}
\label{result3}

We analyzed the \HI images at different resolutions (see Section \ref{sect2}) and detected both \HI emission and absorption in IRAS 17526+3253. The fitted parameters for the line profiles in these regions are presented in Table \ref{tablea2-regionsHI}.

For the medium-resolution \HI image, we selected nine square regions and two circular regions (regions 1-11, see Fig. \ref{fig1}) to extract the \HI line profiles. \HI emission was detected in regions 2-8 and 11 (see Fig. \ref{NS-HIFS}). The northwest regions (2, 3, and 6) show the lowest peak velocity around 7500 \kms, while the southern regions (7 and 8) have peak velocities above 7800 \kms. The middle regions (4 and 5) show peak velocities around 7700 \kms (see Table \ref{tablea2-regionsHI} and Fig. \ref{NS-HIFS}). Regions 5 and 6 also exhibit redshifted \HI absorption, while region 8 shows weak blueshifted emission. The circular region (10) around the NW nucleus has no detectable emission or absorption, likely due to overlapping \HI emission and absorption in the same area, as absorption was detected in the high-resolution \HI image. The circular region (11) around the SE nucleus shows \HI emission peaking at 7753 \kms. The low-resolution \HI image allowed us to extract the total \HI emission (region 12), shown in Fig. \ref{HItotal}. The emission spans an elongated region less than 2 arcminutes in size, with two peaks in the line profile at 7495 and 7806 \kms. 
In the high-resolution \HI image, \HI absorption was detected toward both the NW and SE nuclei in elliptical regions (13 and 14, see Figs. \ref{HI-FS+XS} and \ref{HIabs_line}). Region 13 shows two peaks at 7631 and 7818 \kms, while region 14 presents a broad absorption profile with peaks at 7636 \kms and an FWHM of around 1353 \kms (see Table \ref{tablea2-regionsHI}).

The SoFiA software identifies a region in the \HI absorption image that is consistent with our moment map created by integrating channel maps over selected velocity ranges (see Figs. \ref{momentxs} and \ref{HI-FS+XS}). The velocity map (Fig. \ref{momentxs}) suggests that the NW nucleus may contain two distinct velocity structures: NWa, with a lower velocity, and NWb, with a higher velocity. The \HI absorption spectra for NWa, NWb, and the other selection regions are shown in Fig. \ref{HIabs_line2}, and their fitted parameters are listed in Table \ref{tablea2-regionsHI}. NWa and NWb exhibit a velocity offset of approximately 380 \kms. The naming convention follows \cite{2019MNRAS.486.3350S}, who also reported a similar velocity offset.  

The \HI emission region identified by SoFiA extends in both the SE and NW directions, with a total extent of approximately $4'$ (see Figs.\ref{momentfs}, \ref{momentfs2}, and \ref{desiimg}). This is nearly twice the extent of the optical emission region seen in the HST image (see Fig.\ref{HItotal}). However, the extension toward the southeast is consistent with the optical structure (likely to be a tidal tail) visible in DESI optical image (see Fig.\ref{desiimg} and the online color image \footnote{\href{https://www.legacysurvey.org/viewer/desi-edr/?ra=268.62255&dec=32.88742&zoom=14}{DESI color image (grz-band) of IRAS17526+3253}}). The extension toward the northwest includes a region of about $0.6'$ without an obvious optical counterpart in the DESI image, and may require further confirmation. The SoFiA-derived velocity centroid and PV diagram maps reveal two distinct velocity components, likely separated by a boundary between the NW and SE nuclei. The NW nucleus exhibits a velocity of $\sim7500$ \kms, while the SE nucleus peaks around $\sim7800$ \kms (see Figs.~\ref{momentfs}, \ref{momentfs2} and \ref{pvdiagram}). The \HI emission at a velocity of $\sim7800$ \kms, primarily from the SE galaxy, is over three times stronger than that at $\sim7500$ \kms, which mainly originates from the NW galaxy (see Fig. \ref{HItotal}). Similar results can also be seen in the \HI emission intensity map generated by the SoFiA software, as shown in Fig. \ref{desiimg}.

\begin{figure*}
   \centering
\includegraphics[width=9cm,height=9cm]{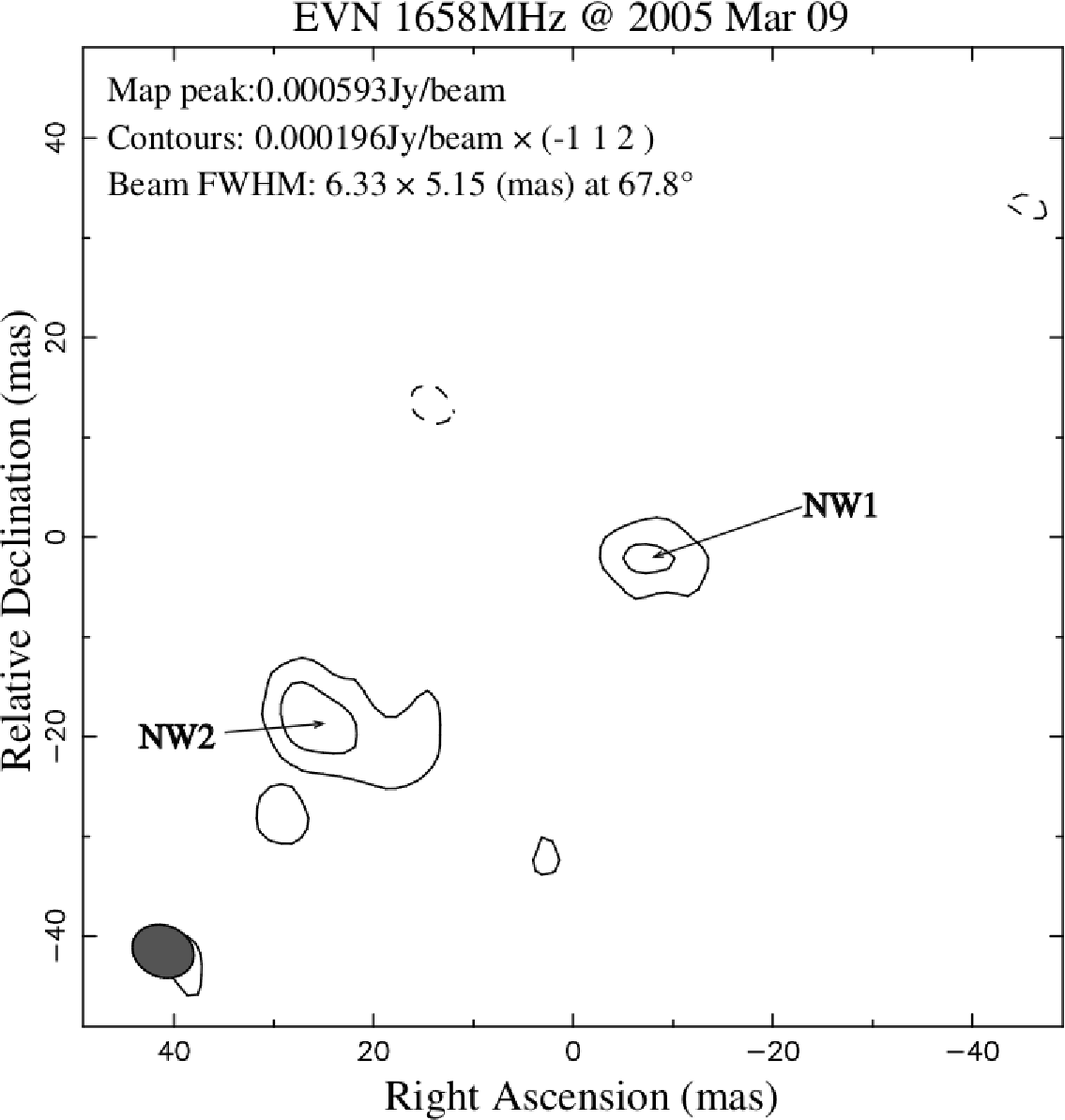}
\includegraphics[width=9cm,height=9cm]{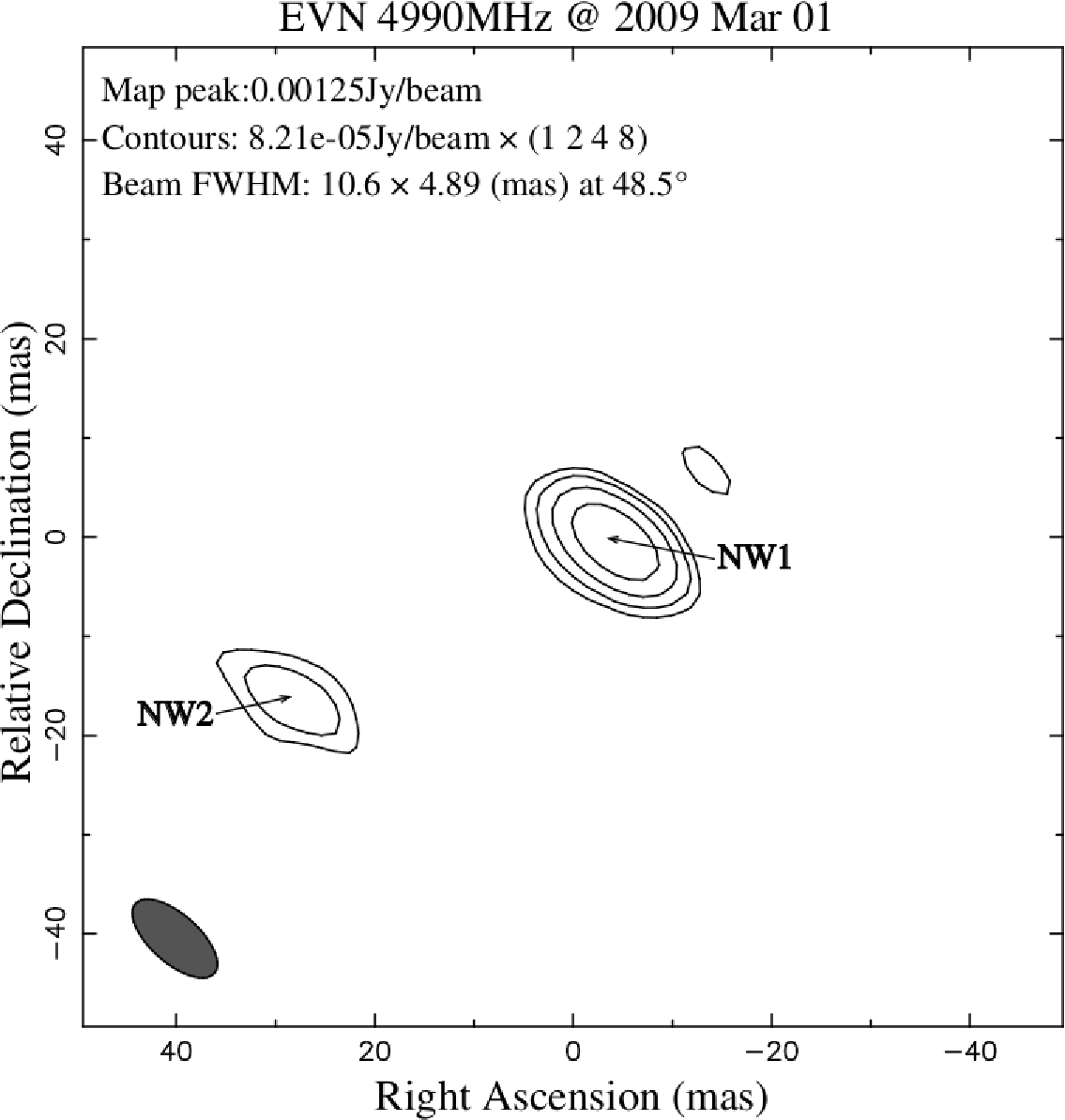}
\includegraphics[width=9cm,height=9cm]{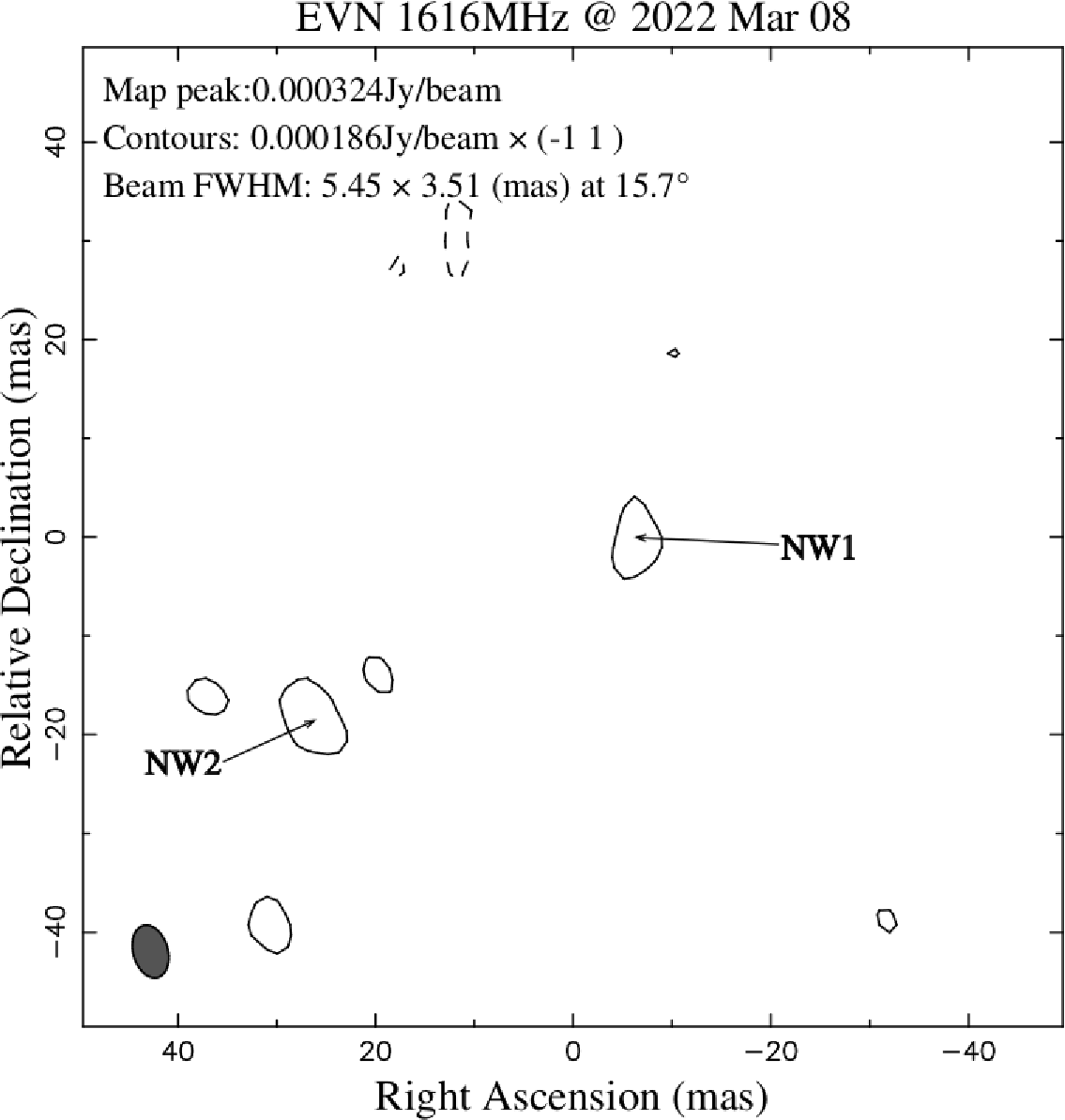}
\includegraphics[width=9cm,height=9cm]{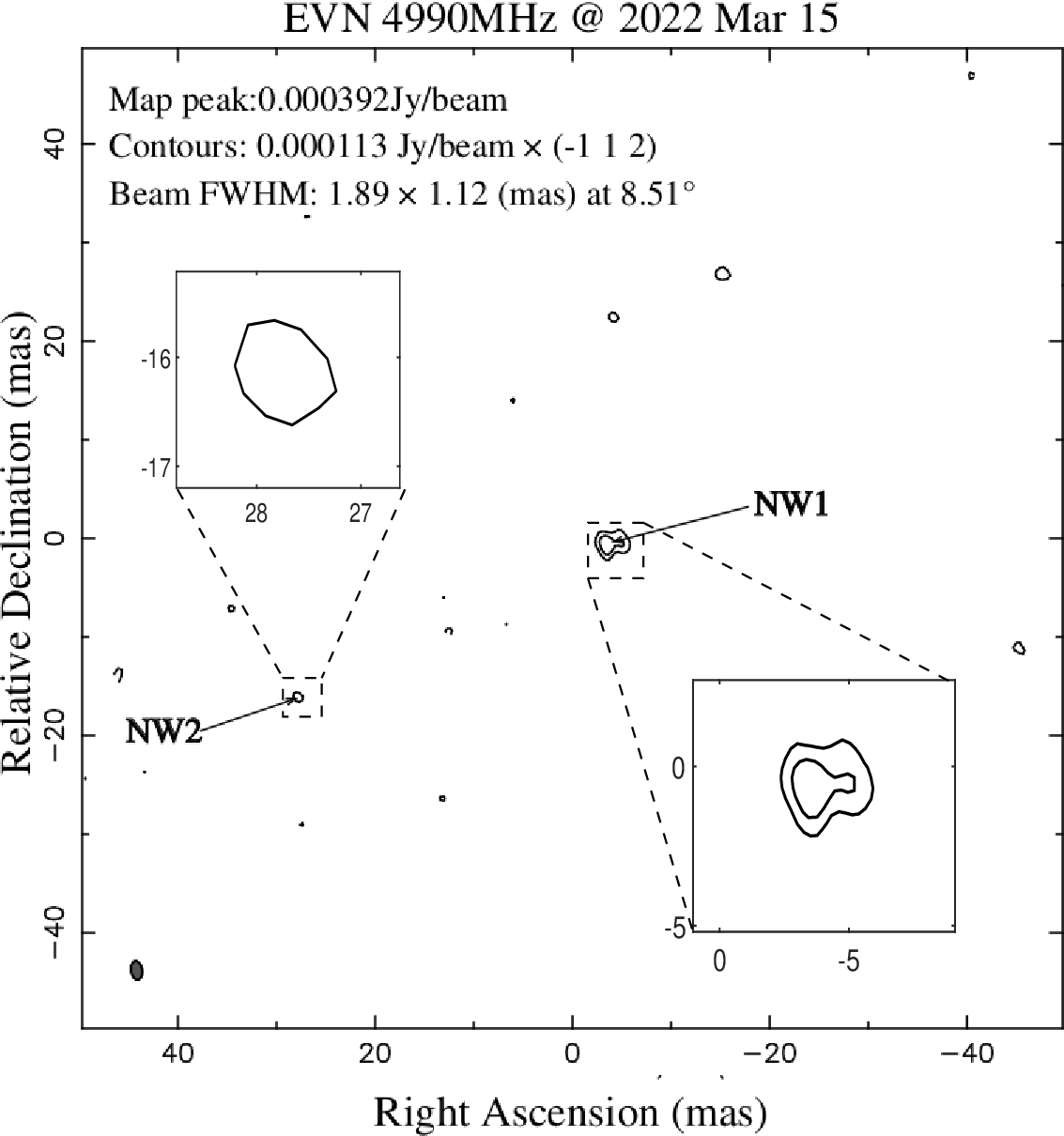}
      \caption{EVN L- and C-band radio continuum images of the NW nucleus of IRAS 17526+3253 using natural weighting. The plots are centered at R.A. = 17:54:29.411, Dec = 32:53:14.696. The map peak intensity, contour levels, and beam FWHM are labeled in each image. The first contour in all images corresponds to a SNR of approximately 3 (3$\sigma$). }
      \label{evnNW}%
\end{figure*}

\begin{figure*}
   \centering
\includegraphics[width=9cm,height=9cm]{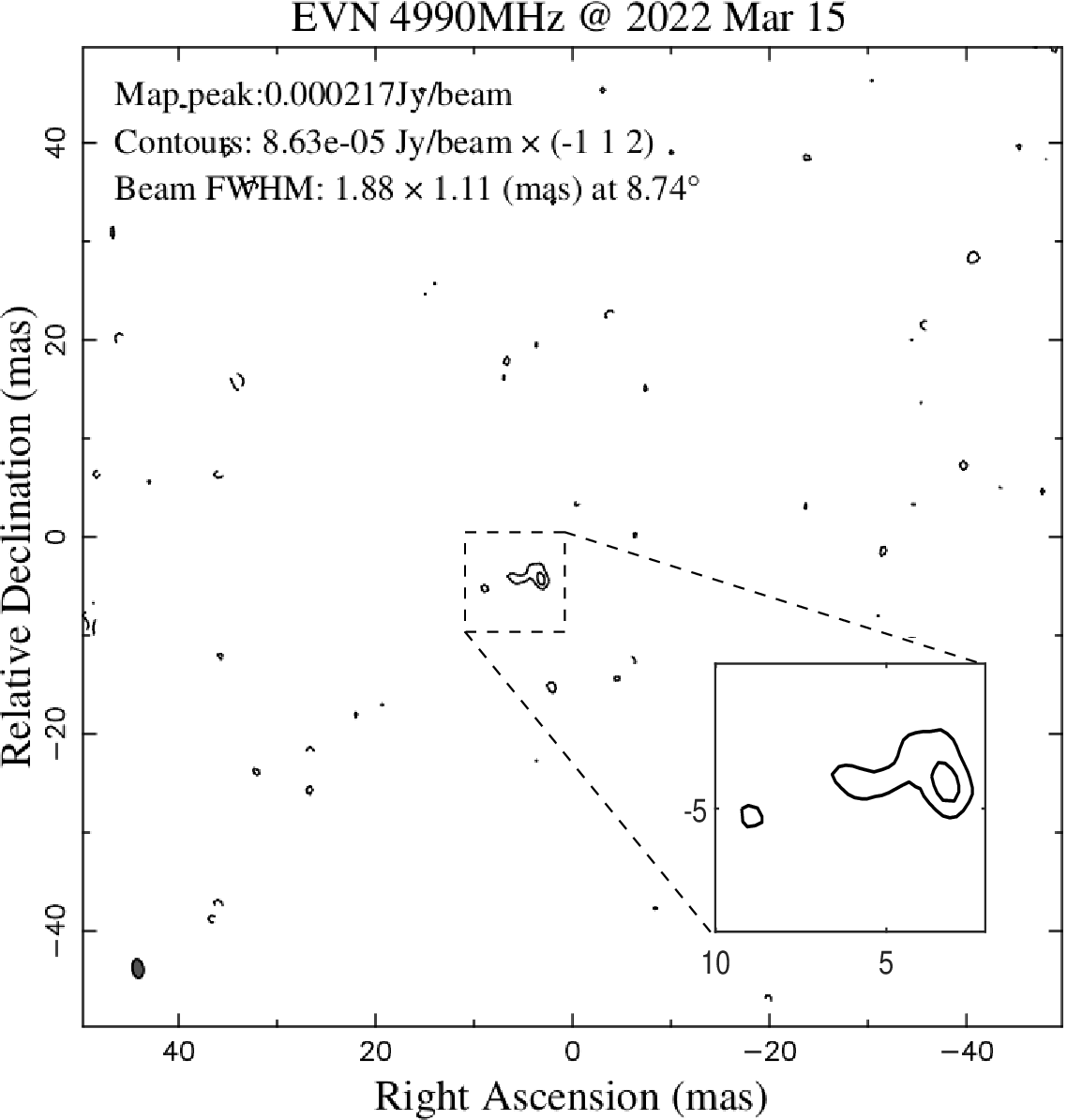}
\includegraphics[width=9cm,height=9cm]{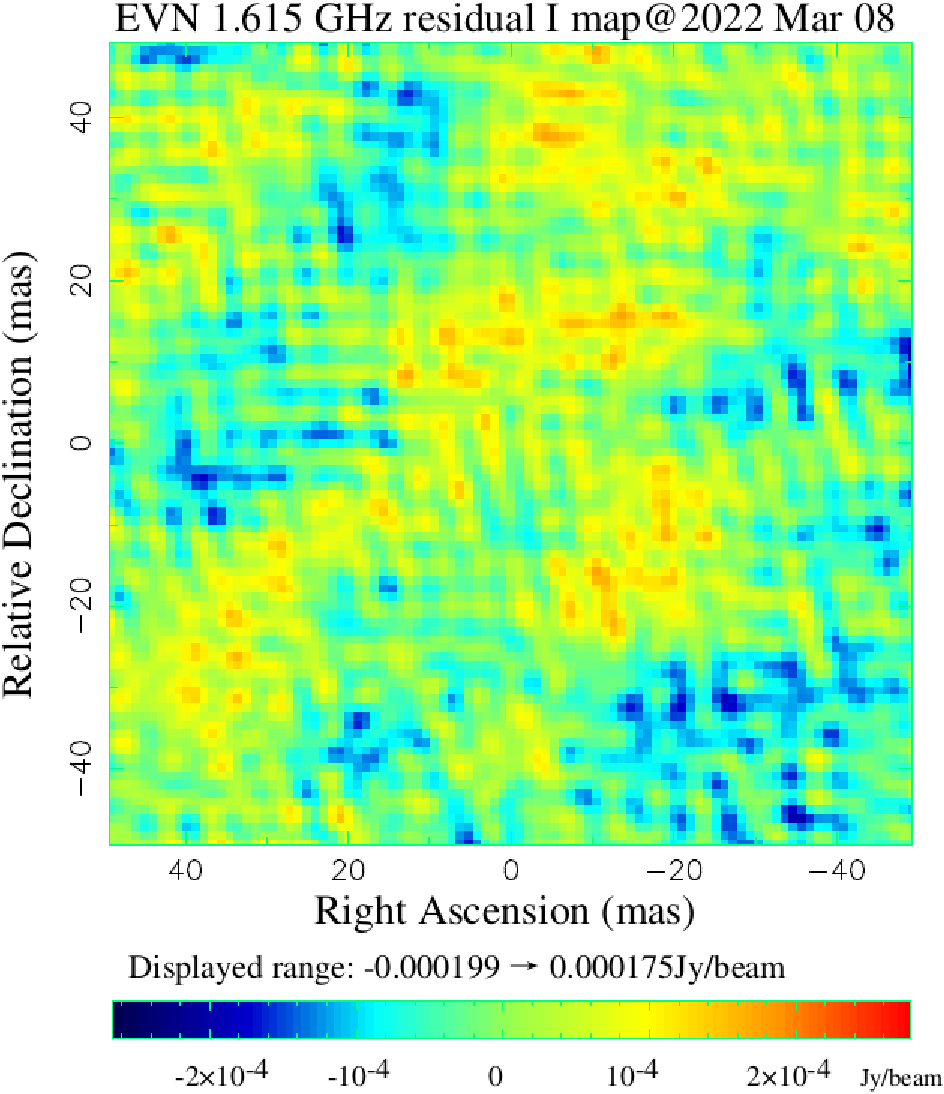}
      \caption{EVN L- and C-band radio continuum images of the SE nucleus of IRAS 17526+3253 using natural weighting. The plots are centered at R.A. = 17:54:30.554, Dec = 32:53:06.09. The map peak intensity, contour levels, and beam FWHM are labeled in the EVN C-band image (left panel). The first contour in all images corresponds to a SNR of approximately 3 (3$\sigma$). The color image in the right panel shows the L-band map, where no significant structure is detected above the 3$\sigma$ noise level (approximately 0.18 mJy/beam).}
      \label{evnSE}%
\end{figure*}

\subsection{The multi-band arcsecond-scale radio continuum emission}
\begin{figure*}
   \centering
\includegraphics[width=11cm,height=8cm]{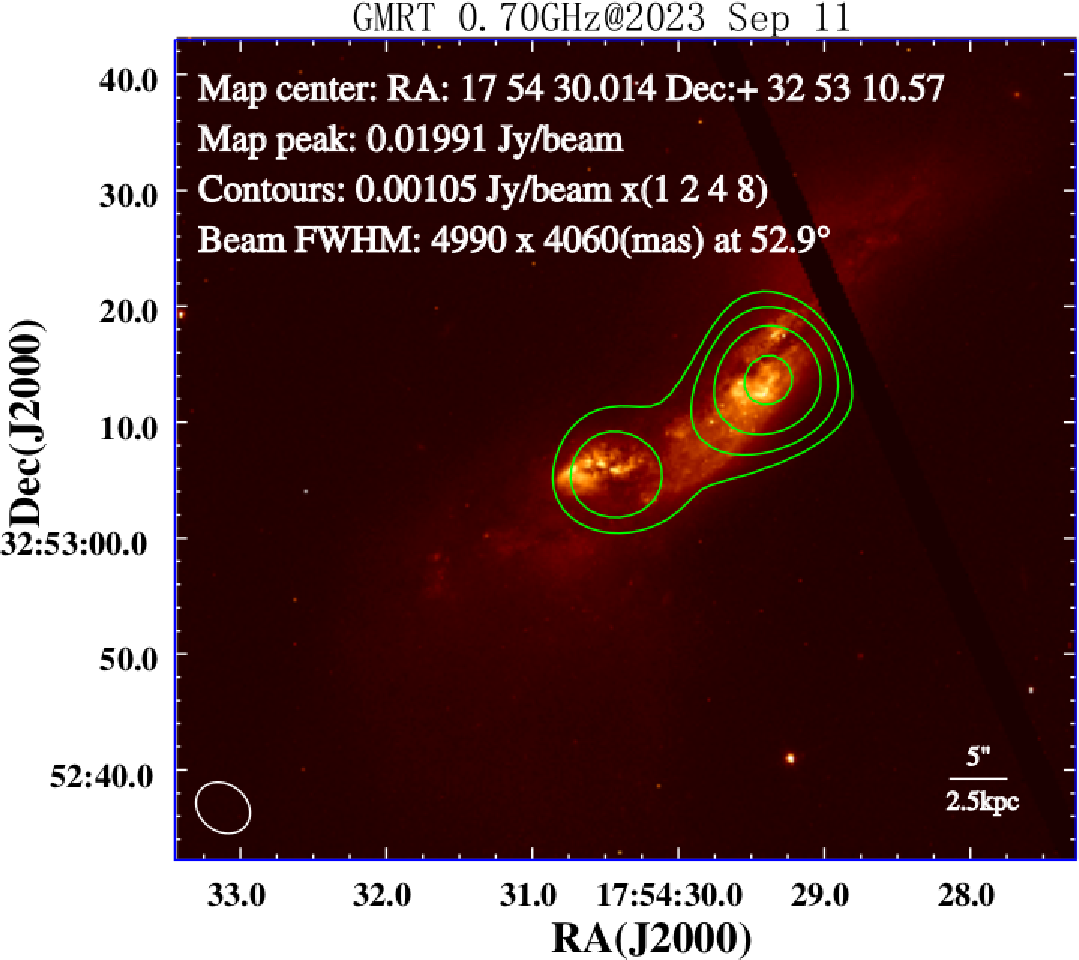}
\includegraphics[width=11cm,height=8cm]{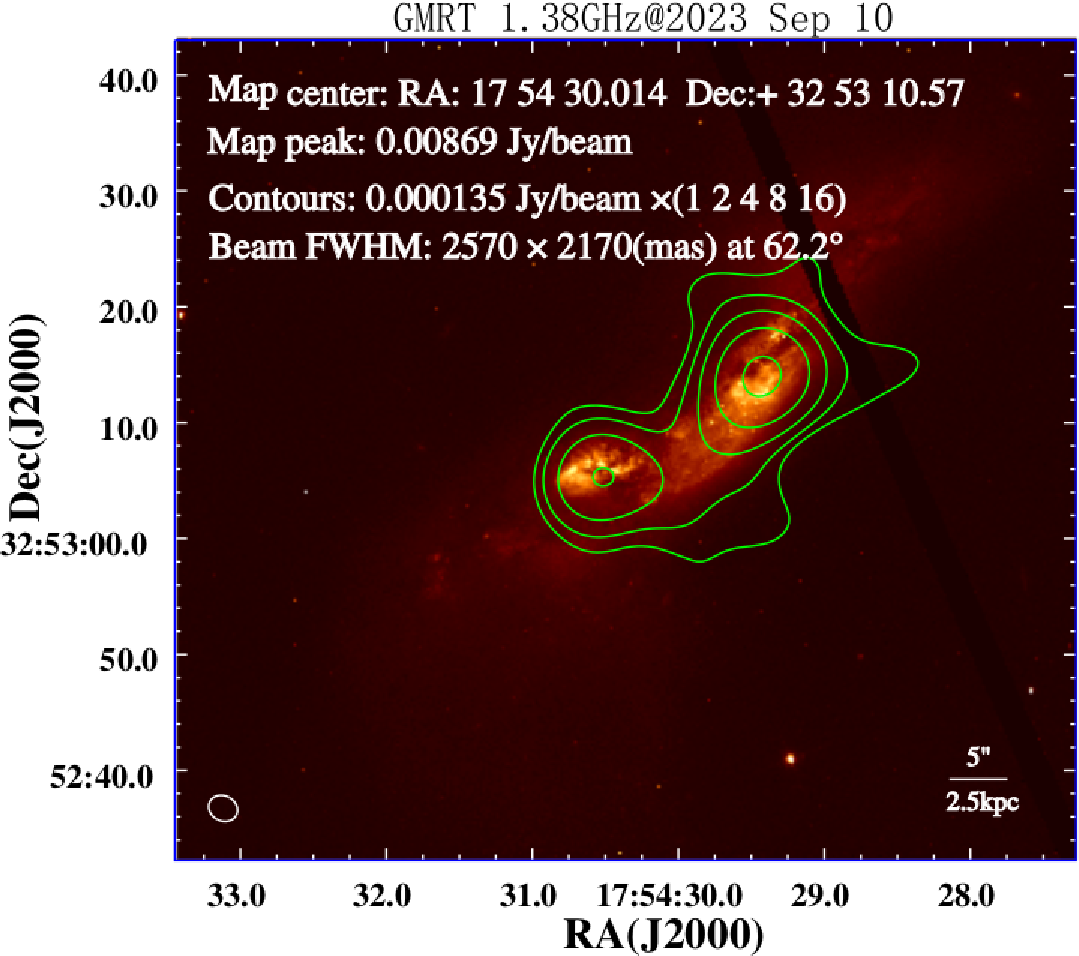}
\includegraphics[width=11cm,height=8cm]{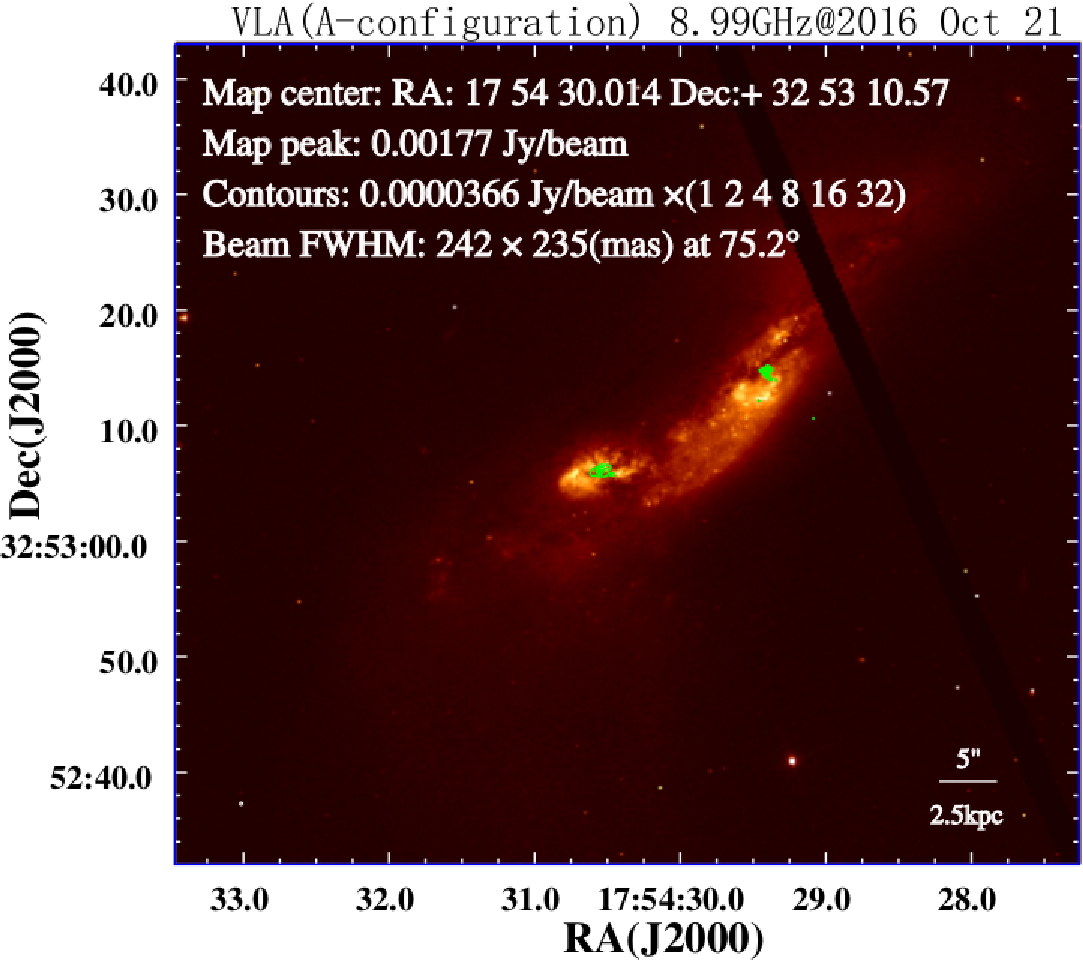}
    \caption{Multi-band radio contour maps of IRAS 17526+3253 from GMRT and VLA data overlaid on the HST/ACS F814W(I) image. The map peak intensity, contour levels, and beam FWHM are indicated in each image. The first contour in all images corresponds to a level of approximately 3 $\sigma$. }
    \label{continuum-contour1}%
\end{figure*}

\begin{figure*}
   \centering
\includegraphics[width=9cm,height=7cm]{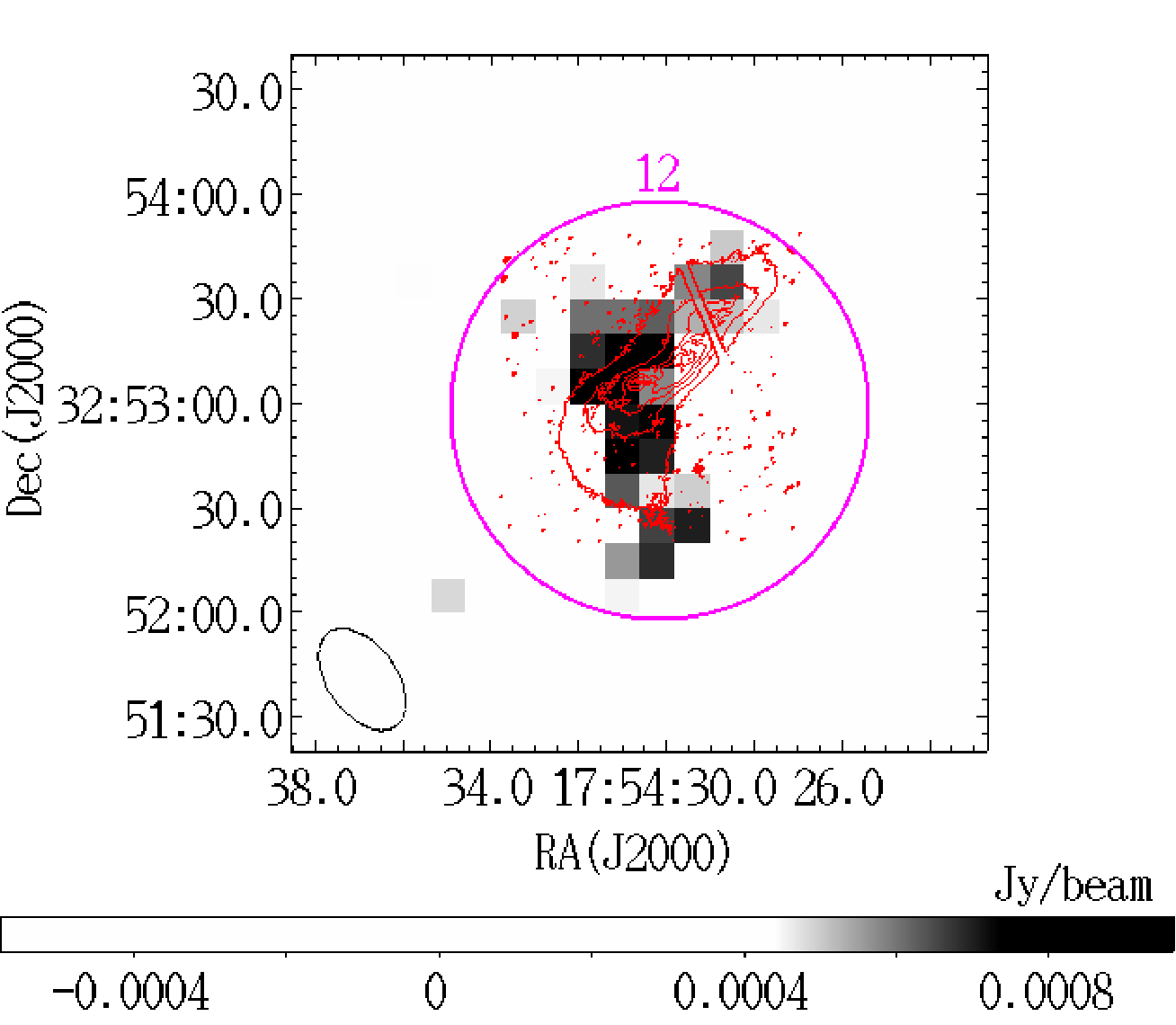}
\includegraphics[width=9cm,height=7cm]{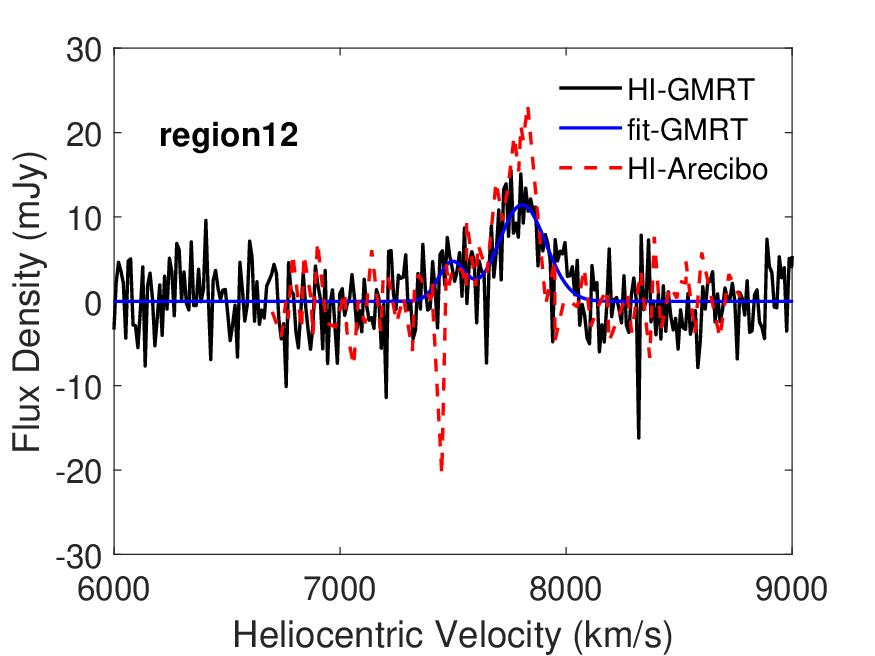}
      \caption{Low-resolution \HI line emission image of IRAS 17526+3253. Left panel: The grayscale image displays the combined continuum-subtracted \HI emission (moment = -1 map) with velocities ranging from 7385 to 7910 \kms. The beam FWHM is 33.25 arcsec $\times$ 19.76 arcsec at a position angle of 35 degrees. The grayscale range is shown at the bottom of the image. The red contour is derived from the colored HST image presented in Fig. \ref{fig1}. Right panel: The black solid line represents the detected \HI spectrum within a 2-arcmin region (region 12) shown in the left panel. The blue line corresponds to the fitted Gaussian line profile, with the parameters listed in Table \ref{tablea2-regionsHI}, while the red dashed line represents the \HI line profile observed with the Arecibo telescope, as reported by \cite{1987ApJ...322...88G}.}
      
    \label{HItotal}%
\end{figure*}

\begin{figure*}
   \centering
  \hfill
   { \includegraphics[width=0.45\textwidth,height=6cm]{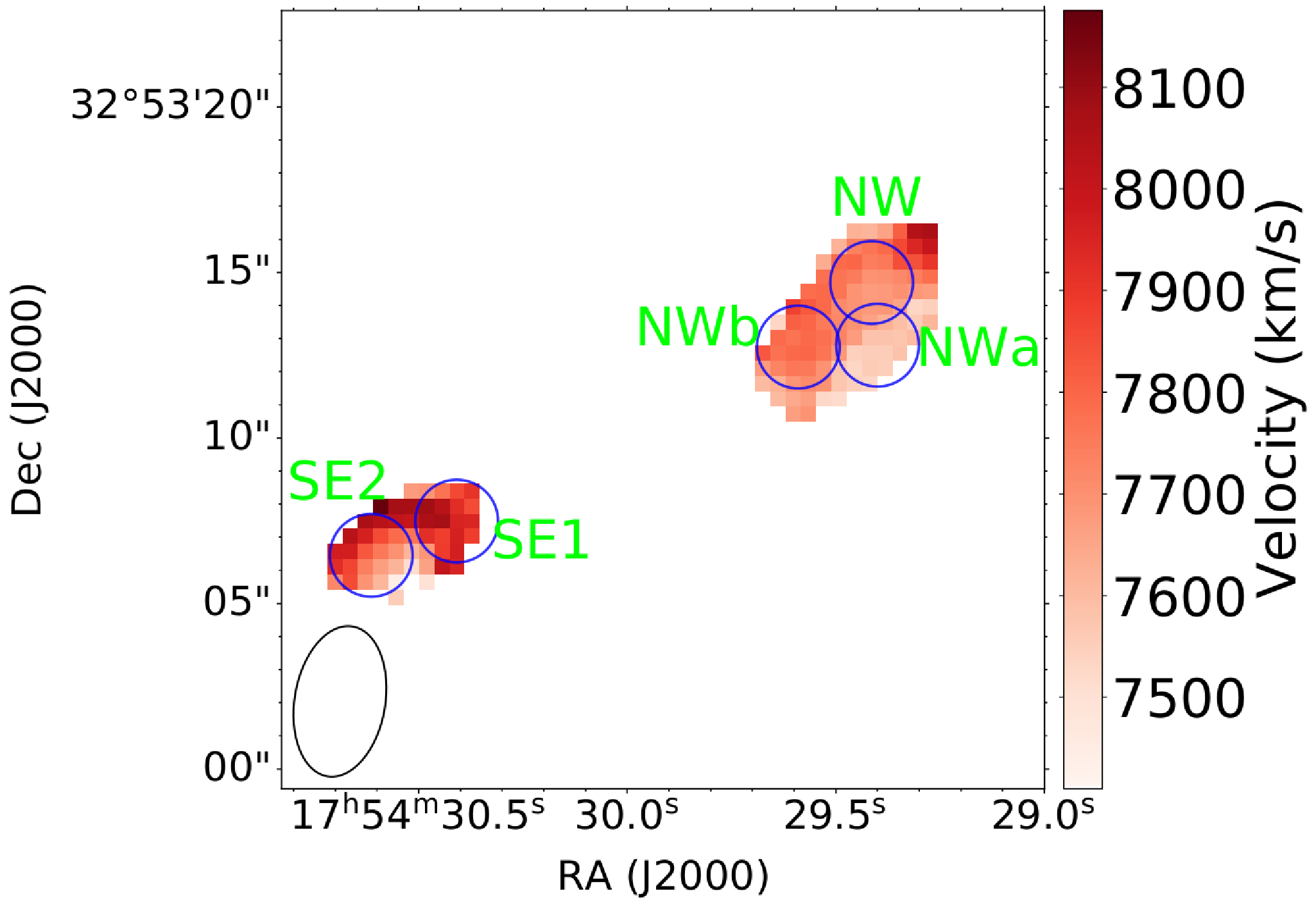}
\includegraphics[width=0.45\textwidth,height=6cm]{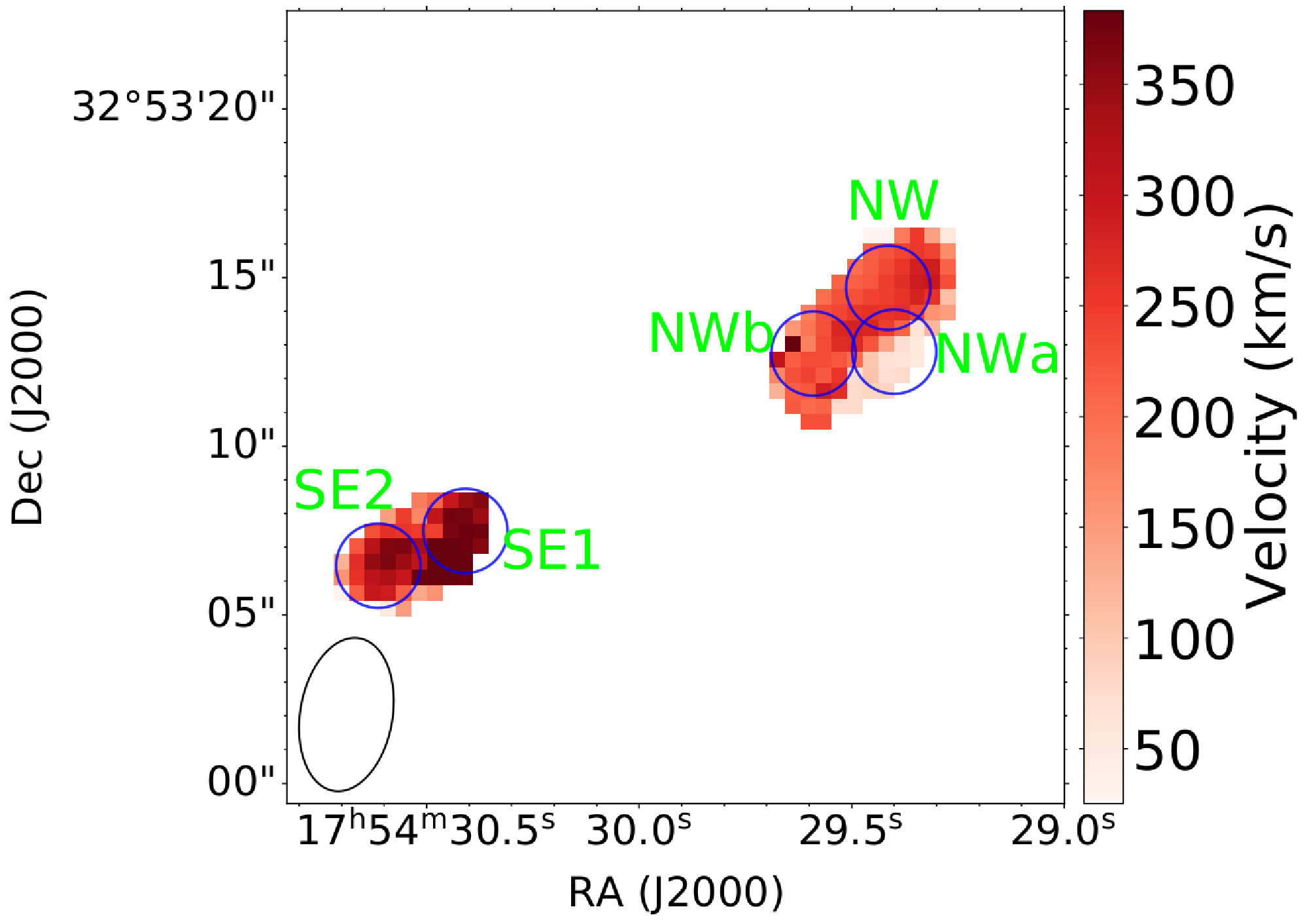}}
   \hfill
         \caption{Velocity centroid (left) and dispersion (right) maps generated using the SoFiA software for the high-resolution \HI image. The blue circles indicate the five regions from which the \HI absorption lines were extracted, each with a size of 2.5" $\times$ 2.5". NW corresponds to the NW1 component in the EVN images shown in Fig. \ref{evnNW}, where "NWa" and "NWb" are defined following \cite{2019MNRAS.486.3350S}. "SE1" and "SE2"  were selected based on the \HI absorption image. The line profiles of these regions are shown in Fig. \ref{HIabs_line2}, and the corresponding parameters are presented in Table \ref{tablea2-regionsHI}.}

    \label{momentxs}%
   \end{figure*}

Based on our GMRT observations at P and L bands, along with VLA archive data, we present the multi-band radio continuum emission of this source. Details of the projects are listed in Table \ref{tablea1:multibandata}, with the images shown in Fig. \ref{continuum-contour1} and \ref{continuum-contour2}. We find that the radio continuum flux originates from both the NW and SE nuclei, with additional diffuse emission around the two nuclei in the P, L, and C-band images. To minimize resolution effects, we measured both integrated and peak flux densities from images restored to uniform cell size and beam (see Section~\ref{sect2}; Table~\ref{tablea1:multibandata}). Using these measurements, we constructed radio SEDs showing both integrated and peak flux densities (see Fig.~\ref{peak}). The results indicate P-band continuum deviates from a simple power law, likely due to low-frequency absorption. Meanwhile, the diffuse radio continuum is resolved in the X-band VLA images, with a resolution of 0.2", leaving only compact emission in the central regions of the NW and SE nuclei.

\section{Discussion}{\label{sec:discussion}}

\subsection{Radio view of the nuclear activity in IRAS 17526+3253}
\label{discuss_radiocon}
Generally, most LIRGs are dominated by intense star formation, though many also show evidence of AGN activity \citep[see][and references therein]{2017MNRAS.471.1634H}. The nuclear activity of IRAS 17526+3253 has been studied in detail, and previous works have found no clear evidence that the galaxy hosts an AGN. Based on multiple diagnostic methods using radio, infrared and optical data, \citet{2006A&A...449..559B} classified IRAS 17526+3253 as a starburst-dominated source.
\citet{2019MNRAS.486.3350S} further investigated the galaxy using Gemini GMOS/IFU, HST, 2MASS, and VLA data. Their results revealed widespread star formation across the galaxy’s envelope on scales of several tens of kiloparsecs. In addition, optical emission-line ratio diagnostics indicated that star formation is the dominant ionization source in the optical regime.
IRAS 17526+3253 is also included in an ongoing project examining a sample of (U)LIRGs, most of which are interacting systems exhibiting OH megamaser activity \citep{2024MNRAS.52710844H}. Notably, it is the only galaxy in that sample that previously showed no evidence for AGN activity \citep{2024MNRAS.52710844H}.

\ho megamaser emission has been detected in IRAS 17526+3253 \citep[see][]{2013A&A...560A..12W}. In general, the presence of an AGN is considered essential for the production of \ho megamaser emission \citep{2012IAUS..287..323T}. \citet{2019MNRAS.486.3350S} investigated the optical emission-line profiles in six regions near the NW nucleus of this galaxy. One of these, region (d), is located approximately 1.9" north of the brightest optical knot and corresponds to a faint optical region near the galaxy's center \citep[see details in][]{2019MNRAS.486.3350S}. This region shows features that could be interpreted as evidence for AGN activity, though they might also result from shock ionization. Specifically, the optical emission lines in region (d) exhibit broader profiles compared to other regions, and H$\alpha$ is only slightly stronger than [N II] $\lambda$6583. In all three BPT diagnostic diagrams \citep{1981PASP...93....5B}, this region lies near the boundary between the \HII region and AGN, or between the composite and AGN classifications, making it a notable outlier in the system.

\subsubsection{The existence of a radio core and a jet component in NW nuclei of IRAS 17526+3253?}

Our EVN observations show that the brightness temperature of NW1 and NW2 does not align with typical OH megamaser galaxies, where the brightness temperature is on the order of $10^{6}$ K often explained as due to clustered supernova remnants and/or luminous radio supernovae \citep[see][and references therein]{2006ApJ...653.1172M,2005ApJ...618..705P}. On the other hand, it is consistent with the brightness temperature of the AGN core and jet components in the literature, such as the radio AGN in the LIRG NGC 6240, which has a brightness temperature around $ 10^{7}$ K \citep{2011AJ....142...17H}. The VLBI images of the radio AGN in IC 883 \citep{2012A&A...543A..72R} and NGC 7469 \citep{2003ApJ...592..804L}, as well as Arp 299A \citep{2010A&A...519L...5P,2009A&A...507L..17P}, also support this comparison.

Generally, a core-jet morphology is strong evidence of a radio AGN. We found that the NW1 component shows a high brightness temperature (see Table \ref{table2:evncontinuum}) and exhibits an inverted spectrum, suggesting a turnover frequency between 1.7 and 5.0 GHz, as expected for the core of an AGN whose radio emission is partially self-absorbed \citep{2010A&A...519L...5P}. Therefore, it is likely that NW1 is the core and NW2 is the jet component, although no connecting ridges or jet structures were found between the two components, which are separated by about 17 pc.

\cite{2010A&A...519L...5P} estimated the L and C-band radio luminosity of the brightest AGN core components of Arp 229 is about 1.8$\times$$10^{27}$ erg/s/Hz and 2.0$\times$$10^{27}$ erg/s/Hz, respectively. We find that the radio luminosity of NW1, NW2 and SE all have values larger than the brightest AGN core component in Arp 299 (see Table \ref{table2:evncontinuum}). The C-band radio luminosity ($\nu$L) at epoch 2009 Mar 01 of NW1+NW2 is about 1.4$\times$$10^{38}$ erg/s, which is about 7 times higher than the radio luminosity of core and jet components in Arp 299 (about 1.9 $\times$$10^{37}$ erg/s), typical of LLAGNs \citep[see][]{2010A&A...519L...5P}. And also higher than the radio luminosity of SNe/SNRs in the well-studied starburst galaxy Arp 220 by \citep{2019A&A...623A.173V}. However, the radio power of NW1 and NW2 is comparable but still slightly lower than some extreme radio supernova, e.g. SN 1986J in NGC 891 with a maximum radio power Log $P_{r}$ about 21.15 [W/Hz] and a radio supernova in Mrk 297 with 21.5 [W/Hz] \citep[see][and reference therein]{1998ApJ...492..137S}. The brightness temperature and radio power of the compact component in SE nuclei is likely consistent with the starburst origin as compared with the SNe/SNRs in Arp 220 by \cite{2019A&A...623A.173V} and in Arp 299 by \cite{2009A&A...507L..17P}, additional evidence is needed to further confirm the presence of AGN in this nuclei.

\citet{2019MNRAS.486.3350S} reported the coordinates of the brightest optical emission-line and continuum knot at RA (J2000): 17:54:29.4, Dec (J2000): 32:53:12.8. In our EVN observations, we find that the peak radio continuum emission is located at RA (J2000): 17:54:29.41070 $\pm$0.2 mas, Dec (J2000): 32:53:14.6955 $\pm$0.1 mas (NW1), which is approximately 1.9" north of the brightest optical component. This radio peak aligns with a faint optical region near the center of the galaxy, as shown in Fig. \ref{continuum-contour1}, where the 8.9 GHz radio continuum contours are overlaid on the HST image. The optical region closest to the EVN continuum peak corresponds to region (d) in \citet{2019MNRAS.486.3350S}. The high brightness temperatures and radio luminosities of NW1 and NW2 are most consistent with an AGN origin \citep[see][]{2013ApJ...779..173N}. Our results therefore suggest that the properties observed in region (d) may be driven by AGN activity. Although \citet{2019MNRAS.486.3350S} proposed that the observed ionization characteristics could be due to shock excitation, it is also possible that such shocks are themselves induced by a radio AGN. This is supported by the scenario proposed by \citet{2018MNRAS.474.5319H}, in which radio-emitting plasma interacts with the surrounding interstellar medium via shocks, contributing to gas excitation. However, we note that the flux densities of NW1 and NW2 in our 2022 EVN observations are lower than those in previous epochs (see Table \ref{table2:evncontinuum}). Continued high-sensitivity VLBI monitoring of IRAS 17526+3253 is therefore essential to rule out alternative interpretations such as the presence of two extremely luminous radio supernovae.

\subsubsection{The multi-band arcsecond-scale observations}

The integrated flux density at 1.38 GHz from GMRT observations for the NW and SE nuclei is about 27 mJy and 12 mJy, respectively (see Table \ref{tablea1:multibandata}). The total flux density of the two nuclei is consistent with the L-band NVSS flux of about 45 mJy. The total flux density remains higher than the peak radio flux from the L-band GMRT observation after restoring the beam to 5"$\times$ 5" (see Table \ref{tablea1:multibandata}). This indicates that the radio continuum emission is diffusely distributed around and between the two nuclei (see Fig. \ref{continuum-contour1}), consistent with the interpretation of \cite{2019MNRAS.486.3350S} that the radio continuum emission originates from star-forming regions across the galaxy.

Generally, the radio spectra of LIRGs and ULIRGs below 1.4 GHz are rarely a simple power-law and spectral turn-overs or bends are often shown in their radio spectra \citep{2010MNRAS.405..887C,2018MNRAS.474..779G}. The radio spectrum of IRAS 17526+3253 also shows a deviation below 1.4 GHz, consistent with these characteristics. Since synchrotron self-absorption requires extremely high brightness temperatures to be significant, free-free absorption is considered the primary mechanism for shaping the spectral characteristics \citep{2010MNRAS.405..887C}. We fitted the spectrum using the equation from \cite{2018MNRAS.474..779G}, modeling the radio continuum as the sum of two distinct components: one representing the steep-spectrum non-thermal synchrotron emission, and the other describing the flat-spectrum thermal free-free emission, following this form:

\begin{equation}
\begin{split}
S_\nu = \left(1-e^{-\tau}\right)\left[B+A	\left(\frac{\nu}{\nu_{t}}\right)^{0.1+\alpha}\right]\left(\frac{\nu}{\nu_{t}}\right)^2,
\label{eq2}
\end{split}
\end{equation}
\noindent where $A$ and $B$ are free parameters representing the synchrotron and free-free normalization components, respectively. The free parameter $\alpha$ represents the synchrotron spectral index, which defines the slope of the synchrotron emission spectrum. The optical depth $\tau$ is described as $\tau = (\frac{v}{v_t})^{-2.1}$, where $v_t$ is the turnover frequency. For the fitting process, we set $\alpha$ in the range of -1.4 to -0.5 \citep{2018MNRAS.474..779G}.

The fitting results show a steep spectral index for both the integrated and peak flux densities ( $\alpha_{\mathrm{int}}$ and $\alpha_{\mathrm{peak}} \approx -1.4$; see Fig. \ref{peak}). The total and peak flux densities from the VLA-A 9 GHz observations were excluded because they fall well below the model prediction, suggesting significant flux loss due to resolution effects. Even after restoring to a larger beam, the missing flux remains unrecovered. A similar trend was reported by \citet{2006A&A...449..559B}, who found partial resolution of C-band emission compared to L-band data. This is consistent with our EVN results, where most arcsecond-scale emission is resolved out. Such behavior is typical for AGN in LIRGs, where compact emission on $\sim$100 pc scales is traced by VLBI \citep{2003ApJ...592..804L}, while larger-scale emission can originate from circumnuclear star formation. \citet{2019MNRAS.486.3350S} also found that star-forming regions align with the arcsecond-scale radio features. These results suggest that both a central AGN and nuclear star formation contribute to the observed radio continuum in IRAS 17526+3253.

\subsection{Implications about the merging and evolution stage from \HI images}

\subsubsection{Characterizing the \HI emission and absorption gas
in IRAS 17526+3253}

We detected both \HI absorption and emission in IRAS 17526+3253 using three \HI image with different beam sizes (see Section~\ref{result3}). The \HI absorption is confined to two compact regions associated with the NW and SE nuclei, while the emission is distributed across a much larger area, approximately 2 arcminutes or more in extent. 
\HI absorption profiles have been detected toward both nuclei (see Table \ref{tablea2-regionsHI} and Fig. \ref{HIabs_line} and \ref{HIabs_line2}). The absorption line in region 13 consists of two components, at velocities of 7631 and 7818 \kms. In contrast, the line profile in region 14 shows a broad feature centered around 7636 \kms, which is consistent with the first component observed in region 13. Based on the radio continuum and the \HI absorption line profiles, we estimate the peak optical depths ($\tau$) for the NW and SE nuclei to be 0.13 and 0.11, respectively—indicating similar absorption depths in both nuclei.

Our findings thus support the conclusion by \cite{2019MNRAS.486.3350S} that a velocity discontinuity exists in the NW nucleus. However, the spatial resolution of our \HI absorption data is lower than that of the optical spectroscopy used by \cite{2019MNRAS.486.3350S}, and the absorption region spans less than two synthesized beams in the NW nucleus and less than one beam in the SE nucleus (see Fig. \ref{momentxs}). Therefore, we can only determine the velocity centroids for NWa and NWb, without resolving a detailed velocity field. We also note that the S/N in the moment map of \HI absorption is about 3–6 toward the NW nucleus, while that for the SE nucleus is only slightly above 3 (see Fig. \ref{HI-FS+XS}). This suggests that higher-sensitivity observations will be needed to confirm the absorption feature and velocity structure of the SE nucleus. 

Based on optical imaging and spectral line analysis, \cite{2019MNRAS.486.3350S} proposed that this source is a mid-stage major merger, with two main galactic nuclei separated by approximately 8.5 kpc, embedded in an elongated, tidally distorted, irregular envelope with complex structure. We investigated the velocity distribution of \HI gas around this galaxy and found that the peak velocities increase from the northwest to the southeast, ranging from 7500 to 7800~\kms (see Fig. \ref{NS-HIFS}). The \HI emission line profiles in regions 2, 3, 6, and 8 exhibit broader FWHM compared to those in regions 4, 5, and 7 (see Table~\ref{tablea2-regionsHI}). These broader line profiles may result from gas mixing between the two merging nuclei or from turbulence induced by the interaction, consistent with the characteristics of mid-stage mergers where galaxy disks are disturbed and overlapping \citep{2016ApJ...825..128L,2021A&A...649A.137P}. Regions 4 and 7 may be less influenced by the merger, as indicated by their narrower line profiles. The narrow \HI emission line in region 5 may have a different origin, possibly affected by \HI absorption, as suggested by the detection of absorption in the high-resolution \HI image. We also identified redshifted \HI gas in regions 5 and 6 and blueshifted components in region 8, consistent with the ongoing merging activity.

The CO(2--1) emission in IRAS 17526+3253 peaks at 7500 and 7800 \kms \citep{2008A&A...477..747B}, consistent with the fitted velocities of the total \HI emission (region 12; see Fig. \ref{HItotal} and Table \ref{tablea2-regionsHI}). Additionally, the total \HI emission from GMRT observations matches well with the \HI line profiles obtained from Arecibo observations. However, there is a notable discrepancy: a deep absorption feature seen in the Arecibo data is not present in the GMRT spectra (–20 mJy, see Fig.\ref{HItotal}). If real, this feature would imply an unusually high optical depth ($\sim$0.6), given the $\sim$45 mJy diffuse continuum from both nuclei reported in NVSS. However, \citet{1987ApJ...322...88G} noted that interference spikes near 1385 MHz ($\sim$7500 km/s) appeared in several line profiles in their sample. The non-detection of this feature in our GMRT observations confirms that the apparent absorption is most likely caused by an RFI spike.

\subsubsection{Comparison with the characters of \HI emission in (U)LIRGs}
The \HI 21-cm line emission in (U)LIRGs exhibits two main characteristics, as summarized by \cite{2015ApJ...805...31L}:

First, (U)LIRGs generally have lower atomic than molecular gas content, attributed to the conversion of \HI\ into molecular gas, with the molecular-to-atomic gas ratio increasing with infrared excess. Based on 2MASS K-band data, \cite{2019MNRAS.486.3350S} suggested that the two nuclei in IRAS 17526+3253 have similar bulge masses, though the NW nucleus appears larger and brighter in both optical and radio images. CO(2–1) observations by \cite{2008A&A...477..747B} revealed two velocity components at $\sim$7500 and $\sim$7800 \kms, likely corresponding to the NW and SE galaxies \citep{2019MNRAS.486.3350S}, as confirmed by our GMRT \HI\ data. However, the \HI\ emission from the NW nucleus is three times weaker than that from the SE. \cite{2025MNRAS.537.1597A} showed that the molecular-to-atomic gas ratio remains roughly constant across Hubble types, suggesting that the difference in \HI\ luminosity here may reflect different evolutionary stages. In early to mid-stage mergers, atomic gas is often funneled inward and converted into molecular gas \citep{2016ApJ...825..128L}, which may explain the NW nucleus’s brightness despite its weak \HI\ emission. The potential detection of a radio AGN in the NW nucleus of IRAS 17526+3253 could be linked to this process, as merger-driven inflows may fuel both star formation and AGN activity \citep[see][and references therein]{2021MNRAS.506.5935R}.

Second, the \HI\ distribution in (U)LIRGs often displays complex, merger-induced structures such as tidal tails and gas streams \citep{2015ApJ...805...31L}. In IRAS 17526+3253, the \HI\ extent aligns well with deep optical images from HST and DESI. According to the merger classification by \cite{2024A&A...691A..82C}, the system exhibits clear tidal features consistent with Stage III, particularly Stage IIIb, characterized by a projected nuclear separation $\leq$10 kpc—here, $\sim$8.5 kpc. Our \HI\ velocity dispersion and PV diagram maps reveal two distinct kinematic components and elevated velocity dispersion between the nuclei (Figs.\ref{momentfs}, \ref{momentfs2} and \ref{pvdiagram}), further supporting this classification. The broad \HI\ absorption lines suggest intense interaction in the central regions (Figs.\ref{HIabs_line}, \ref{HIabs_line2}), while the absence of a common \HI\ envelope implies that large-scale gas mixing has not yet occurred. Combined with widespread star formation in the extended envelope \citep{2019MNRAS.486.3350S}, these findings indicate that IRAS 17526+3253 is likely in a transitional phase between Stage IIIb and IIIc.

\subsection{The prospect about being a dual megamaser galaxy}
In the literature, two commonly adopted thresholds are used to classify OH megamasers: 1 $L_{\odot}$ \citep[see][]{2009A&A...502..529S,2011A&A...525A..91T,1990A&A...229..431H} and 10 $L_{\odot}$ \citep[e.g.][]{2002AJ....124..100D,2016ApJ...816...55W}. \cite{1989CRASB.308..287M} reported an isotropic OH luminosity of log($L_{\rm OH}$/$L_{\odot}$) = 0.99 for IRAS 17526+3253, although no accompanying flux density or spectral information was provided. Based on this luminosity, the source could be classified either as an OH kilomaser or nearly a megamaser (if the threshold for megamasers is 10 $L_{\odot}$), or as an OH megamaser (if the threshold is 1 $L_{\odot}$), depending on the criterion adopted in the literature. Our EVN observations did not detect OH line emission. Similarly, we also investigated the pipeline calibrated archive EVLA A-array data (project 16B-063) also yielded a non-detection, with a comparable noise level ($\sim$1.1 mJy) and noticeable RFI contamination. Given the limitations of both datasets, we cannot conclusively determine the cause of the non-detection—whether it is due to RFI, the OH line emission being resolved out, or possible variability in the OH line emission. Future observations with higher sensitivity will be necessary to accurately measure the OH line flux in this source. 

\begin{figure*}
   \centering
   \hfill
      { \includegraphics[width=0.45\textwidth,height=6cm]{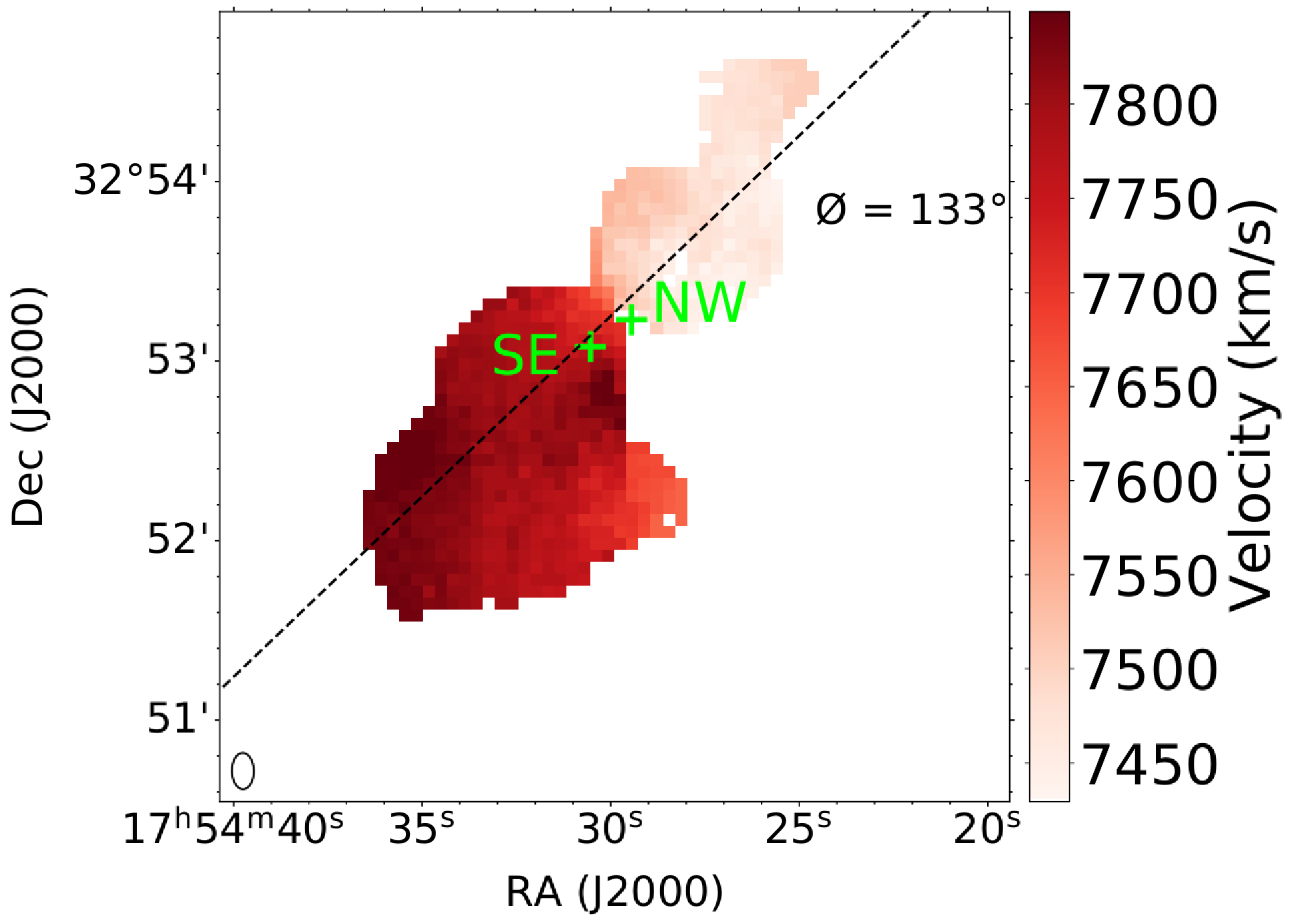}
\includegraphics[width=0.45\textwidth,height=6cm]{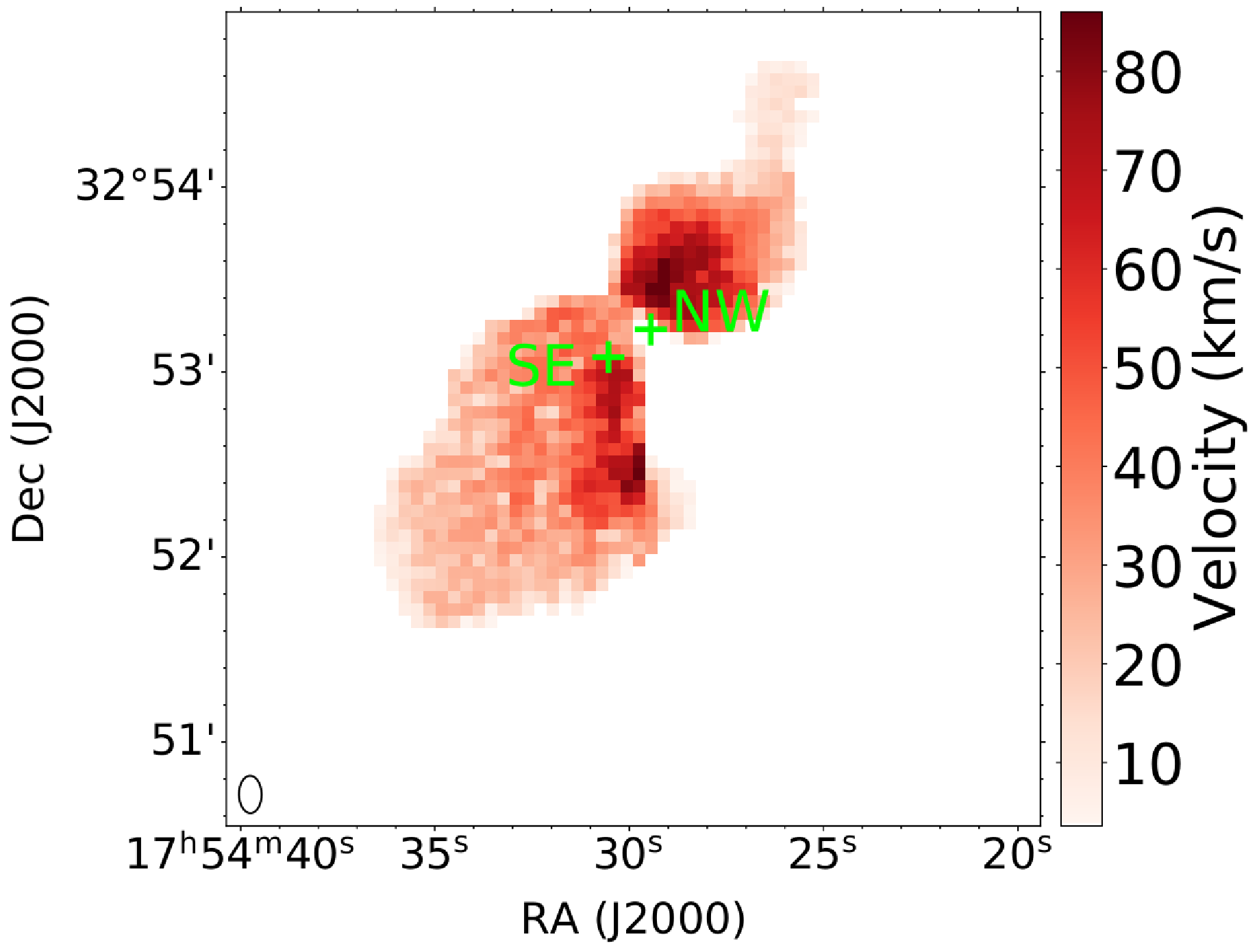}
}

  \hfill
         \caption{Velocity centroid (left) and dispersion (right) maps generated using the SoFiA software for the medium-resolution \HI image. The green crosses labeled “NW” and “SE” mark the positions of the NW and SE nuclei as seen in the EVN radio continuum images (Figs. \ref{evnNW} and \ref{evnSE}). Here, “NW” corresponds to the NW1 component in those EVN images. The black dashed line indicates the major axis identified by the SoFiA software for extracting the PV diagram. The corresponding dynamical center is shown in Fig.~\ref{pvdiagram}.}

    \label{momentfs}%
   \end{figure*}

The isotropic luminosity threshold for water megamaser is
traditionally 10 $L_{\odot}$ \citep[e.g.][]{2005A&A...436...75H}.
The \ho maser emission in IRAS 17526+3253 has a luminosity of approximately 360 $L_{\odot}$, clearly qualifying it as a megamaser \citep{2013A&A...560A..12W}. If we adopt the first option (1 $L_{\odot}$) for the OH megamaser threshold, IRAS 17526+3253 could then be considered as a dual megamaser candidate. The water maser line profile shows three components: two broad features at 7797 and 7810 \kms, and a narrow line at 7808 \kms \citep[see][]{2013A&A...560A..12W}. Based on these peak velocities, \cite{2019MNRAS.486.3350S} suggested that the water maser emission is associated with the southeastern (SE) nucleus. By comparing these velocities with our \HI emission velocity map (see Figs. \ref{momentfs} and \ref{momentfs2}), it is likely that the maser peaks correspond to the SE nucleus. However, the \HI emission near and between the two nuclei shows broader velocity dispersion (see the same figures), so the possibility that the water maser originates from the northwestern (NW) nucleus cannot be excluded. High spatial resolution observations of the water maser line are essential to accurately determine its origin.

To date, only three dual megamaser galaxies have been reported in the literature. Two of these—Arp 299 and II Zw 096—were identified by \citet{2016ApJ...816...55W}, while the third is the well-known luminous OHM galaxy Arp 220, whose \ho megamaser emission was confirmed by \citet{2017A&A...602A..42K}. Among them, Arp 220 is a late-stage merger with a nuclear separation of approximately 0.37 kpc \citep{2017ApJ...841...44P}, whereas Arp 299 and II Zw 096 are classified as intermediate-stage mergers \citep{2022A&A...661A.125W}. Based on \HI, CO, and OH line data, \citet{2022A&A...661A.125W} suggested that II Zw 096 may be in a slightly earlier merger stage compared to Arp 299. Our \HI velocity measurements of IRAS 17526+3253 show distinct velocity structures for the NW and SE nuclei, whereas the \HI emission in II Zw 096 appears mixed and complex on large scales \citep[see][]{2022A&A...661A.125W}. These results suggest that IRAS 17526+3253 is likely in an earlier merger stage than II Zw 096. This interpretation is also consistent with the larger nuclear separation observed in IRAS 17526+3253. Among the four systems, only Arp 299 and IRAS 17526+3253 show possible evidence of a radio jet \citep{2010A&A...519L...5P}. No clear radio AGN signatures have been detected in II Zw 096 or Arp 220, although both galaxies host massive black holes in their nuclear regions \citep[e.g.,][]{2022A&A...661A.125W, 2019A&A...623A.173V,2007A&A...468L..57D}. High-resolution \ho observations of Arp 299 and Arp 220 have shown that the emission is associated with both nuclei \citep{2011A&A...525A..91T, 2017A&A...602A..42K}. Similar high-resolution observations of \ho emission in IRAS 17526+3253 and II Zw 096 may help pinpoint the emission location, identify the nucleus of origin, and determine whether the \ho maser emission arises from both nuclei, as in other dual megamaser systems.

\section{Summary}
\label{summary}
We present high-resolution radio observations of OH, \HI, and multi-band, multi-resolution radio continuum emission from IRAS 17526+3253. Our analysis focuses on the properties of the radio continuum and \HI gas in this merging galaxy.

We detect radio continuum emission from both the northwestern (NW) and southeastern (SE) nuclei of IRAS 17526+3253. The NW nucleus hosts two bright, compact components. L- and C-band EVN observations yield logarithmic brightness temperatures (log $T_\mathrm{b}$) of approximately 7.4 K. The radio luminosities of these components exceed $10^{28}$ erg/s/Hz, significantly higher than those typically associated with known radio supernovae (SNe). Moreover, the radio continuum emission is located in an optically obscured region of the galaxy, and the associated optical spectral line ratios lie near the boundaries between the ‘\HII region’–‘AGN’ and ‘composite’–‘AGN’ classifications across all three BPT diagrams. These results support the presence of a radio AGN in the NW nucleus. The EVN observation of the OH line emission achieved a noise level of approximately 0.9 mJy/beam in RFI-free channels, while RFI-affected channels exhibited higher noise levels, ranging from 1.4 to 5 mJy/beam. No OH line emission was detected in either case.

Additionally, we detect \HI emission and absorption in the GMRT observations. All \HI features reported in this work are detected with a SNR level between 3 and 6-7. \HI absorption lines are observed toward both nuclei. The \HI absorption velocity field of the NW nucleus shows a possible two-component structure, consistent with optical results reported by \cite{2019MNRAS.486.3350S}. The \HI emission velocity field reveals a two-velocity system originating from the NW and SE galaxies, with peak velocities around 7500 and 7800 \kms, respectively. The integrated \HI line profile is in good agreement with previous Arecibo observations. The total \HI emission from the SE galaxy is found to be over three times higher than that from the NW galaxy. Redshifted \HI components are identified in two central regions (regions 5 and 6), while a blueshifted component is seen in the southern region (region 8), indicating complex gas kinematics likely associated with merger-driven dynamics. The large-scale \HI emission appears to extend both southeastward and northwestward. The southeastern extension aligns with the tidal tail seen in DESI optical images, while the northwestern extension lacks an optical counterpart and may require further confirmation.

\begin{acknowledgements}
We thank the referee for the constructive comments and suggestions, which helped improve this paper.
This work is supported by the grants of NSFC (Grant No. 12363001).The European VLBI Network is a joint facility of European, Chinese, and other radio astronomy institutes funded by their national research councils.The Giant Metrewave Radio telescope (GMRT) is the most sensitive synthetic aperture Radio telescope in the meter band. It is operated by a department of the Tata Institute of Fundamental Research – the National Radio Astronomical Center (NCRA) in India. The National Radio Astronomy Observatory is operated by Associated Universities, Inc., under a cooperative agreement with the National Science Foundation.

\end{acknowledgements}
\bibliographystyle{aa}
\bibliography{175261}

\begin{appendix}
\section{Online materials}
\twocolumn
\counterwithin{figure}{section}
\begin{figure*}
   \centering
   \includegraphics[width=16cm,height=12cm]{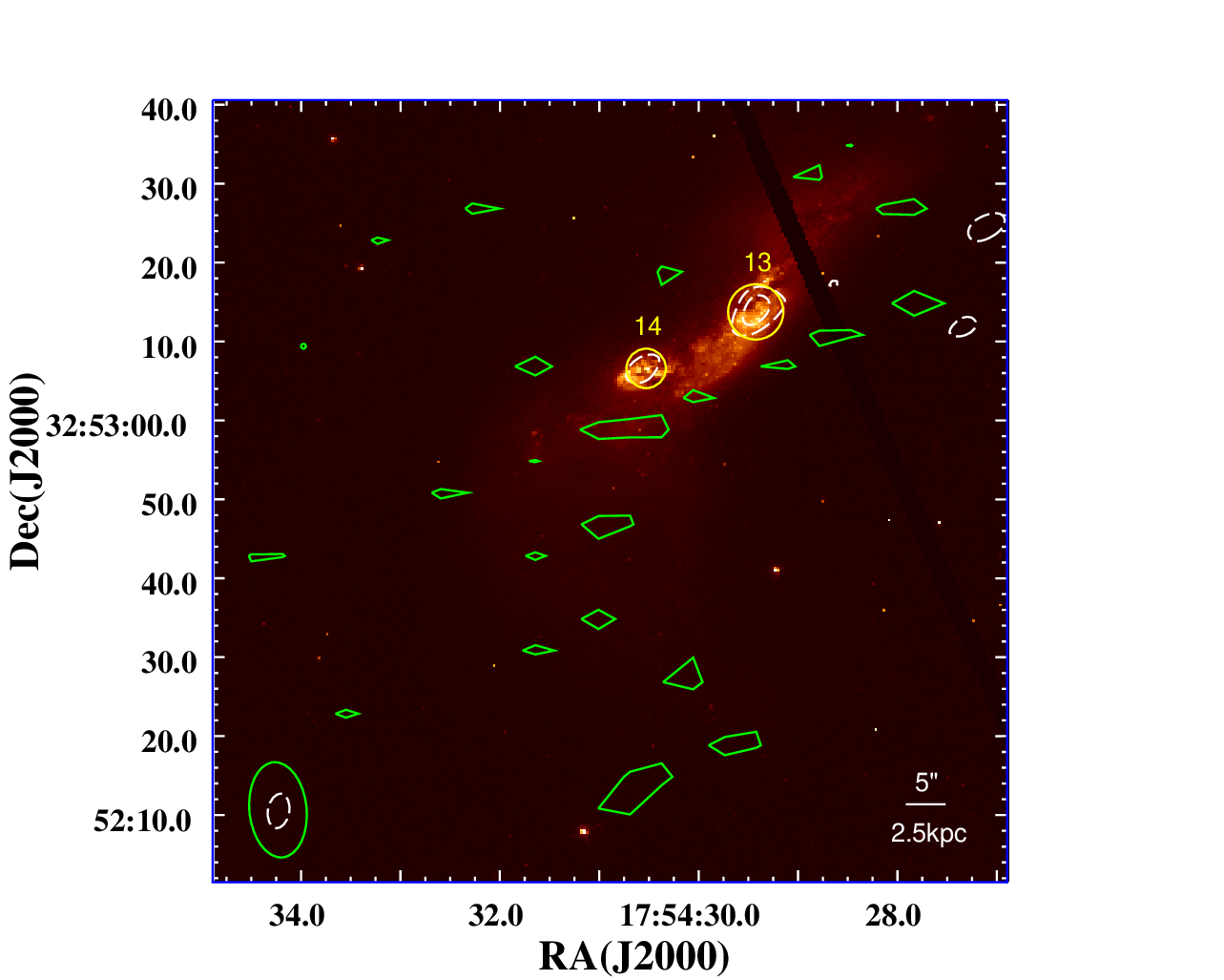}  
    \caption{Contour maps of \HI emission and absorption at medium and high resolutions overlaid on HST /ACS F814W(I) image. The white contours are derived from the \HI image with a beam size of 4.1 arcsec $\times$ 2.8 arcsec at a position angle of -9.74 degrees. The 1$\sigma$ noise level is approximately 0.15 mJy/beam, combining channels within the velocity range of 7076 to 7900 \kms. The green contours are from an image produced with a beam size of 12.1 arcsec $\times$ 7.3 arcsec at a position angle of 5.73 degrees. The 1$\sigma$ noise level is about 0.18 mJy/beam, also combining channels within the velocity range of 7076 to 7900 \kms. The two yellow circles (regions 13 and 14) with diameters of 7" and 5" mark the areas where the \HI absorption lines of the two regions were extracted.}
    \label{HI-FS+XS}%
\end{figure*}

\begin{figure*}
   \centering
  
   \subfigure[EVN continuum subtracted OH line channel image at V=7500 \kms ($\Delta$ V = 12 \kms) for both the NW and SE nuclei.]{ \includegraphics[width=0.485\textwidth,height=6.2cm]{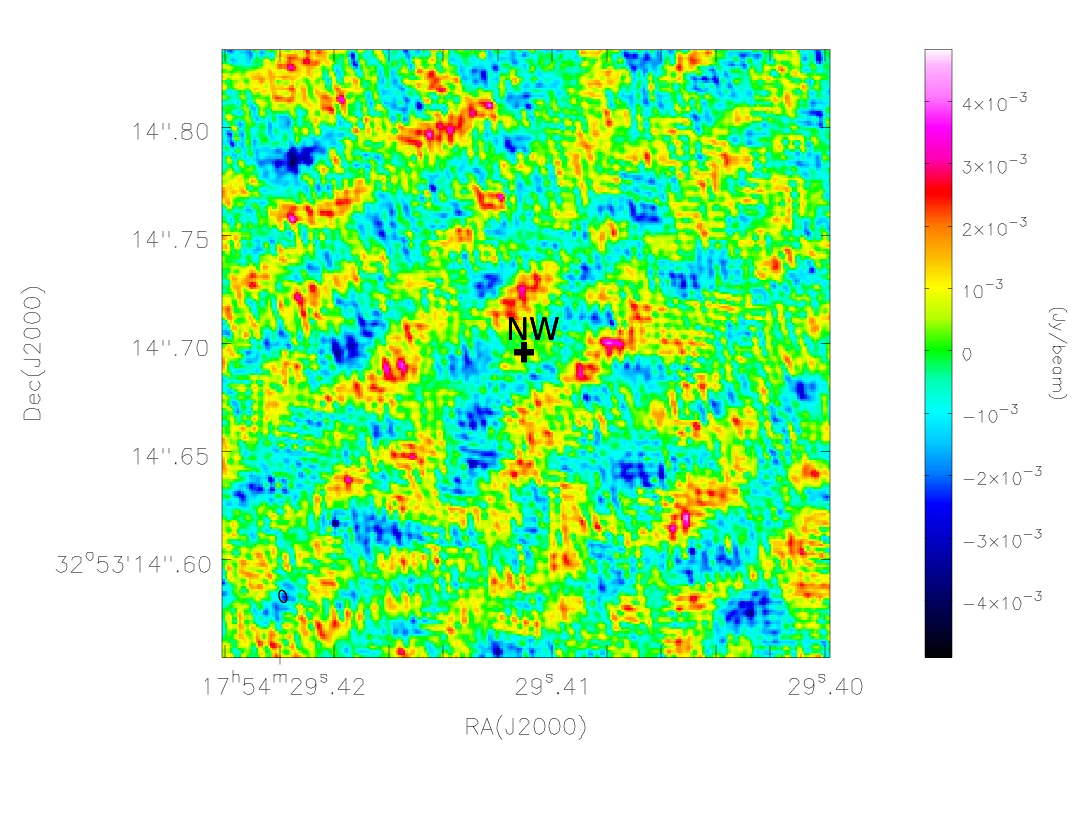}
\includegraphics[width=0.485\textwidth,height=6.2cm]{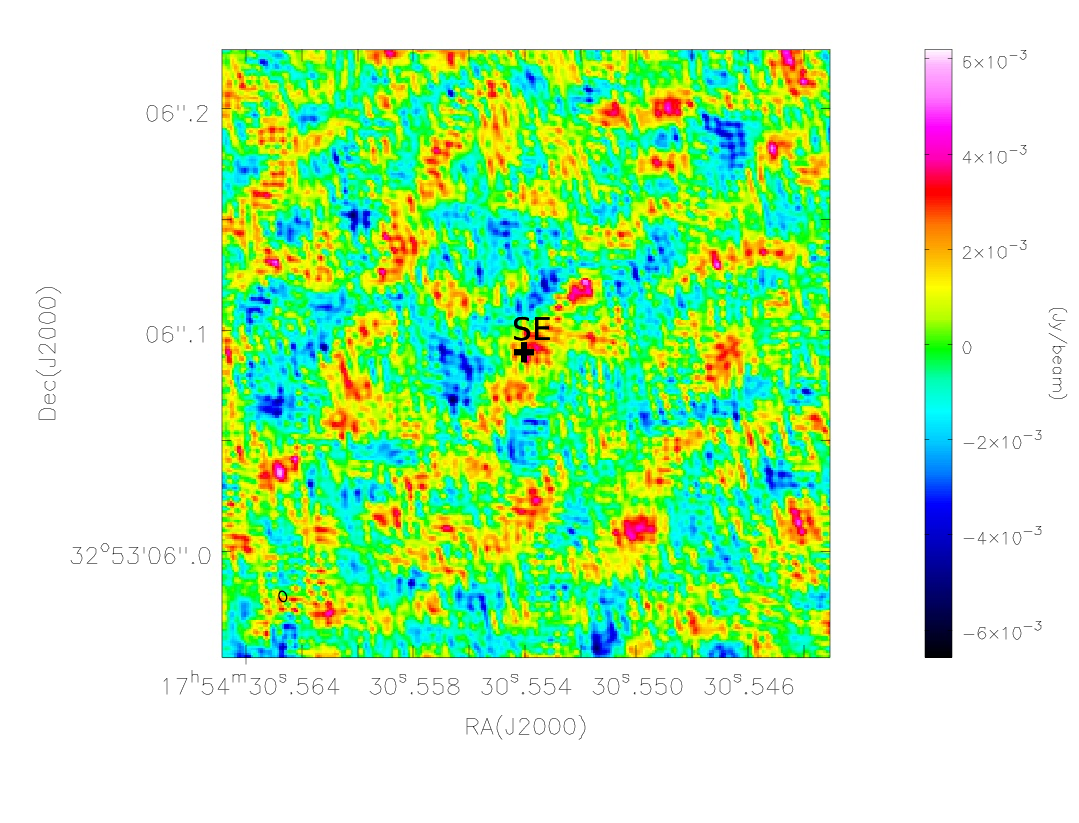}}
   
   \hfill
      \subfigure[EVN continuum subtracted OH line integrated channel image encompassing velocities from 7000 to 8000 \kms]{ \includegraphics[width=0.485\textwidth,height=6.2cm]{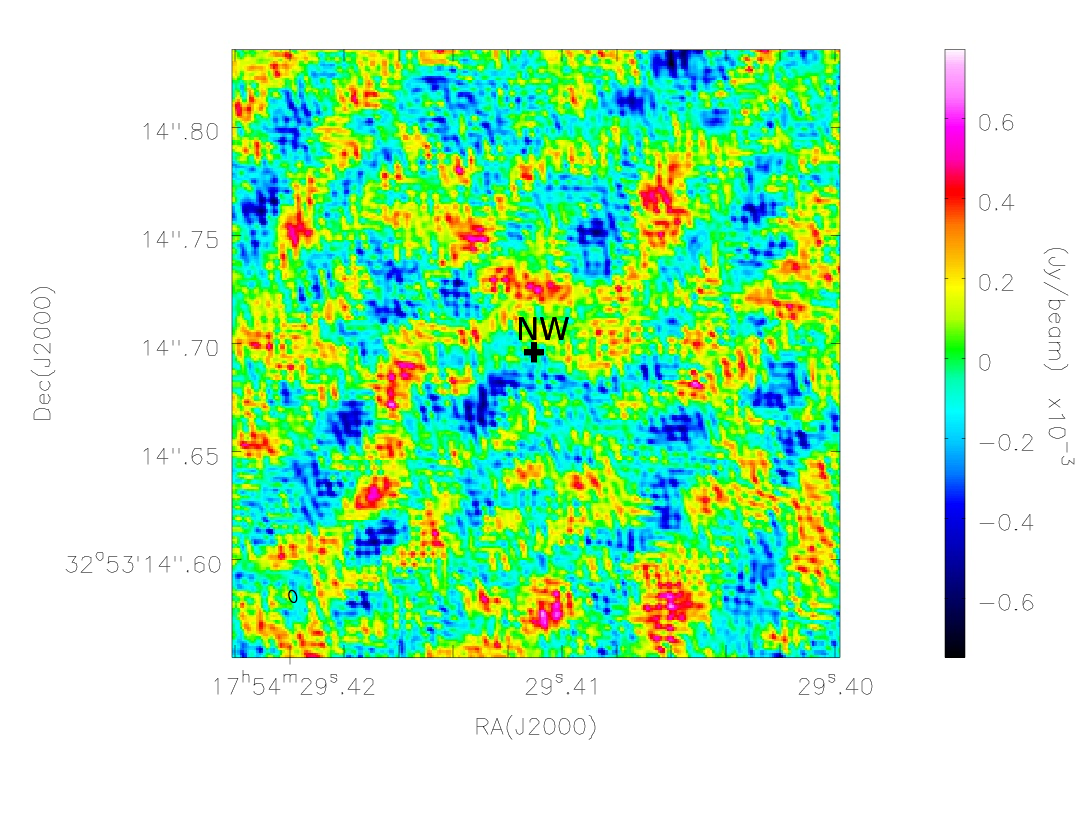}
\includegraphics[width=0.485\textwidth,height=6.2cm]{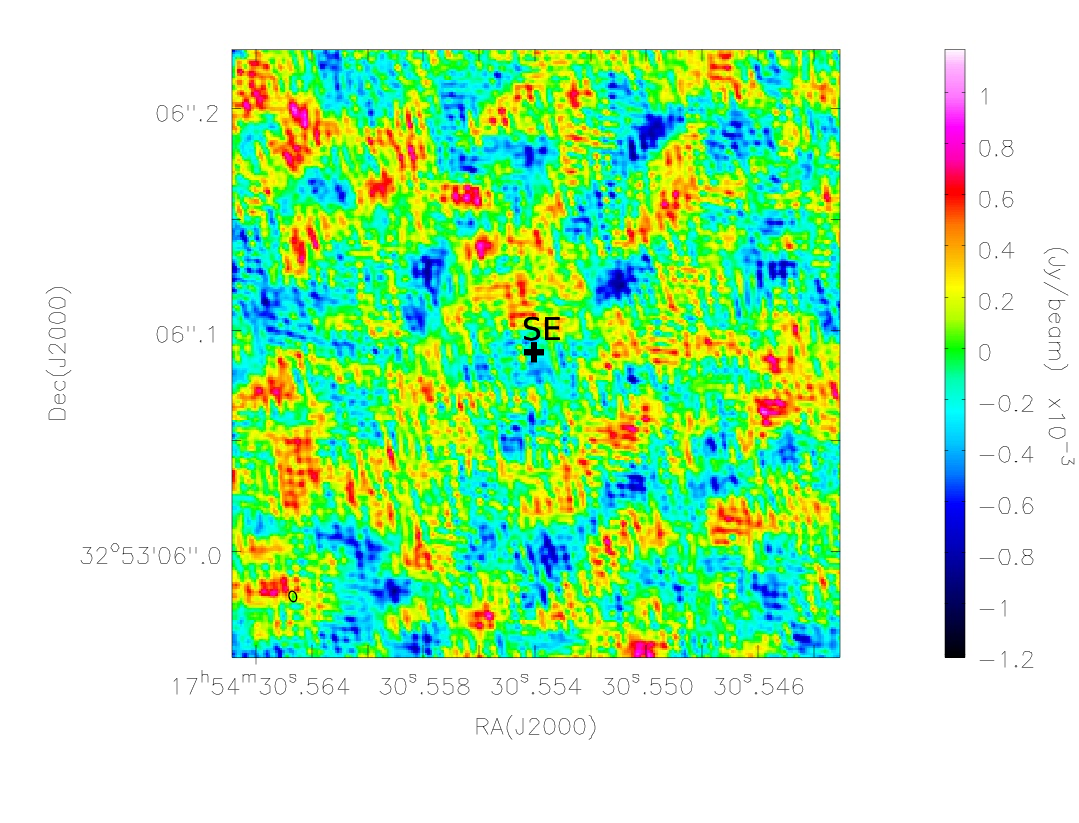}}  
  \hfill
         \caption{High resolution OH 1667/1665 MHz line emission observed by IRAS17526 EVN project. The upper two images show the velocity range of 7500 \kms in the NW and SE kernels, while the lower two images show the velocity distance of 7000 $\sim$ 8000 \kms in the NW and SE kernels. The center coordinates of NW and SE nuclei are RA: 17 54 29.411, Dec:+32 53 14.696 and RA: 17 54 30.554, Dec:+32 53 06.09, respectively.}
    \label{EVNOHdirty}%
   \end{figure*}
   
\begin{figure*}
   \centering
\includegraphics[width=9cm,height=6cm]{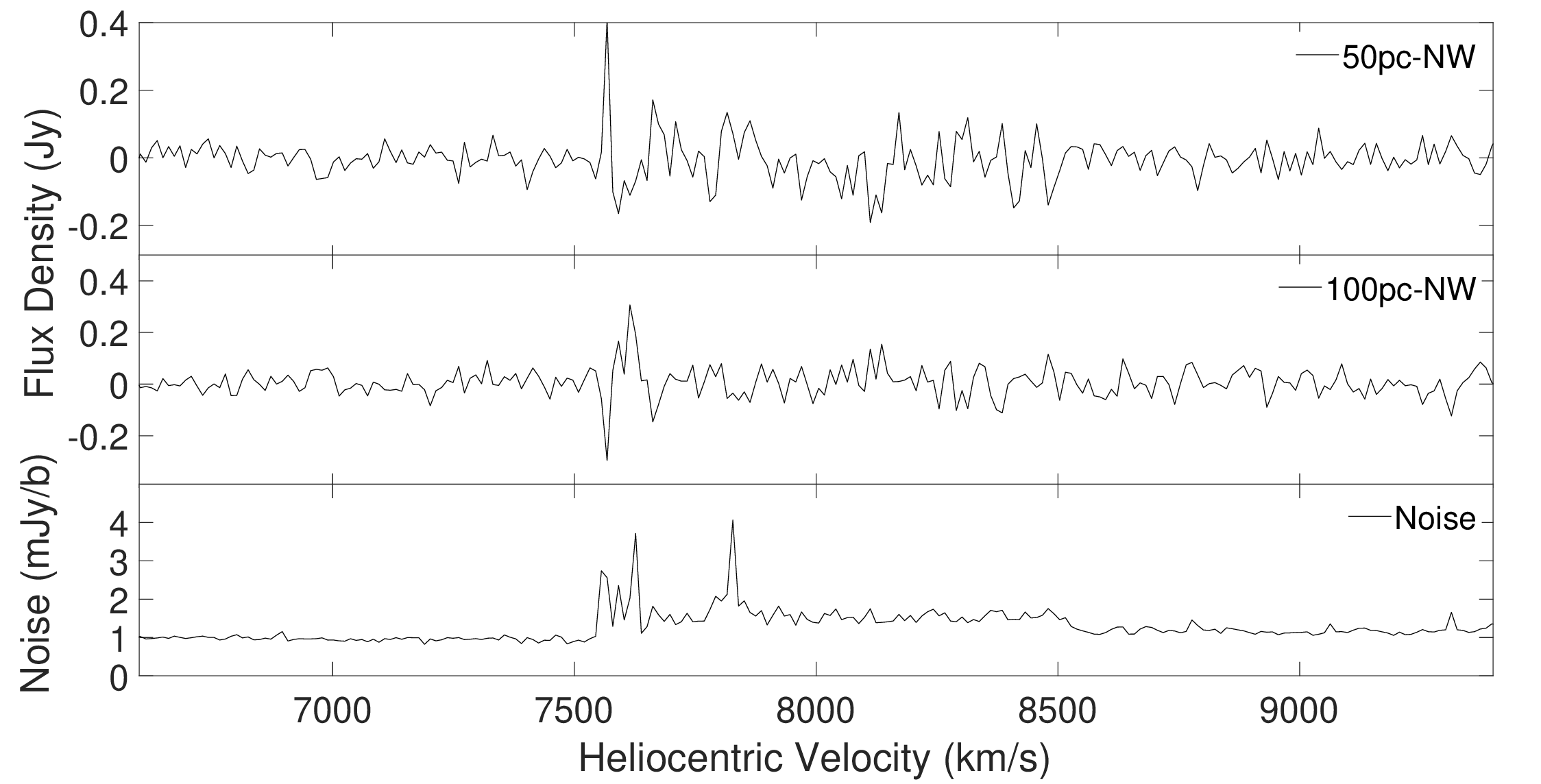}
\includegraphics[width=9cm,height=6cm]{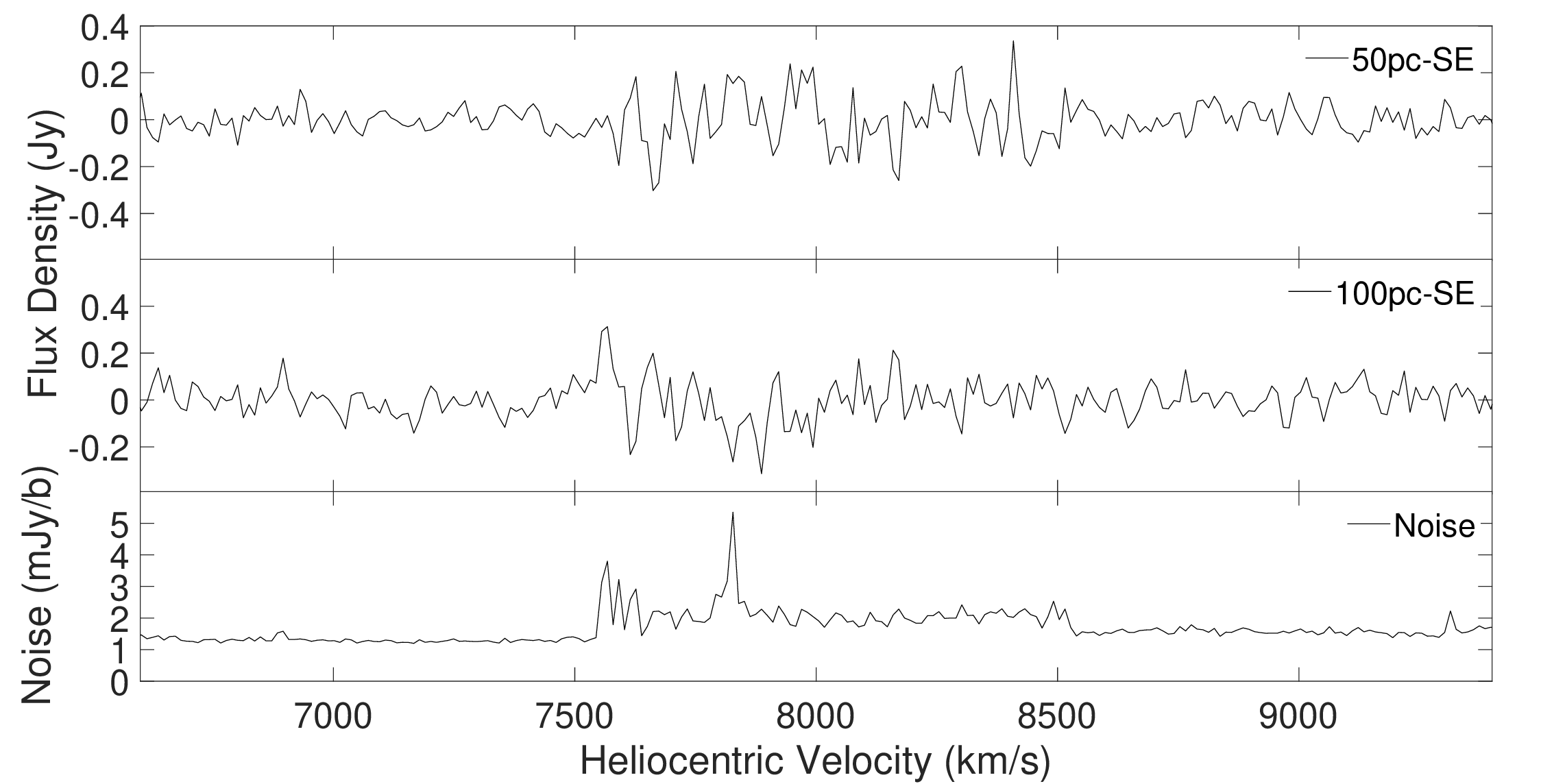}
    \caption{OH line profiles extracted from the EVN line cube image near the NW and SE nuclei. The top and middle panels show the line profiles extracted from regions approximately 50 pc and 100 pc in size, centered on the NW and SE nuclei, respectively (see Fig. \ref{EVNOHdirty}). The bottom panel displays the background 1$\sigma$ noise estimated from off-source regions for each channel image.}.
    \label{EVNOHline}%
\end{figure*}

 \begin{figure*}
   \centering
\includegraphics[width=9cm,height=7cm]{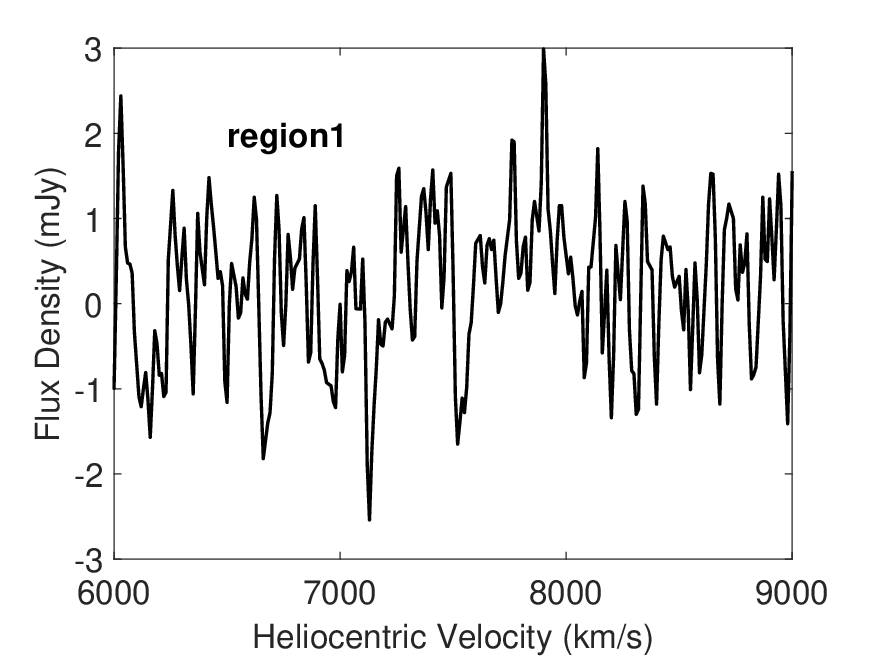}
\includegraphics[width=9cm,height=7cm]{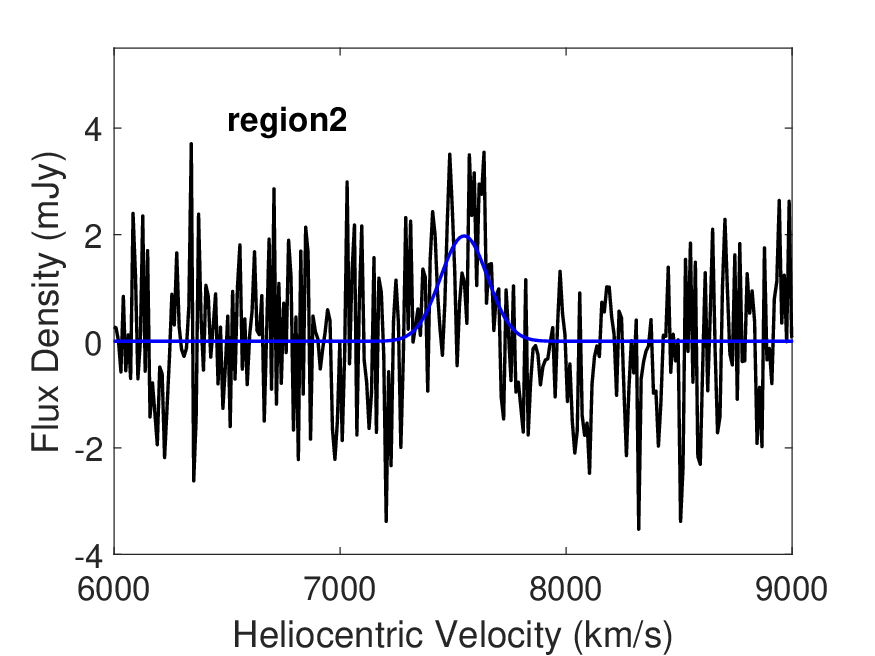}
\includegraphics[width=9cm,height=7cm]{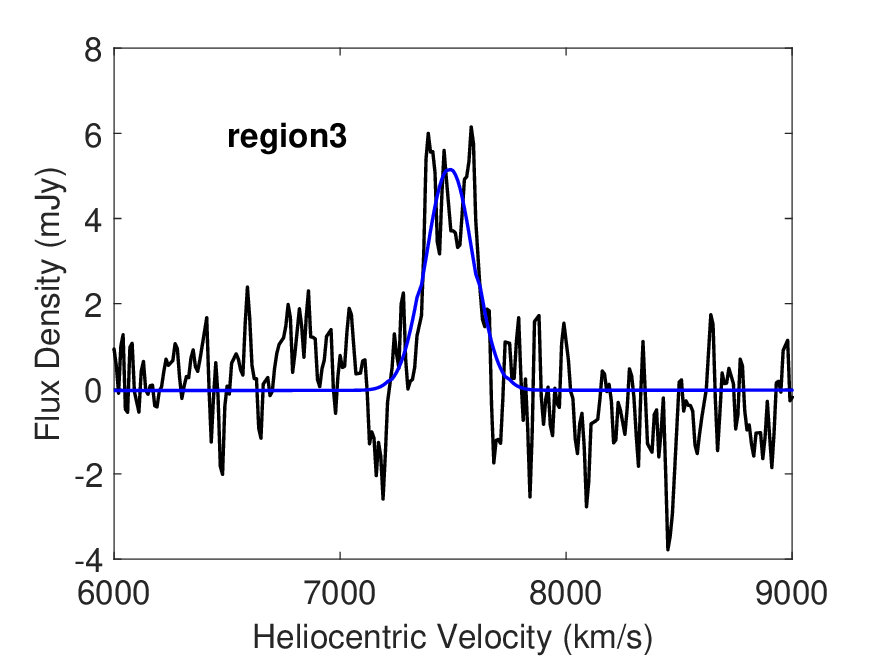}
\includegraphics[width=9cm,height=7cm]{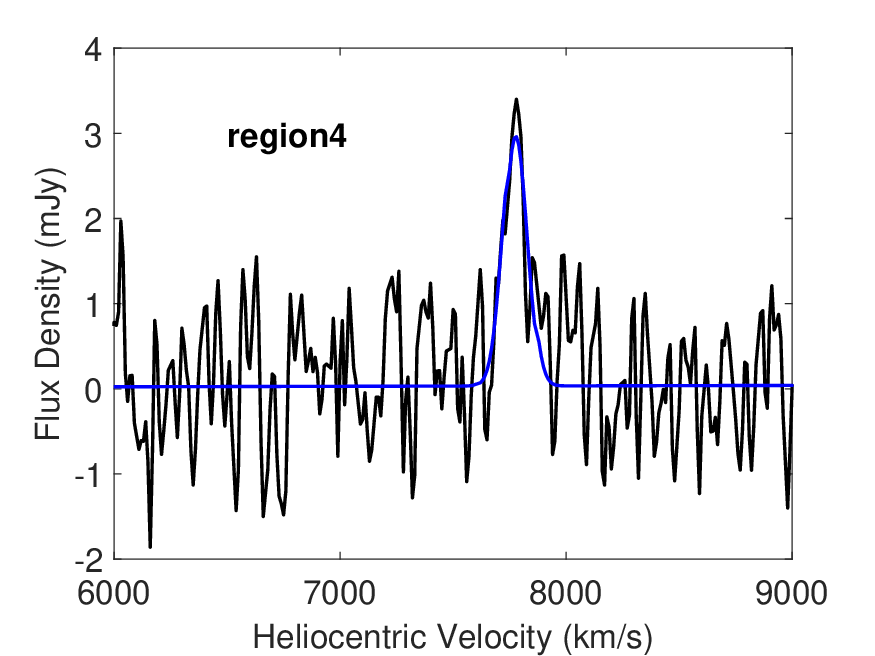}
\includegraphics[width=9cm,height=7cm]{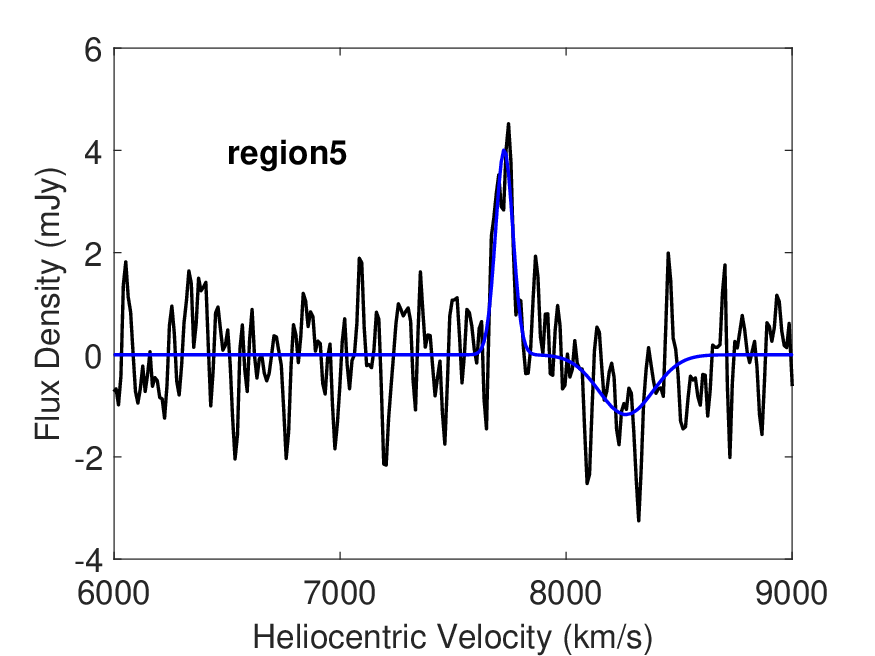}
\includegraphics[width=9cm,height=7cm]{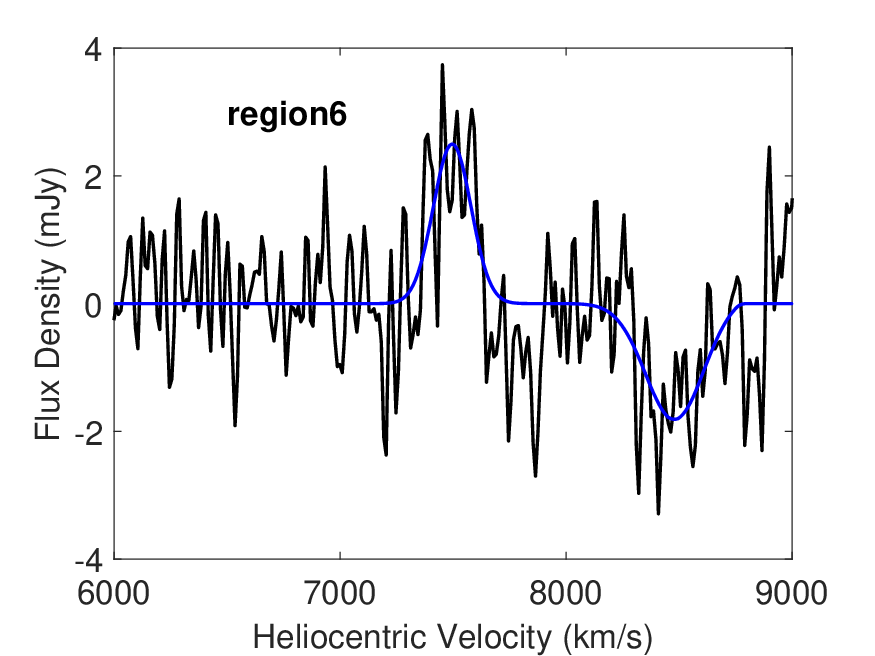}
\caption{\HI emission lines extracted from nine regions around IRAS 17526+3253 (see Fig. \ref{fig1}). Each \HI line profile was fitted with one or two Gaussian components, with the corresponding parameters listed in Table \ref{tablea2-regionsHI}. }
 \label{NS-HIFS}%
\end{figure*}
\addtocounter{figure}{-1}
 \begin{figure*}
   \centering
\includegraphics[width=9cm,height=7cm]{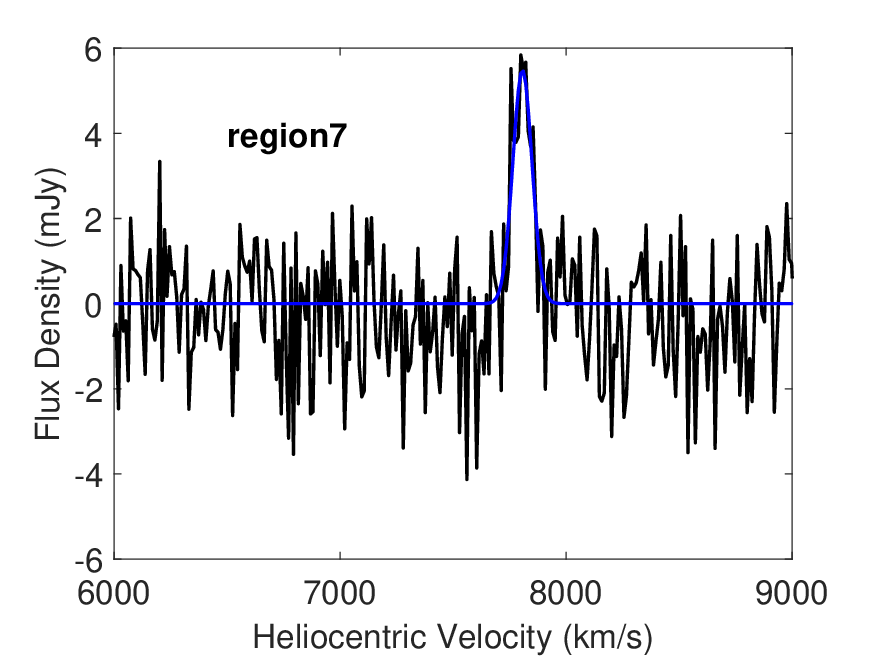}
\includegraphics[width=9cm,height=7cm]{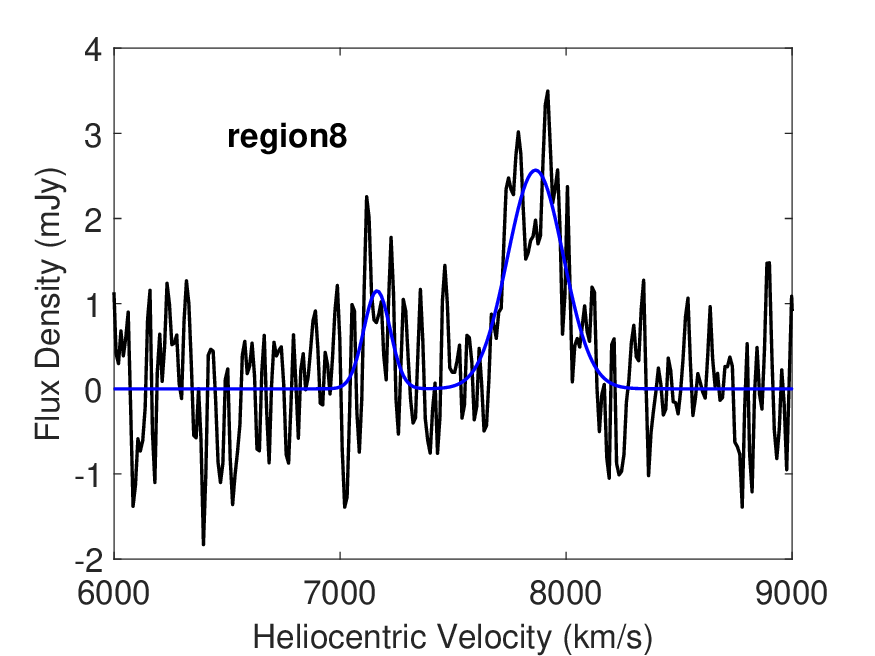}
\includegraphics[width=9cm,height=7cm]{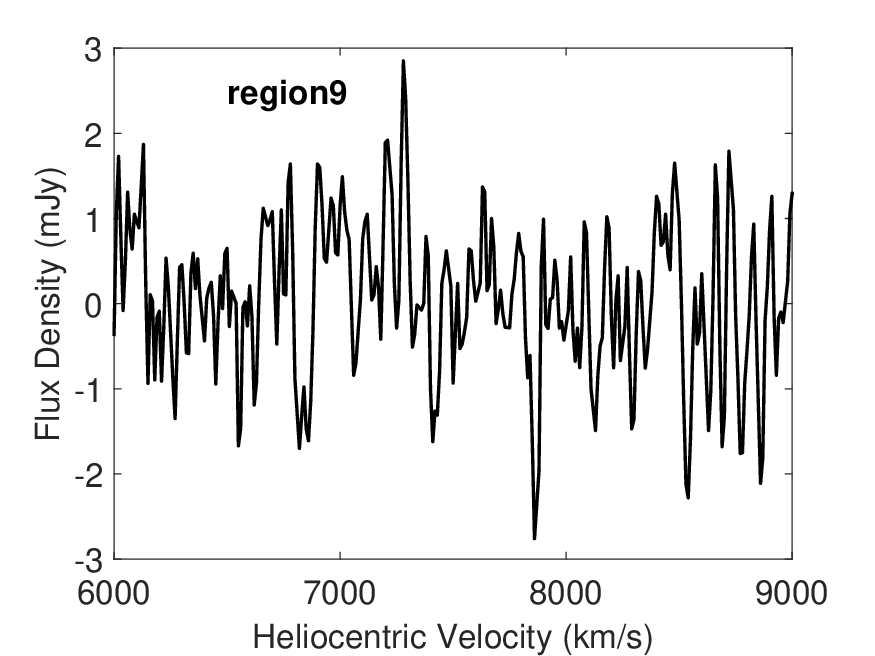}
\includegraphics[width=9cm,height=7cm]{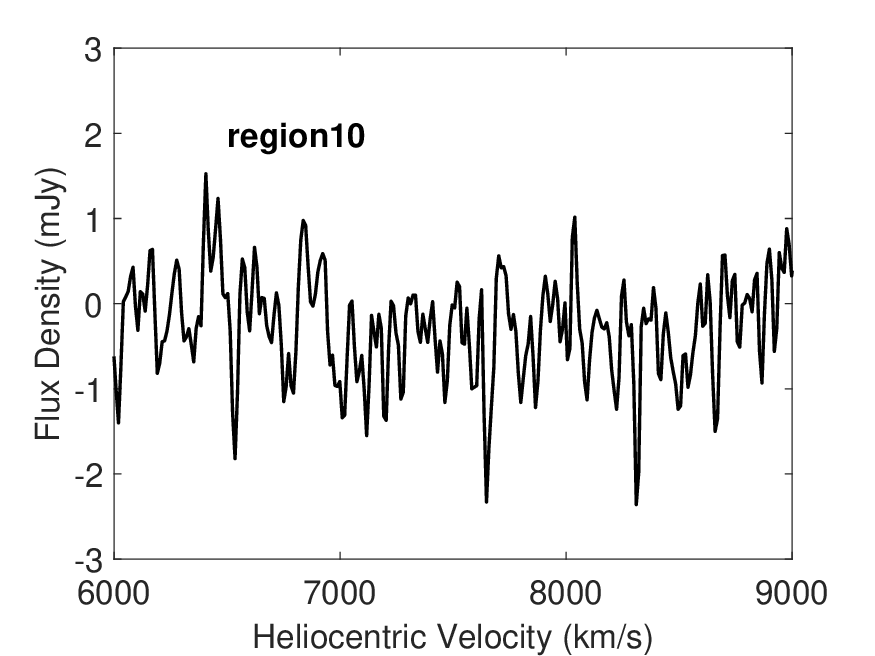}
\includegraphics[width=9cm,height=7cm]{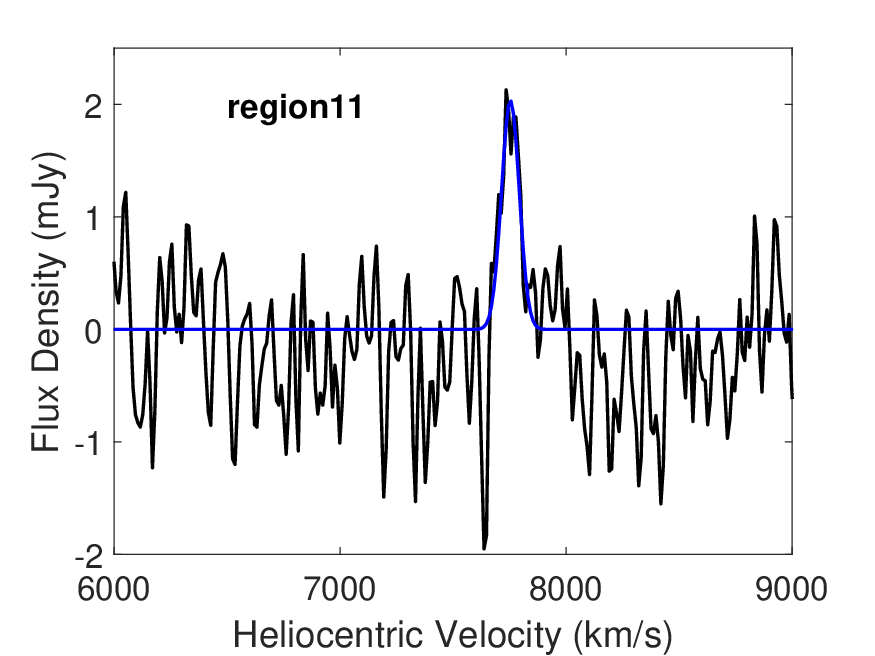}

\caption{Continued. }
\end{figure*}

 \begin{figure*}
   \centering
\includegraphics[width=9cm,height=7cm]{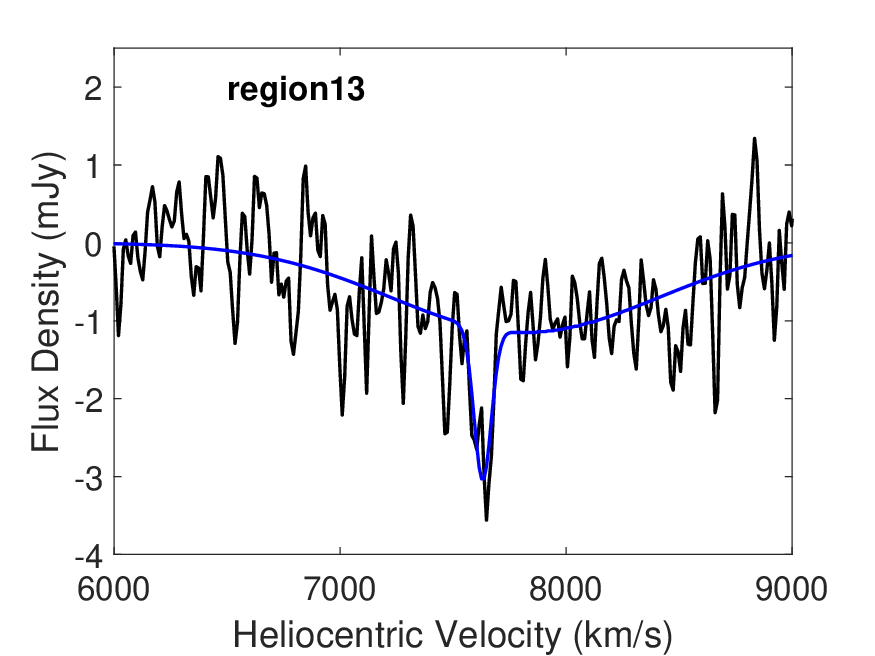}
\includegraphics[width=9cm,height=7cm]{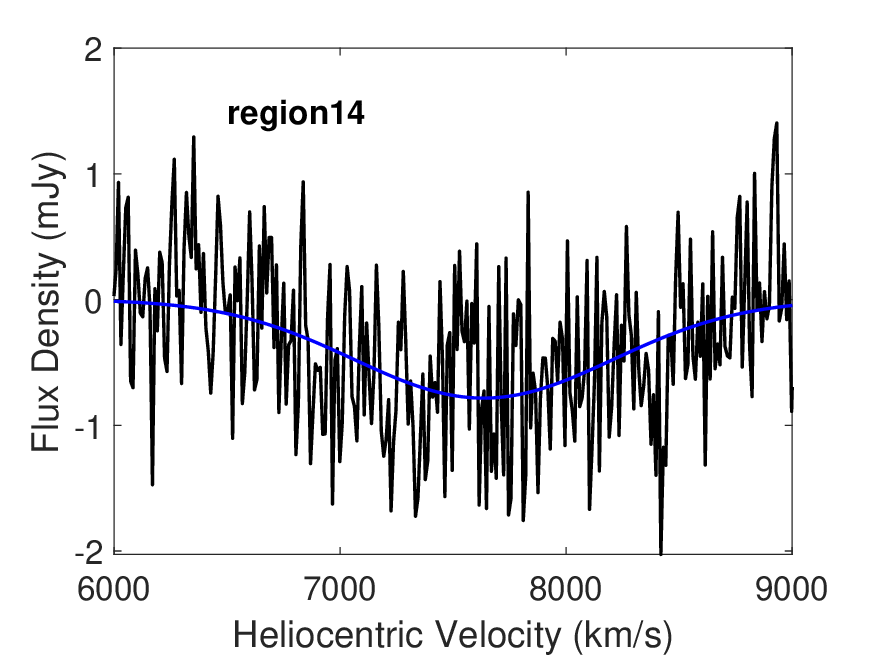}
 \caption{\HI absorption lines extracted from regions 13 and 14 (see Fig. \ref{HI-FS+XS}). The \HI line profiles were fitted with one or two Gaussian components, as detailed in Table \ref{tablea2-regionsHI}.}
 \label{HIabs_line}%
\end{figure*}

\begin{figure*}
   \centering
\includegraphics[width=9cm,height=7cm]{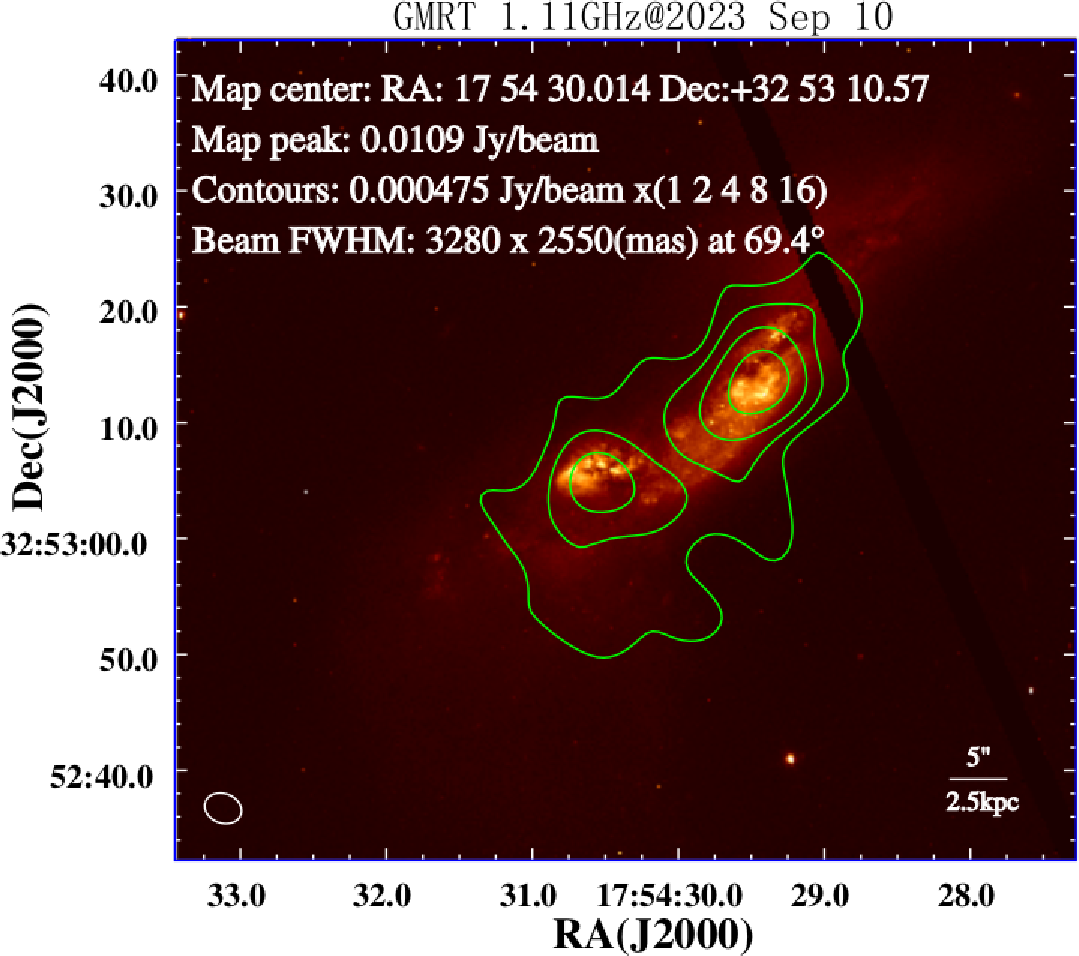}
\includegraphics[width=9cm,height=7cm]{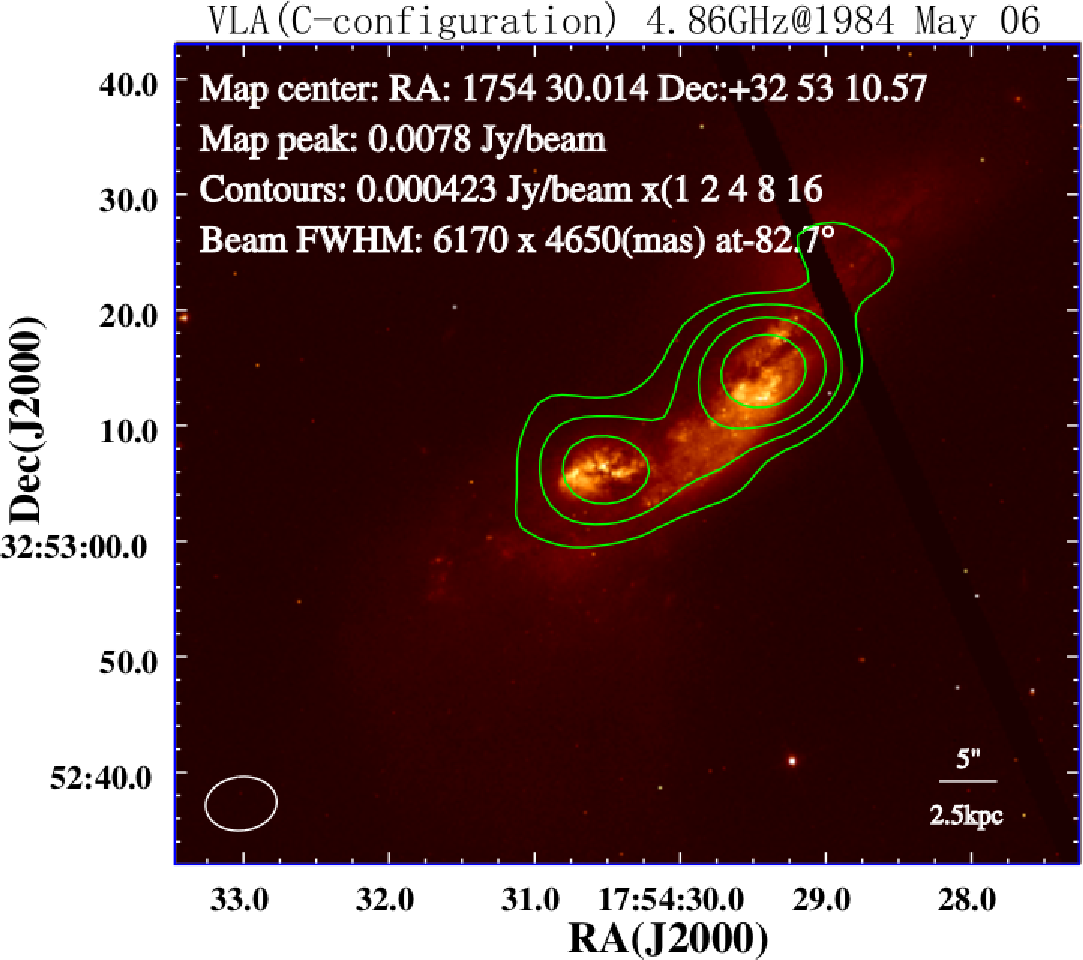}
    \caption{Multi-band radio contour maps of IRAS 17526+3253 from GMRT and VLA data overlaid on the HST/ACS F814W(I) image. The map peak intensity, contour levels, and beam FWHM are indicated in each image. The first contour in all images corresponds to a level of approximately 3 $\sigma$.}
    \label{continuum-contour2}%
\end{figure*}

 \begin{figure*}
   \centering
\includegraphics[width=9cm,height=7cm]{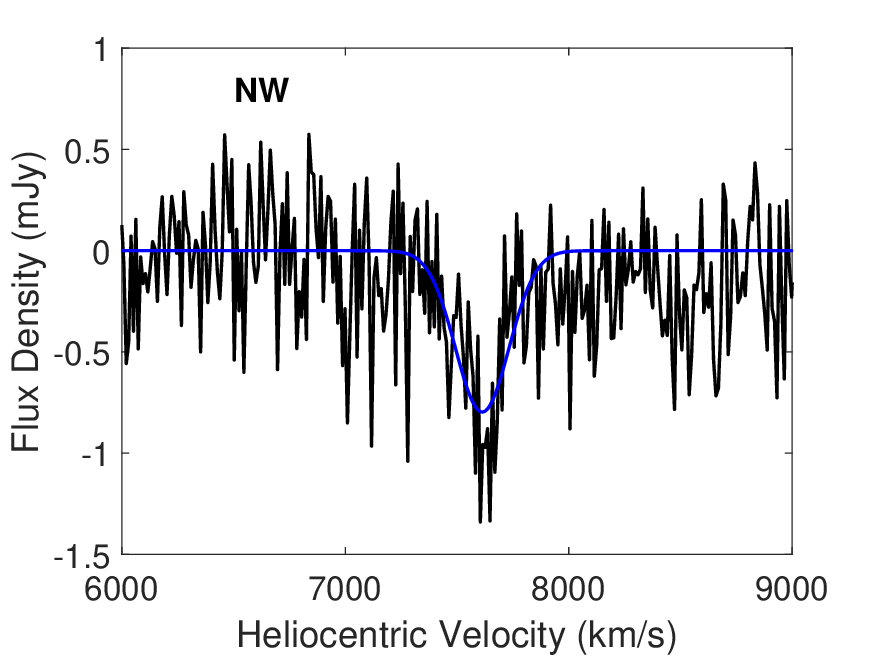}
\includegraphics[width=9cm,height=7cm]{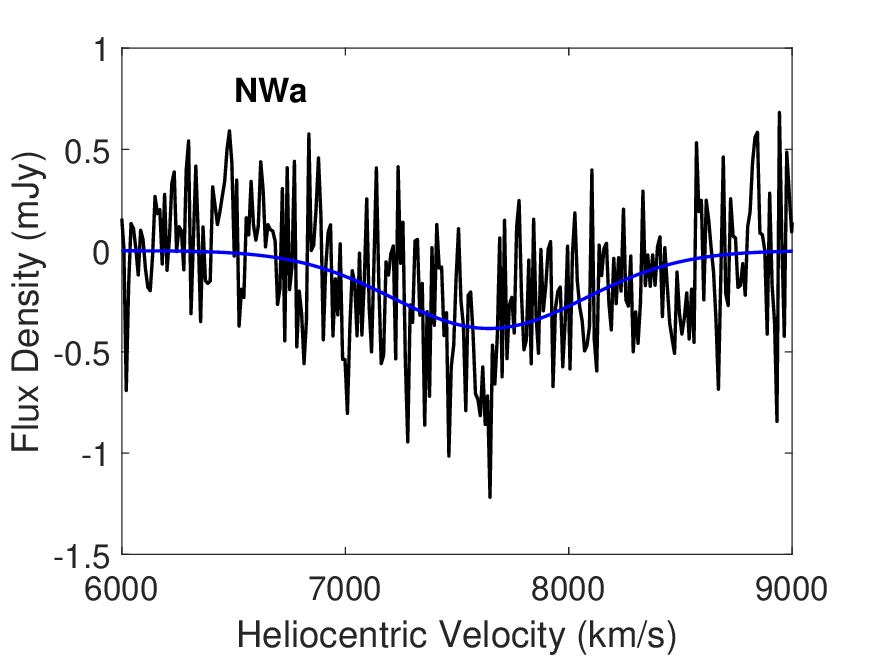}
\includegraphics[width=9cm,height=7cm]{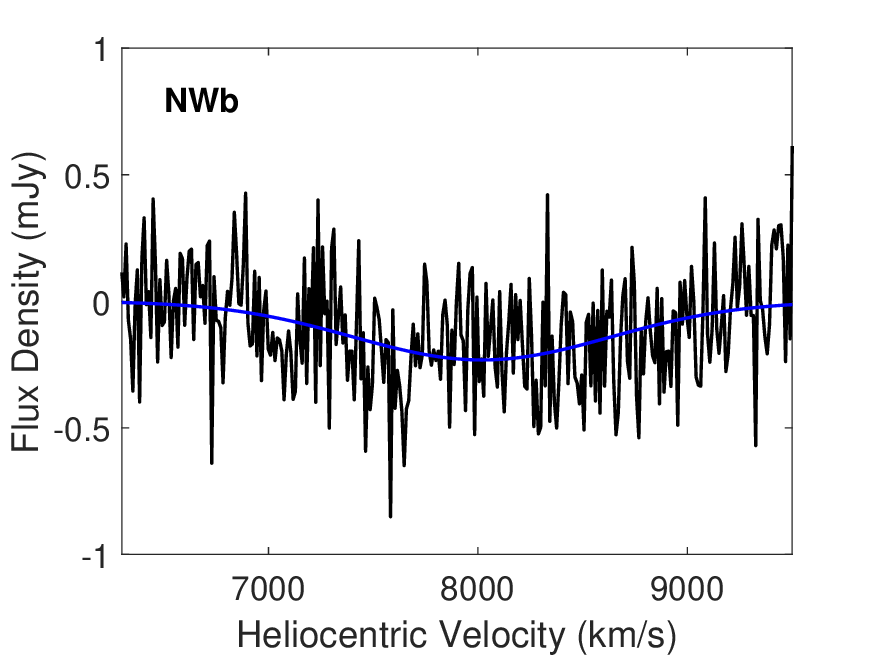}
\includegraphics[width=9cm,height=7cm]{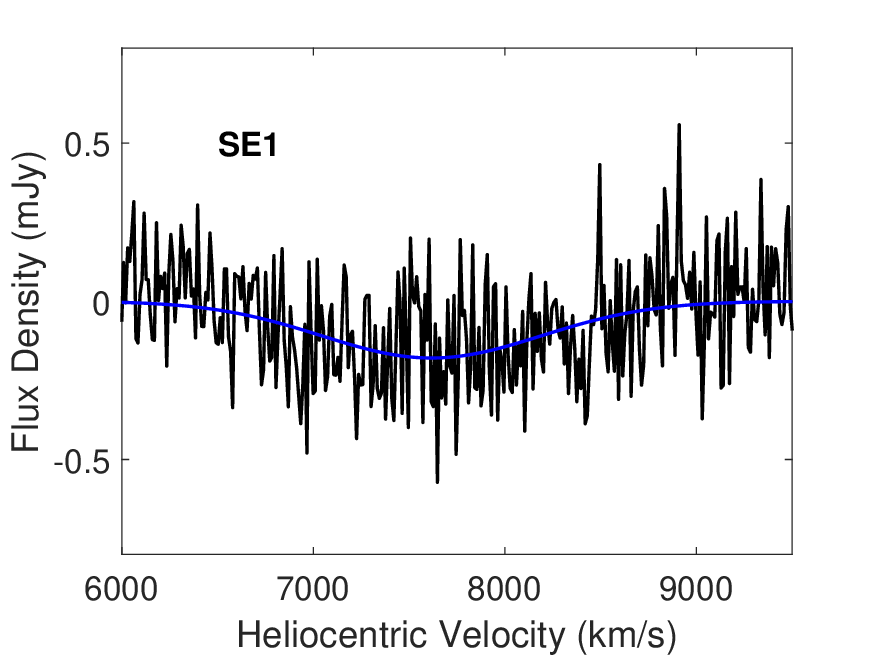}
\includegraphics[width=9cm,height=7cm]{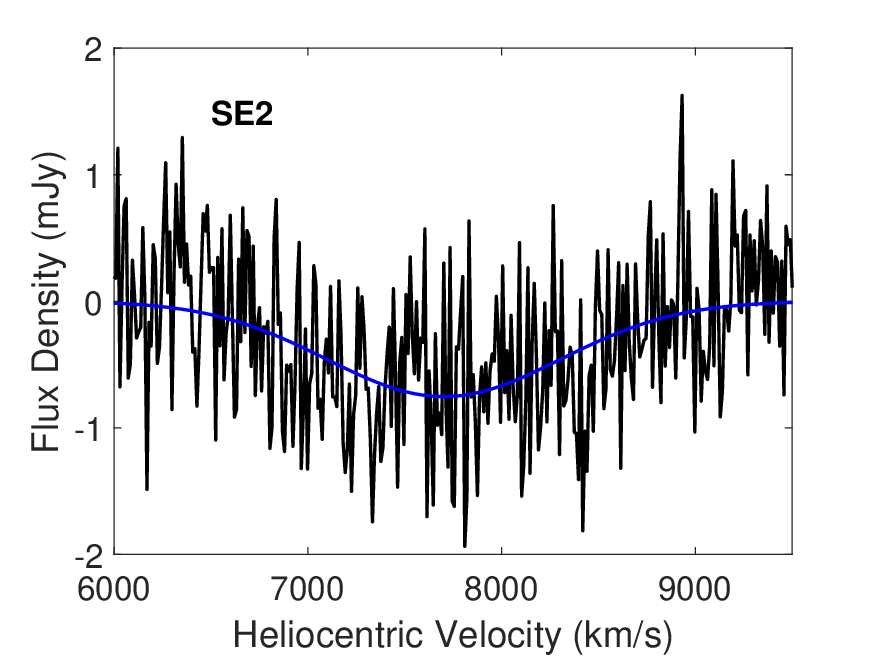}
 \caption{\HI absorption lines extracted from five regions (see Fig. \ref{momentxs}). The \HI line profiles were fitted with one or two Gaussian components, as detailed in Table \ref{tablea2-regionsHI}.}
 \label{HIabs_line2}%
\end{figure*}

\begin{figure*}
   \centering
  \hfill
        { \includegraphics[width=0.45\textwidth,height=6cm]{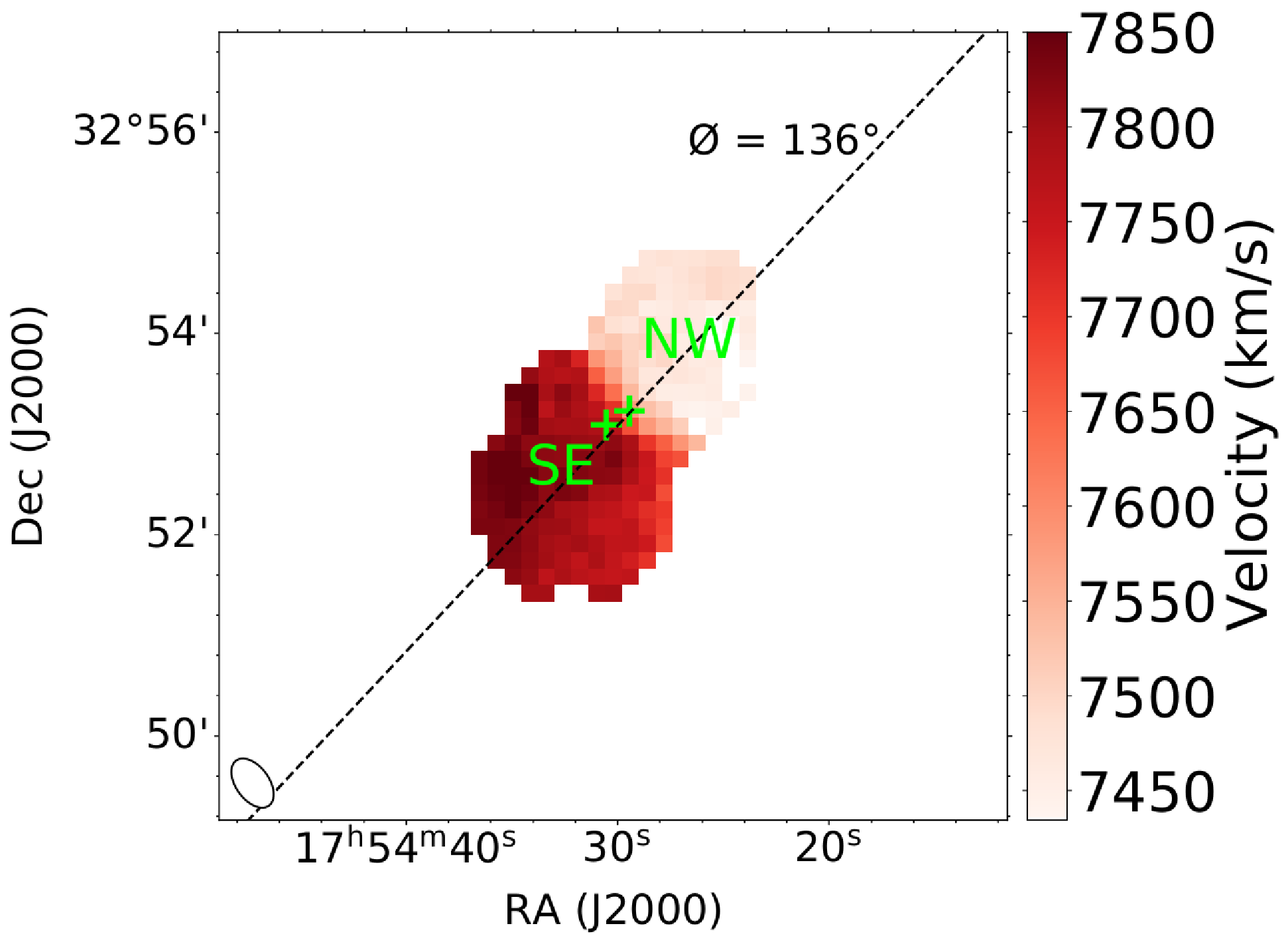}
\includegraphics[width=0.45\textwidth,height=6cm]{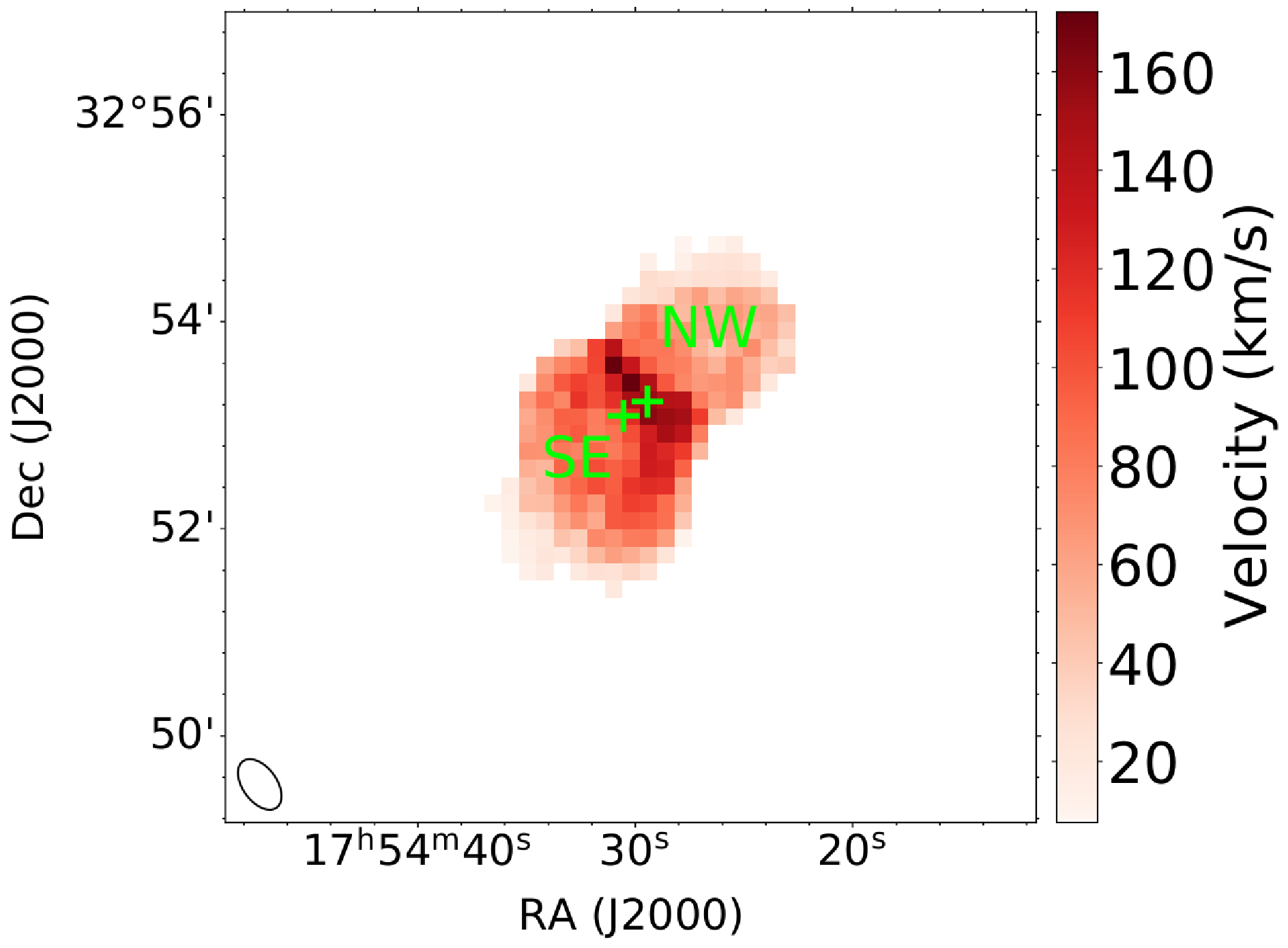}}  
 \hfill
         \caption{Velocity centroid (left) and dispersion (right) maps generated using the SoFiA software for the low-resolution \HI image. The green crosses with name 'NW' and 'SE' is same as defined in Figs. \ref{momentxs}and \ref{momentfs}. The black dashed line indicates the major axis identified by the SoFiA software for extracting the PV diagram. The corresponding dynamical center is shown in Fig.~\ref{pvdiagram}.}
    \label{momentfs2}%
   \end{figure*}

 \begin{figure*}
   \centering
 \includegraphics[width=9cm,height=7cm]{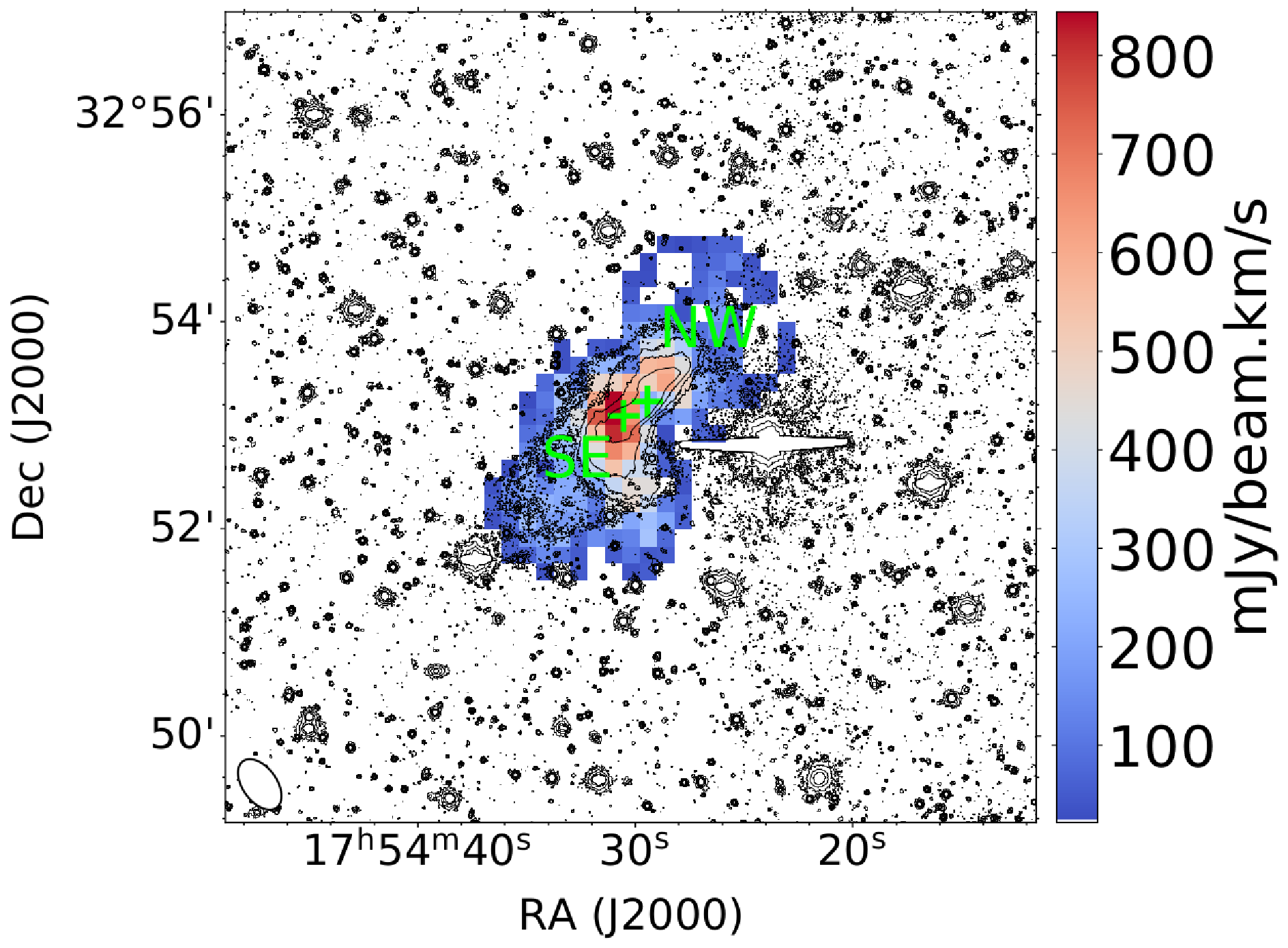}  
\includegraphics[width=9cm,height=7cm]{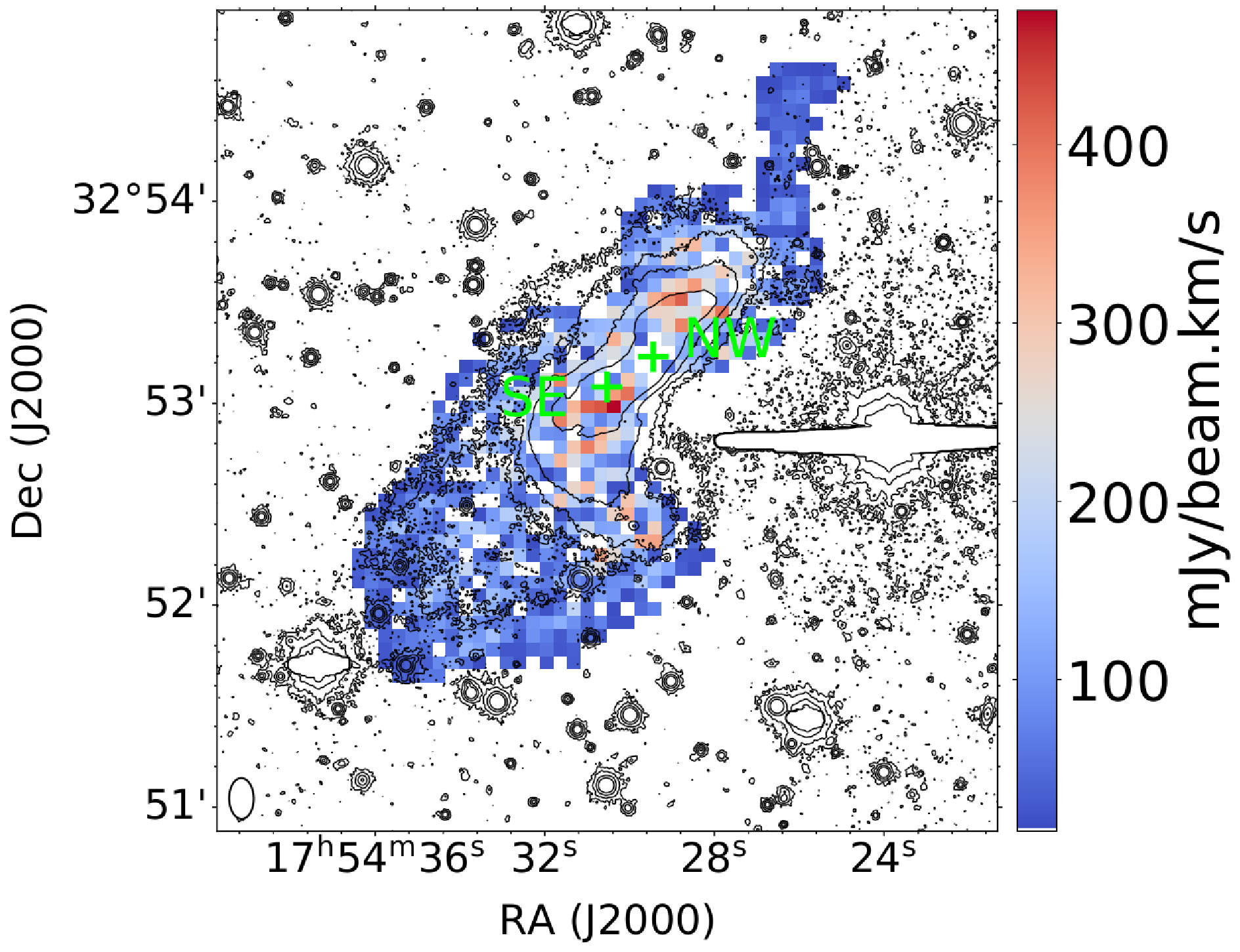}

\caption{Total intensity maps generated using SoFiA software for the low and medium-resolution \HI image. The colored images in the Left and right panel with the colorbar on the right of the two images is for the low and medium \HI images, respectively. The contour map of the two panel images are same which are generated from online DESI optical (grz) image (see section ~\hyperref[sec:discussion]{\ref*{sec:discussion}}).}
 \label{desiimg}%
\end{figure*}

\begin{figure*}
   \centering
        {\includegraphics[width=0.45\textwidth,height=6cm]{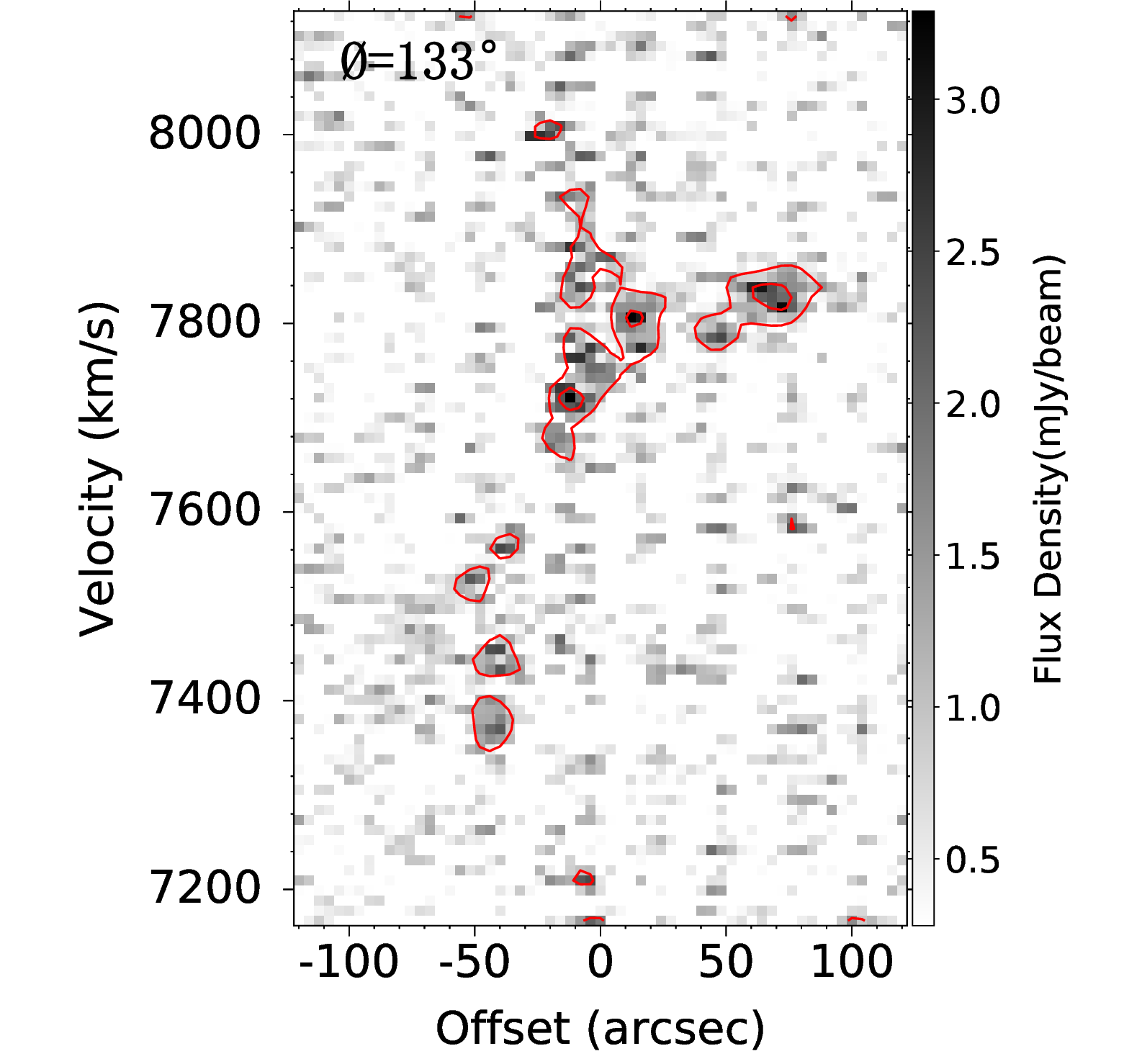} 
                \includegraphics[width=0.45\textwidth,height=6cm]{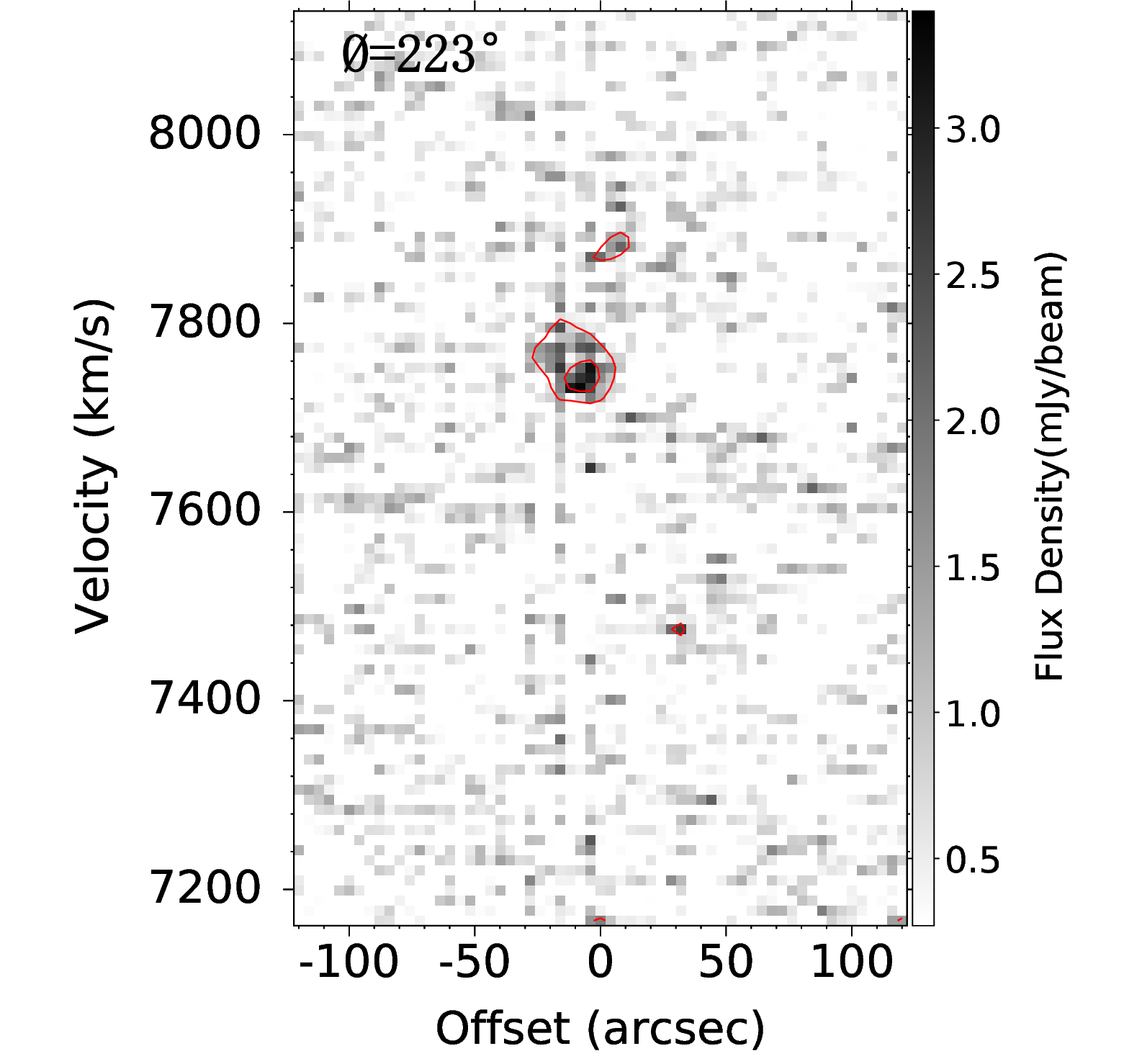} 
        \includegraphics[width=0.45\textwidth,height=6cm]{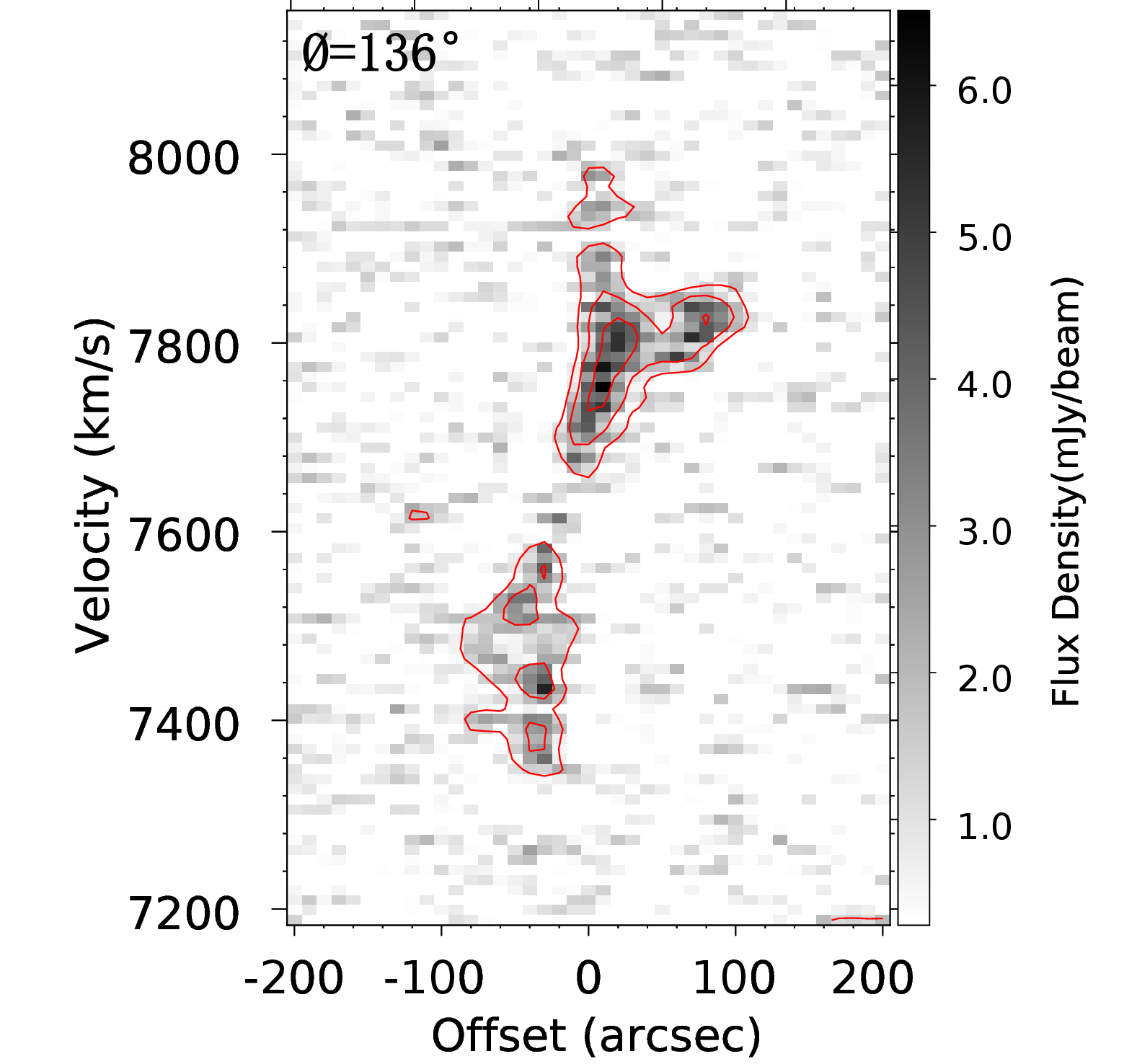}
\includegraphics[width=0.45\textwidth,height=6cm]{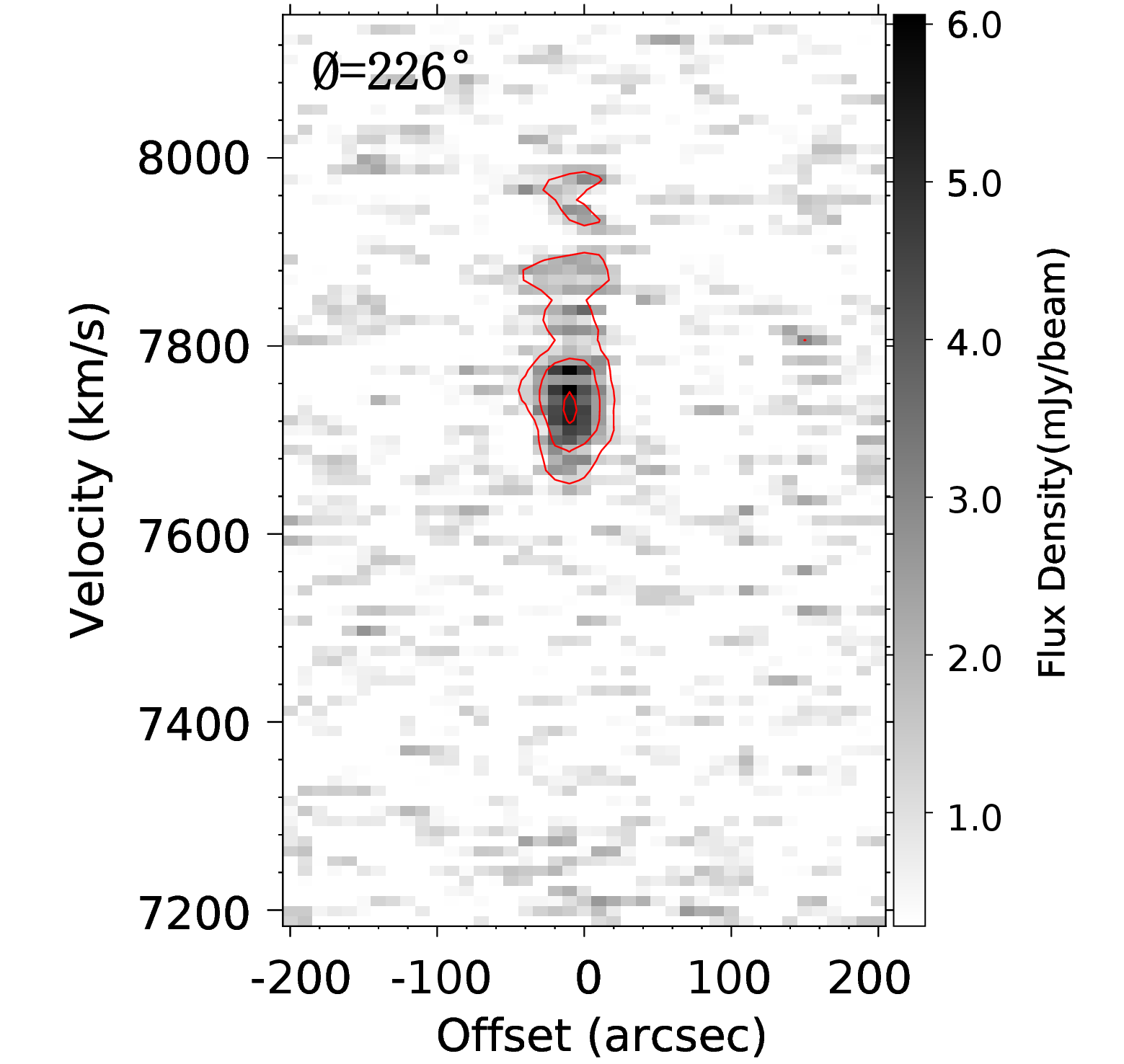}}  

         \caption{Position–velocity (PV) diagrams generated from the medium and low resolution \HI images using the Sofia software. The top two panels are based on the medium-resolution images, centered at RA = 17:54:30.277, Dec = +32:53:01.501, with position angles (PAs) of $133^\circ$ and $223^\circ$, respectively (see Fig.~\ref{momentfs}). The 1$\sigma$ noise levels are 0.58 mJy beam$^{-1}$ (left) and 0.60 mJy beam$^{-1}$ (right), with contour levels at 1.5$\sigma$ and 3$\sigma$. The bottom two panels are from the low-resolution image, centered at RA = 17:54:30.742, Dec = +32:52:55.029, with PAs of $136^\circ$ and $226^\circ$ (see Fig.~\ref{momentfs2}). The 1$\sigma$ noise levels are 0.66 mJy beam$^{-1}$ (left) and 0.84 mJy beam$^{-1}$ (right), with contour levels at 1.5$\sigma$, 3$\sigma$, and 5$\sigma$. }
    \label{pvdiagram}%
   \end{figure*}
\begin{figure*}
   \centering
\includegraphics[width=9cm,height=7cm]{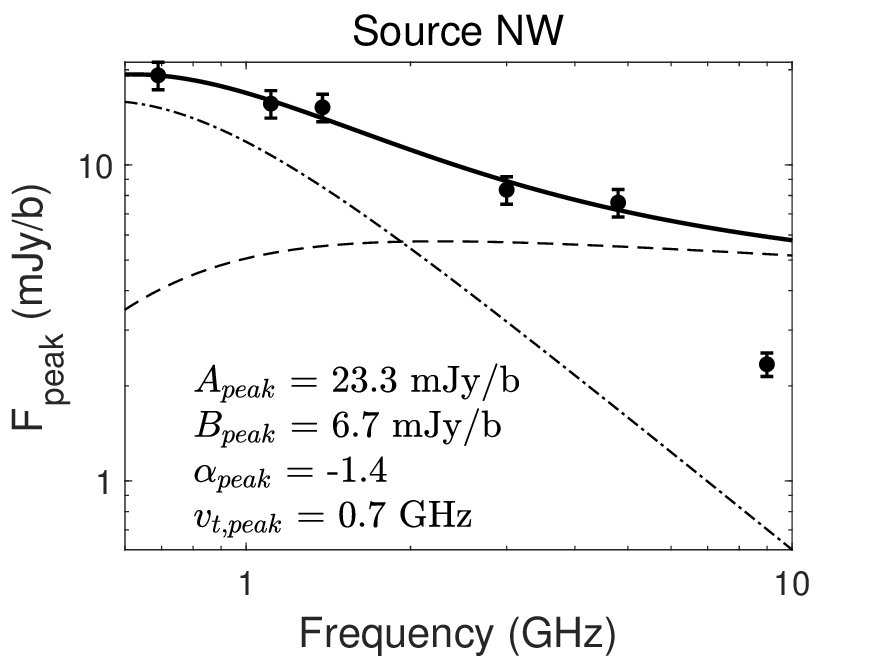}
\includegraphics[width=9cm,height=7cm]{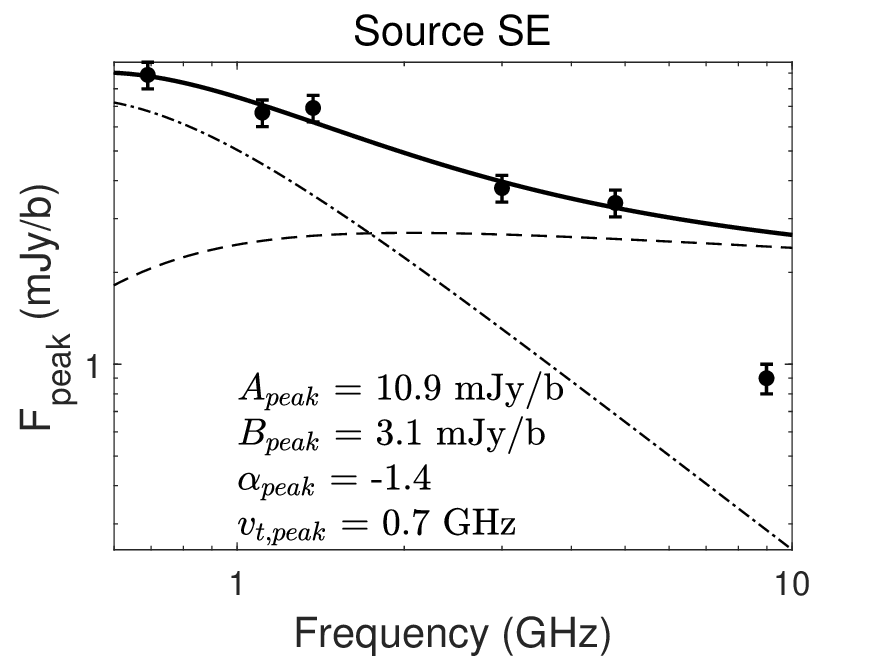}
\includegraphics[width=9cm,height=7cm]{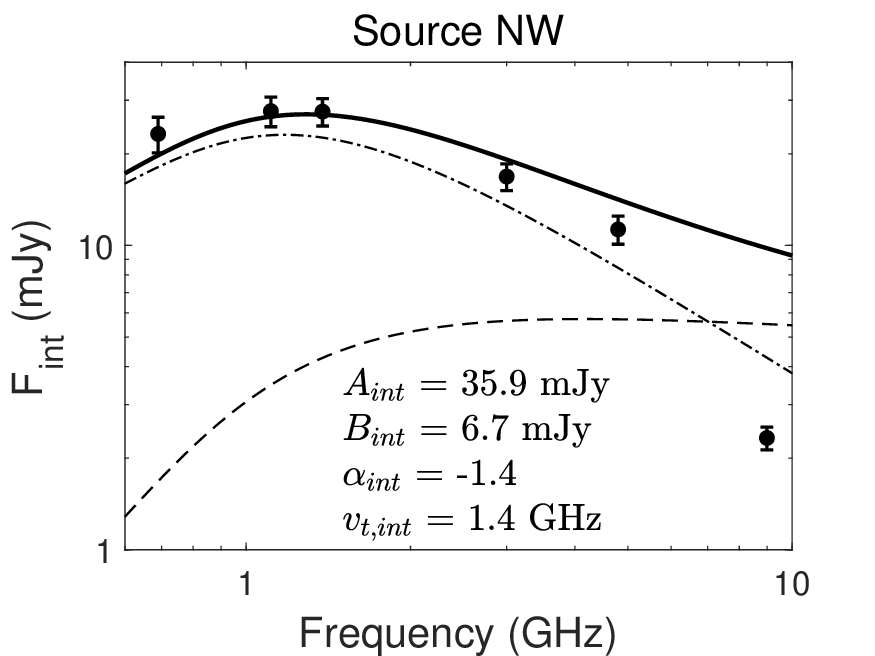}
\includegraphics[width=9cm,height=7cm]{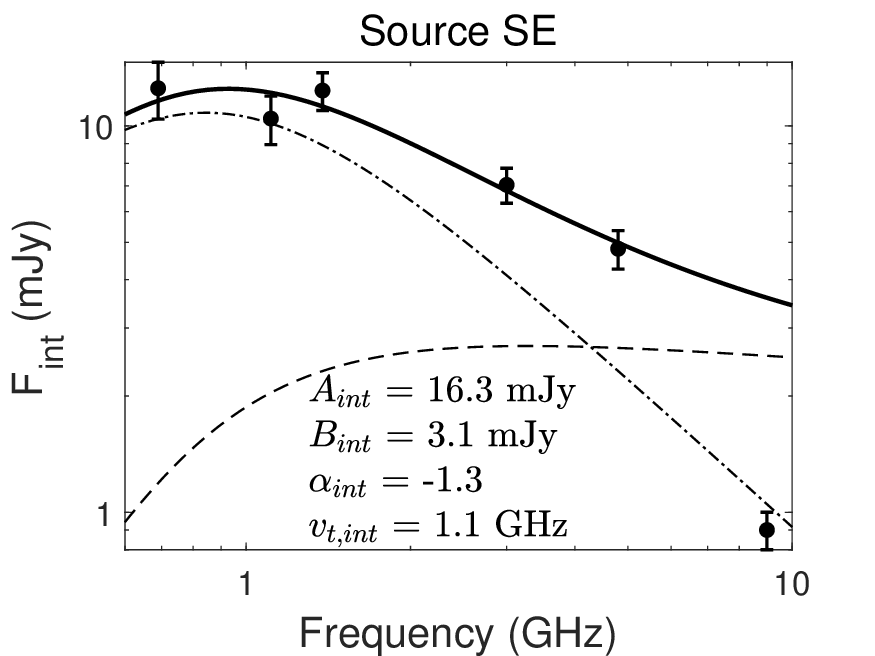}
\caption{Multi-band peak and integrated radio flux densities for the NW and SE nuclei of IRAS 17526+3253, based on VLA and GMRT observations listed in Table \ref{tablea1:multibandata}. The top two panels show the peak flux densities for the NW and SE nuclei, respectively, while the bottom two panels present the integrated flux densities. In each panel, the solid line represents the best-fitting model described by Equation \ref{eq2}. The dashed and dot-dashed lines indicate the thermal (free-free) and non-thermal (synchrotron) components, respectively. During the fitting process, the spectral index $\alpha$ was constrained to lie between -1.4 and -0.5 \citep{2018MNRAS.474..779G}.}
    \label{peak}%
\end{figure*}

  \setlength{\tabcolsep}{0.01in}
  \begin{table*}
       \caption{Parameters of the high-resolution radio continuum observations.}
     \label{tablea1:multibandata}
  \begin{center}
  \begin{tabular}{c c c c l c c c c c c c}     
  \hline\hline
   Epoch& Freq. & Array          & Phase      & Program     & beam  &PA         &rms(a)     &rms(b)  &Comp. &$F_{int}$ &$F_{peak}$  \\
                   &   (GHz)     &  &  Calibrator&& (") $\times$(")    &($\circ$)  &(mJy/beam) &(mJy/beam)  &          &   (mJy)        &(mJy/beam) \\
      \hline
    2023Sep11      & 0.69       & GMRT  &J1753+2848 & 44048  & 5.0$\times$4.1 &53 & 0.35& 0.45         &NW           &    23 $\pm$ 3     & 19   $\pm$ 2   \\
                   &            &       &           &        &                &   &           &   &SE           &    13 $\pm$ 2     &  8.9 $\pm$ 0.9   \\      
    2023Sep10      & 1.11       & GMRT  &...        & 44048  & 3.3$\times$2.6 &70 & 0.16& 0.18         &NW           &    28 $\pm$ 3     & 15   $\pm$ 2   \\
                   &            &       &           &        &                &   &         &     &SE           &    10 $\pm$ 2     & 6.7  $\pm$ 0.6   \\     
    2023Sep10      & 1.38       & GMRT  &...        & 44048  & 2.6$\times$2.2 &62 & 0.05& 0.06         &NW           &    28 $\pm$ 3     &    15$\pm$ 2   \\
                   &            &       &           &        &                &   &       &       &SE           &    12 $\pm$ 1     & 6.9  $\pm$ 0.7   \\  
    2016Oct21      & 3.00       & VLASS &-          & -      & 2.7$\times$2.3 &-86& 0.08& -            &NW      & 16 $\pm$ 1     & 8.3$\rm ^{*}$ $\pm$ 0.8   \\
                   &            &       &           &        &                &   &         &     &SE           & 7.0 $\pm$ 0.7 & 3.8$\rm ^{*}$  $\pm$ 0.4\\  
    1984May06      & 4.86       & VLA-C &1732+389   & AB275  & 6.1$\times$4.6 &83 & 0.14& 0.12         &NW           &     11$\pm$ 1     & 7.6  $\pm$ 0.8   \\
                   &            &       &           &        &                &   &        &      &SE           &  4.8  $\pm$ 0.6   & 3.4  $\pm$ 0.3  \\
    2016Oct21      & 8.99       & EVLA-A&J1753+2848 & 16B-063&0.27$\times$0.26&75 & 0.01& 0.01         &NW           &  2.2  $\pm$ 0.2   &  2.1  $\pm$ 0.1  \\
                  &            &       &           &        &                &   &          &    &SE           &   0.9  $\pm$ 0.1   &  0.9 $\pm$ 0.1  \\

  \hline

       \end{tabular}\\ 
    \end{center}
     \vskip 0.1 true cm \noindent Column (4): The phase calibrator for GMRT and VLA projects. Column (5): The program name. Columns (6) and (7): The beam FWHM and position angle. Column (8): The 1$\sigma$ noise level for the radio continuum image shown in Figs. \ref{continuum-contour1} and \ref{continuum-contour2} before restoring the beam. Column (9): The 1$\sigma$ noise level for the radio continuum image shown in Figs. \ref{continuum-contour1} and \ref{continuum-contour2} after restoring the beam. Column (10): The component name.  Columns (11)-(12): The integrated and peak flux densities measured from restored images with a beam size of 5" $\times$ 5". The integrated flux densities were calculated by summing the flux within circular regions of approximately 15" and 12" for the NW and SE nuclei, respectively, using the 'imstat' task in CASA (mean intensity multiplied by the number of beam areas). The $F_{peak}$ is the maximum flux density shown in the image. An asterisk (*) indicates that the peak flux was calculated by summing the total flux within a 5" $\times$ 5" region, as we could not restore the beam size of the VLASS image. The error is calculated as: $\sqrt{(N \times \sigma^2) + (0.1 \times S_{peak})^2}$, where N is the number of beams within the 3$\sigma$ region.
    \end{table*}


\begin{table*}
       \caption{Parameters of the \HI emission line regions in IRAS 17526+3253}
     \label{tablea2-regionsHI}
  \begin{center}
  \begin{tabular}{c c c c l c c c c}     
  \hline\hline
 Regions(HI) & Position (J2000)       &  Gauss amplitude     & Gauss center          & Gauss FWHM           & Gaussian area     \\
             &                             & mJy                  & \kms                  & \kms                 & mJy*\kms           \\
  \hline
     region 1   
             & 175431.678+325333.794   &  -         & -      & -       & -      \\

     region 2    
             &175430.173+325333.794    &  2.1 $\pm$ 0.2       & 7554 $\pm$ 17      & 277 $\pm$ 40       & 630 $\pm$ 86     \\

     region 3    
             &  175428.644+325333.794  &  5.1 $\pm$ 0.3         & 7484 $\pm$ 8      & 250  $\pm$ 19      &  1381$\pm$96    \\
     
     region 4    
             &  175431.678+325313.937  &  2.9 $\pm$ 0.3         & 7775 $\pm$ 7     & 127 $\pm$ 16      & 397 $\pm$ 48     \\

     region 5     
             & 175430.173+325313.937   &  4.0 $\pm$ 0.4         & 7726 $\pm$ 4      & 90 $\pm$ 11      & 384 $\pm$ 40     \\
             &   &  -1.1 $\pm$ 0.2         & 8265 $\pm$ 28      & 282 $\pm$ 67      & -351 $\pm$ 72     \\
     region 6     
             & 175428.644+325313.937   & 2.5 $\pm$ 0.3         & 7496 $\pm$ 12      & 196$\pm$ 29      & 522$\pm$ 68    \\
             &    & -1.8 $\pm$ 0.3         & 8480 $\pm$ 22      & 302 $\pm$ 53      & -597 $\pm$ 101     \\
             
     region 7    
             & 175431.678+325253.931   &  5.4 $\pm$ 0.4         & 7807 $\pm$ 4      & 100  $\pm$ 8      & 582 $\pm$ 42     \\

     region 8    
             & 175430.173+325253.931   &  1.1 $\pm$ 0.5         & 7163 $\pm$ 16      & 137  $\pm$ 39      & 168 $\pm$ 41     \\
             &    &  2.5 $\pm$ 0.2         & 7865 $\pm$ 11      & 290  $\pm$ 25      & 793 $\pm$ 60    \\
     region 9   
             & 175428.644+325253.931   & -         & -      & -      & -     \\
     region 10    
             & 175429.444+325313.864   & -         & -      & -      & -     \\
     region 11    
             & 175430.538+325305.532   &  2.1 $\pm$ 0.3         & 7753 $\pm$ 6     & 92  $\pm$ 15     & 200 $\pm$ 28     \\
     region 12    
             & 175431.208+325253.885    &  4.6 $\pm$ 1.5         & 7495 $\pm$ 26      & 141  $\pm$ 61     & 692 $\pm$ 267     \\
             &   &  11.4 $\pm$ 1.2         & 7806 $\pm$ 13      & 244  $\pm$ 34     & 2976 $\pm$ 343    \\
     region 13     
             & 175429.426+325313.756   &  -1.1 $\pm$ 0.1         & 7818 $\pm$ 5      & 1402  $\pm$ 123      & -1716 $\pm$ 130     \\
             &    &  -1.9 $\pm$ 0.3         & 7631 $\pm$ 7      & 89  $\pm$ 19      & -184 $\pm$ 37     \\
    region 14    &  175430.528+325306.597   &   -0.78 $\pm$ 0.08         &  7635 $\pm$ 72     &  1353  $\pm$ 169      &  -1128 $\pm$ 122     \\

      NW    
             & 175429.414+325314.696       &  -0.61 $\pm$ 0.08         & 7612 $\pm$ 17     & 249  $\pm$ 40        & -162 $\pm$ 22     \\
      NWa    
             & 175429.400+325312.800       &  -0.29 $\pm$ 0.04         & 7643 $\pm$ 67     & 1016  $\pm$ 157      & -319 $\pm$ 43     \\
      NWb    
             & 175429.590+325312.744       &  -0.23 $\pm$ 0.03         & 8019 $\pm$ 93     & 1452  $\pm$ 218      & -358 $\pm$ 47     \\
      SE1    
             & 175430.409+325307.487       &  -0.18 $\pm$ 0.02         & 7617 $\pm$ 82    & 1354  $\pm$ 195    & -258 $\pm$ 32     \\
      SE2    
             & 175430.614+325306.452       &  -0.24 $\pm$ 0.03         & 7796 $\pm$ 78     & 1436  $\pm$ 184      & -370 $\pm$ 41     \\

  \hline
  \end{tabular}\\

    \end{center}
    \vskip 0.1 true cm \noindent Columns (1): the emission components; (2): RA and Dec of the components; Columns (3)-(6) The Gaussian-fitted model parameters.
    \end{table*}




\end{appendix}

\end{document}